\pdfoutput=1
\documentclass[3p,12pt]{elsarticle}
\usepackage{amsmath,amssymb,bm,booktabs,adjustbox}
\usepackage{array}
\usepackage[T1]{fontenc}
\usepackage{graphicx,placeins,tikz}
\usetikzlibrary{shapes.geometric,arrows.meta,positioning,calc,fit,backgrounds}
\tikzset{
  flow/.style={font=\footnotesize, node distance=3mm and 4mm, >={Stealth[length=2mm]},
    execute at begin node={\thickmuskip=5mu\relax\medmuskip=4mu\relax\hyphenpenalty=10000\relax}},
  fbase/.style={draw, line width=0.5pt, align=center, inner sep=2.2pt, minimum height=7mm},
  proc/.style={fbase, rectangle, text width=#1}, proc/.default=70mm,
  pre/.style={fbase, rectangle, text width=#1, double, double distance=1.1pt}, pre/.default=70mm,
  term/.style={fbase, rounded corners=3.5mm, text width=#1, fill=black!6}, term/.default=70mm,
  dec/.style={fbase, diamond, aspect=2.6, inner sep=0.6pt, text width=#1}, dec/.default=27mm,
  con/.style={draw, circle, line width=0.7pt, inner sep=0.8pt, minimum size=5.5mm, font=\footnotesize\bfseries},
  arr/.style={->, line width=0.5pt},
  call/.style={->, line width=0.5pt, dashed},
  lab/.style={font=\scriptsize\itshape, inner sep=1pt},
}

\usepackage[colorlinks=true,linkcolor=blue,citecolor=blue,urlcolor=blue]{hyperref}
\biboptions{sort&compress}
\providecommand{\Amat}{\mathsf{A}}\providecommand{\Pran}{\mathit{Pr}}
\providecommand{\Kn}{\mathit{Kn}}
\newcommand{\ii}{\mathrm{i}}
\renewcommand{\hat}[1]{\widehat{#1}}
\newcommand{\Ma}{\mathit{Ma}}
\newcommand{\Rey}{\mathit{Re}}
\journal{Journal of Computational Physics}

\begin{document}
\begin{frontmatter}
\title{Kinetic linear stability theory for incompressible parallel shear flows with the full Boltzmann collision operator}
\author{V. Theofilis}
\ead{vassilis@technion.ac.il}
\affiliation{organization={Faculty of Aerospace Engineering, Technion -- Israel Institute of Technology},
            addressline={Technion City}, city={Haifa 3200003}, country={Israel}}
\begin{abstract}
Modal and non-modal linear stability analysis of incompressible shear flows is performed for the first time by linearizing the Boltzmann equation and its full collision operator. In this kinetic linear stability theory (kLST) the hard-sphere operator and its BGK, ES-BGK and Shakhov relaxation models share the same discretization, so that their results can be compared directly. Written in the peculiar velocity about a Chapman--Enskog base state, the collision operators are independent of the cross-stream coordinate and are formed once. Summation-by-parts Legendre collocation, weak diffuse-reflection wall conditions and a moment-consistent velocity derivative yield an energy-stable discrete operator, verified to spectral accuracy by a manufactured solution. Numerical solution of the four-dimensional problem, comparable in size to a TriGlobal problem (Theofilis 2011), is made feasible by the fast spectral method of Gamba, Haack, Hauck and Hu (2017) for the full operator. Modal kLST uses matrix-free shift-invert Arnoldi iteration, preconditioned by an exact relaxation-model inverse. Non-modal kLST is based on an energy norm containing Chu's disturbance energy, a kinetic Reynolds--Orr equation identifying the mean shear as the only energy source. Transient growth and pseudospectra follow from the same preconditioned solves. For plane Couette and Poiseuille flows, the mixing layer and the Blasius boundary layer at Reynolds numbers $10^2$--$10^5$, all four operators recover the Orr--Sommerfeld eigenvalues (Orszag 1971; Mack 1976) in the incompressible limit to a few parts in $10^7$ for the models and $10^6$ for the full operator. The optimal growth of Butler and Farrell (1992) and the resolvent norms of Trefethen et al.\ (1993) are reproduced to $0.1\%$, while the pseudospectra converge where ill-conditioned eigenvalues do not.
\end{abstract}
\begin{keyword}
kinetic linear stability theory \sep Boltzmann equation \sep full collision operator \sep fast spectral method \sep modal analysis \sep non-modal analysis \sep pseudospectra
\end{keyword}
\end{frontmatter}

\section*{Nomenclature}
\newcommand{\nomhead}[1]{\multicolumn{2}{@{}l}{\textit{#1}}\\[1pt]}
\newenvironment{nomtab}{\begin{tabular}{@{}>{\raggedright\arraybackslash}p{0.32\linewidth}>{\raggedright\arraybackslash}p{0.65\linewidth}@{}}}{\end{tabular}\par\smallskip}
{\footnotesize\setlength{\tabcolsep}{3pt}\renewcommand{\arraystretch}{0.87}\noindent
\begin{minipage}[t]{0.485\textwidth}\vspace{0pt}
\begin{nomtab}
\nomhead{Coordinates and distribution functions}
$x,\,y,\,z;\ t$ & streamwise, cross-stream, spanwise; time\\
$\boldsymbol v,\ \boldsymbol c$ & molecular velocity; peculiar velocity $\boldsymbol v-U(y)\boldsymbol e_x$\\
$\boldsymbol C$ & velocity relative to the local mean, $\boldsymbol v-\boldsymbol u$\\
$f,\ \bar f,\ \hat f$ & distribution function; base state; perturbation amplitude\\
$\hat f_m$ & manufactured amplitude function\\
$M$ & Maxwellian at rest, unit density and temperature\\
$\bar g$ & shear function of the base state, $\bar f=M+U'\bar g$\\
\end{nomtab}
\begin{nomtab}
\nomhead{Flow and parameters}
$U(y),\ U,\ U_\infty$ & base velocity profile; reference speed; free-stream speed\\
$h,\ \ell,\ \delta^*$ & half-gap; length scale of a profile; displacement thickness\\
$y_\infty$ & position of a free boundary\\
$a(y)$ & body force per unit mass, $-U''/\nu_\mu$\\
$\rho_0,\,T_0,\,p_0,\ \mu_0$ & reference density, temperature, pressure; viscosity\\
$\nu_\mu,\ \nu,\ \eta$ & collisional parameter $p_0/\mu_0$; collision frequency; exponent of $\mu\propto T^{\eta}$\\
$\Rey,\,\Ma,\,\Kn,\,\Pran$ & Reynolds, Mach, Knudsen and Prandtl numbers; $\gamma=5/3$\\
$n,\ \boldsymbol u,\ T,\ \boldsymbol P,\ \boldsymbol q$ & density, velocity, temperature, pressure tensor, heat flux of $f$\\
$\hat\rho,\ \hat{\boldsymbol u},\ \hat T,\ \hat P_{ij},\ \hat{\boldsymbol q}$ & their perturbation amplitudes, moments of $\hat f$\\
\end{nomtab}
\begin{nomtab}
\nomhead{Collision operators}
$\mathcal{J}$ & collision operator\\
$\mathcal{Q}(\cdot,\cdot),\ \kappa$ & hard-sphere collision integral; its scaling\\
$\mathcal{M},\ \mathcal{E},\ \mathcal{S},\ \boldsymbol\Lambda$ & targets of the BGK, ES-BGK and Shakhov models; ES-BGK tensor\\
$\mathcal{L}_M,\ \mathcal{L}_{\bar g}$ & $\mathcal{J}$ linearized about $M$; part due to $U'\bar g$\\
$\phi$ & ES-BGK and Shakhov part of $\mathcal{L}_{\bar g}$\\
$\psi_m,\ \Pi_h$ & collision invariants and moment weights; hydrodynamic projection\\
$\nu_{\mathrm{rel}};\ \boldsymbol b_m,\ \boldsymbol p_m$ & relaxation-model loss rate; moment weights, gain functions\\
\end{nomtab}
\par\smallskip\noindent{\raggedright\textit{Abbreviations}: BGK, Bhatnagar--Gross--Krook model; ES-BGK, ellipsoidal-statistical BGK model; HS, hard spheres; (k)LST, (kinetic) linear stability theory; SAT, simultaneous approximation term; GMRES, generalized minimal residual method.\par}
\end{minipage}\hfill
\begin{minipage}[t]{0.485\textwidth}\vspace{0pt}
\begin{nomtab}
\nomhead{Stability problem}
$\alpha,\ \beta,\ k$ & streamwise, spanwise wavenumbers; $k=(\alpha^2+\beta^2)^{1/2}$\\
$\omega=\omega_r+\ii\omega_i$ & eigenvalue, $\hat f\propto\mathrm{e}^{-\ii\omega t}$: $\omega_r$ circular frequency, $\omega_i$ amplification ($>0$) or damping ($<0$) rate\\
$c=c_r+\ii c_i$ & complex phase speed $\omega/(\alpha U)$: $c_r$ phase speed, $c_i$ amplification ($>0$) or damping ($<0$), at the rate $\alpha c_i$\\
$c_0,\,c_1,\,c_2,\,c_4$ & coefficients of the fits in $\Ma$; $c_0$: limit $\Ma\to0$\\
$\mathcal{A},\ \mathcal{A}^{\dagger}$ & kinetic stability operator, $\partial_t\hat f=\mathcal{A}\hat f$; its adjoint\\
$\langle\cdot,\cdot\rangle,\ E$ & energy inner product, weighted by $1/M$; disturbance energy\\
$\mathcal{P},\ \mathcal{D}_c,\ \mathcal{D}_w$ & production; collisional and wall dissipation\\
$G(t),\,G_{\max},\,t_{\max}$ & transient growth; its maximum and the time of the maximum\\
$\Lambda_\varepsilon,\ R(\omega)$ & $\varepsilon$-pseudospectrum; resolvent norm\\
$\varkappa$ & condition number of an eigenvalue\\
\end{nomtab}
\begin{nomtab}
\nomhead{Discretization and solver}
$Q,\ \Delta v$ & velocity nodes per direction; their spacing\\
$R_v,\ v_{\mathrm{cut}}$ & truncation radius; radius of the velocity ball\\
$n_b$ & velocity nodes in the ball\\
$N;\ \xi_j,\ w_j$ & intervals in $y$ ($N+1$ points); Legendre--Gauss--Lobatto points and weights\\
$y_{1/2},\ b_y$ & parameters of the mappings in $y$\\
$\mathsf D,\ \mathsf D_{c_x}$ & differentiation matrices in $y$ and in $c_x$\\
$N_r$ & radial nodes of the collision quadrature\\
$\mathsf K,\ \mathsf A,\ n$ & discrete distribution function; discrete operator; unknowns $(N+1)n_b$\\
$\mathsf S_w,\ \mathsf W$ & boundary penalty; weights of the discrete energy\\
$z_0,\ n_m$ & shift; rank of the relaxation-model moments\\
$\mathsf A_{\mathrm p},\ \mathsf B,\ \mathsf U\mathsf V^{T}$ & preconditioning operator; its velocity-diagonal and low-rank parts\\
$\mathsf S,\ r$ & capacitance matrix; its order\\
\end{nomtab}
\end{minipage}
}

\clearpage

\section{Introduction}\label{sec:intro}

Linear stability theory (LST) describes small-amplitude perturbations of laminar flows through the solution of eigenvalue or initial-value problems, which predict modal and non-modal laminar--turbulent transition scenarios, respectively. Modal disturbances grow exponentially and their growth rates are integrated in space in the so-called $e^N$ method, an engineering tool for the prediction of laminar--turbulent transition \citep{Reed1996}, commonly attributed to \citet{SmithGamberoni1956} and \citet{vanIngen1956}, although it was first proposed by \citet{Schlichting1933} as an $e^9$ criterion. Errors in the computed eigenvalues are also amplified exponentially and may lead to unreliable predictions of the transition location, which makes a strong case for accurate numerical methods for the solution of the eigenvalue problem \citep{Orszag1971}. In its present-day form, laminar--turbulent transition prediction rests entirely on the continuum, Navier--Stokes, description of the gas, which becomes increasingly inadequate as rarefaction phenomena appear.

Two routes have been followed to obtain the stability of a flow, in the sense of eigenvalues and growth rates, from kinetic theory. In the first, linear instabilities are extracted from a time-dependent simulation: direct simulation Monte Carlo (DSMC) computations of the B\'enard and Taylor--Couette problems \citep{CercignaniStefanov1992,StefanovCercignani1993,StefanovRoussinovCercignani2002} and a BGK solution of the B\'enard problem \citep{SoneAokiSugimoto1997}, whose thresholds have served as reference for slip-corrected Navier--Stokes analyses \citep{ManelaFrankel2005,ManelaFrankel2007,BenAmiManela2019}, and the extraction of the linear operator from particle-simulation time series \citep{GarciaPenland1991}. The route has since been applied to interfacial and free-shear instabilities and molecular-level turbulence \citep{Gallis2016,Gallis2018,Mohan2022,Mohan2023,Kudryavtsev2019,Shershnev2024}, to hypersonic separated flows and shocks \citep{Tumuklu2018a,Tumuklu2018b,Tumuklu2019,Sawant2021,Sawant2022}, and, with DSMC base flows in classical and BiGlobal analysis \citep{Theofilis2011}, to a global mode at a compression corner that is absent from the Navier--Stokes spectrum \citep{Klothakis2022,Karpuzcu2026}; the lattice Boltzmann method (LBM) has been used in the same way \citep{Shan1997,Mendoza2014}. This route delivers the leading instability, at an accuracy bounded by statistical noise or by the time over which exponential behavior can be isolated, but not a spectrum, without which neither the synchronization of discrete modes in hypersonic boundary layers \citep{Fedorov2011} nor transient growth \citep{TrefethenScience} can be studied.

On the second route, the kinetic equation is linearized and an eigenvalue problem is solved. About a global Maxwellian this gives the spectrum of the linearized collision operator \citep{Grad1963,Grad1963PoF,Cercignani1988,KogelbauerKarlin2024,KogelbauerKarlin2025}, the hydrodynamic eigenvalues \citep{EllisPinsky1975} and the dispersion of sound \citep{WangChangUhlenbeck1952,SirovichThurber1965}; \citet{Lees1959} treated linearized Couette flow across the rarefaction regimes with a moment method, although not as a stability problem. Linearization about a non-equilibrium state has been carried out for unbounded uniform shear flow in the peculiar velocity \citep{LeeDufty1997,Garzo2006}, with rigorous and asymptotic results for the B\'enard and Couette problems \citep{ArkerydEspositoMarraNouri2010,SoneDoi2004}, and the LBM has been used for global modal analysis \citep{Perez2017,Jiang2022,Luo2023}. Numerical eigenvalue problems on a kinetic equation for a wall-bounded or non-uniform shear flow are, however, recent, and all but one rest on a relaxation model of the collision operator.

\citet{Zou2022} were the first to pose such a problem, for plane Poiseuille flow on the BGK equation with a force term: an isothermal model with the spanwise velocity integrated out, two-dimensional disturbances, a numerical kinetic base flow, and a full QZ solution of the matrix problem. The eigenvalue of \citet{Orszag1971} is approached to $2\times10^{-5}$, or $0.6\%$, in the growth rate, without monotone convergence. For plane Couette flow the eigenvalues near the junction of the spectral branches were found never to converge with the number of velocity nodes \citep{Zou2023}; the unit Prandtl number of BGK was identified as its main shortcoming and the Shakhov model used instead for compressible Couette, Rayleigh--B\'enard and supersonic Couette flow \citep{Zou2024,WuC2024,Bi2025}; and the BGK formulation was taken to the Blasius boundary layer and the mixing layer, reproducing the modes of \citet{Mack1976} to within $1.3\%$ \citep{Ma2024}. Independent implementations have been reported for channel flows and a weak shock \citep{KarpuzcuLevin2025,KarpuzcuLevinTheofilis2026,Aharoni2026}.

\citet{Wu2025} presented a kinetic linear stability methodology designed for three-dimensional hypersonic rarefied flows, for which the stability matrix cannot be formed, and demonstrated it on plane sound waves, with the Shakhov model and the linearized Boltzmann operator for three potentials, and on plane Couette flow with the BGK model at rarefaction parameters $\delta=1$ to $100$. The eigenvalue nearest a shift is found by a shifted inverse power method whose inner iteration on the kinetic equation is accelerated by a synthetic scheme driven by moment equations, reaching an eigenpair in a few tens to a few hundred sweeps. For the sound waves the operators differ markedly when the mean free path is comparable to the wavelength; for the Couette flow, at Reynolds numbers of at most order one, the least stable mode and a few nearby modes were computed to an estimated $10^{-4}$. As \citet{Wu2025} notes, the method returns a few modes rather than the spectrum and converges slowly when the eigenvalue sought is not well separated; the three-dimensional application has yet to be reported. Two of its test cases are recomputed here in \ref{app:wu}.

These studies leave several questions open. The full Boltzmann collision operator has not been used in a stability eigenvalue problem for a shear flow, and in particular not at the Reynolds numbers of practical interest, $O(10^2)$ to $O(10^5)$ here, where the stiffness of the collision term is greatest. An iteration that returns one eigenvalue per shift is ill suited to the part of a shear-flow spectrum with many closely spaced eigenvalues and nearly parallel eigenvectors, the origin of transient growth \citep{TrefethenScience,Reddy1993}, and the non-convergence near the junction of the branches \citep{Zou2023} remains unexplained in the context of kinetic linear stability analyses, though it is well understood in continuum theory on account of the non-normality of the discretized stability matrix \citep{Butler1992,TrefethenScience,Reddy1993}. Non-modal stability theory has not been formulated for a kinetic description of a gas: no norm for the energy of a kinetic disturbance, no kinetic Reynolds--Orr equation and no adjoint of the kinetic stability operator have been stated, and no transient growth or pseudospectra computed. The ES-BGK model has not been used in a kinetic stability analysis, the Shakhov model not for low-speed shear flows, and no two collision models have been compared on the same flow. The analyses are two-dimensional, so that the Squire family is absent, and mostly isothermal; no systematic convergence study in the velocity and wall-normal resolutions and in the Mach number has been reported; and the classical benchmarks are reproduced to $10^{-5}$ to $10^{-2}$ rather than to the accuracy of continuum stability theory \citep{Orszag1971}. Finally, hybrid continuum/kinetic transition prediction \citep[e.g.][]{Klothakis2022,Karpuzcu2026} requires base profiles that are not solutions of a kinetic equation to enter a kinetic stability operator.

The present work answers these questions within kinetic linear stability theory (kLST), with the full linearized Boltzmann collision operator for hard spheres and its BGK, ES-BGK and Shakhov simplifications, for four flows: two wall-bounded ones, whose Orr--Sommerfeld spectrum is discrete, and two open ones, whose Orr--Sommerfeld spectrum also has a continuous part. The modal and non-modal stability of plane Poiseuille flow at the test case of \citet{Orszag1971}, $\Rey=10^4$, and plane Couette flow at $\Rey=10^3$, at Mach numbers from $0.2$ to $0.01$, is studied first, followed by the hyperbolic-tangent mixing layer of \citet{Michalke1964} from $\Rey=10^2$ to $10^5$ and the Blasius boundary layer at the test case of \citet{Mack1976}, $\Rey=580$, the last two at Mach numbers from $0.1$ to $0.025$. The eigenvalues are extrapolated to $\Ma\to0$ for comparison with Orr--Sommerfeld results. The base profile may be given in closed form or as nodal values; the base state is this profile with a Chapman--Enskog correction, held steady by a body force, which confines the method to near-continuum, nearly incompressible base flows, $\Kn\ll1$ and $\Ma^2\ll1$, while the perturbation is fully compressible and kinetic. The kLST eigenvalue problem is written in the peculiar velocity, is discretized in the same way for all four collision operators, and keeps all three molecular velocity components, the temperature perturbation and both streamwise and spanwise wavenumbers. Boundary conditions are imposed on the distribution function alone and approach those of the continuum limit as $\Kn\to0$. The problem is solved by shift-invert Arnoldi iteration, which returns several eigenvalues per shift and is preconditioned by the exact inverse of a relaxation-model operator, obtained from the Woodbury identity \citep{Woodbury1950,Hager1989}.

The coupled discretization of three velocity dimensions and one spatial dimension makes the problem comparable in size to that of TriGlobal linear stability theory (LST) based on the continuum equations \citep{Theofilis2011}. Consequently, the discretized kLST stability matrix is never formed, while the collision operators are formed once as dense matrices in velocity space, the full one by the fast spectral method of \citet{GHHH2017}, which makes the present linearization of the full Boltzmann operator affordable. The dense factorizations in the single inhomogeneous direction do not extend to three-dimensional flows, as the iterative approach of \citet{Wu2025} is designed to, making the two approaches complementary. Legendre collocation with summation by parts, weakly imposed wall conditions and a moment-consistent velocity derivative make the discrete operator inherit the energy inequality of the continuous one. The eigenvalues of all four operators are converged in the velocity and wall-normal resolutions and in the Mach number. Finally, a non-modal kLST is formulated, in which the kinetic-theory energy norm contains the disturbance energy of \citet{Chu1965} as its hydrodynamic part. A kinetic Reynolds--Orr equation is written for the first time and shows that the mean shear is the only source of disturbance energy, while collisions and walls only remove energy. Because collisions do not dissipate perturbations in local equilibrium, the energy of a suitably chosen initial perturbation grows at every Reynolds number, and no counterpart of the energy-stability Reynolds number of \citet{Joseph1976} exists in this norm. Pseudospectra and transient energy growth are computed from the same preconditioned shifted solves as the eigenvalues, without time stepping or eigenfunction expansions. Unlike the individual eigenvalues, the pseudospectra are insensitive to the ill-conditioning of the modes near the junction of the branches, and resolve the non-convergence reported by \citet{Zou2023}.

The paper is organized as follows. Section~\ref{sec:theory} states the kinetic
equation and the collision operators, constructs the base state, derives the kLST
eigenvalue problem with its boundary conditions and formulates the non-modal kLST. Section~\ref{sec:methods} describes
the discretization and the solver. Section~\ref{sec:verification} verifies the discrete operator and the solver against a
manufactured solution and checks the structural properties of the discrete operator. Section~\ref{sec:results} presents the four test cases, with their
convergence in the resolutions and in the Mach number, and compares modal and non-modal kLST results with continuum theory. Section~\ref{sec:conclusions}
discusses the results and points to open questions. \ref{app:linops} gives the linearized collision operators,
\ref{app:precond} the preconditioner, and \ref{app:wu} a cross-check against published rarefied test cases.

\section{Theory}\label{sec:theory}
In the kinetic description, the state of a monatomic gas is given by the molecular velocity distribution function $f(\boldsymbol x,\boldsymbol v,t)$, the density of molecules at position $\boldsymbol x$ with velocity $\boldsymbol v$ at time $t$. Its velocity moments are the macroscopic density, velocity, temperature, stress and heat flux, and its evolution is governed by the Boltzmann equation, in which the free streaming of the molecules is balanced by their binary collisions. The gas flows parallel to the $x$ axis, with $y$ the cross-stream and $z$ the spanwise coordinate.

Velocities are scaled by $\sqrt{RT_0}$, density by $\rho_0$, temperature by $T_0$ and lengths by a reference length, so that all quantities are dimensionless and the equation of state is $p=\rho T$. The equilibrium pressure is then $p_0=1$, $\gamma=5/3$, and the rest Maxwellian is $M(\boldsymbol v)=(2\pi)^{-3/2}\exp(-\tfrac12|\boldsymbol v|^2)$. The distribution function satisfies the Boltzmann equation with a body force $a(y)$ per unit mass along $x$,
\begin{equation}
 \frac{\partial f}{\partial t}+\boldsymbol v\cdot\frac{\partial f}{\partial\boldsymbol x}
 +a\,\frac{\partial f}{\partial v_x}=\mathcal{J}(f).
 \label{eq:boltz}
\end{equation}
The collision operator $\mathcal{J}$ is scaled so that its shear viscosity at the reference state is $\mu_0$, and the single collisional parameter is $\nu_\mu=p_0/\mu_0$.

\subsection*{Collision operators}
The full Boltzmann operator for hard spheres, the full operator for short, is compared with three relaxation models, each of which replaces the collision integral by relaxation, at the collision frequency $\nu$, towards a target distribution built from the moments of $f$. The four operators are listed in Table~\ref{tab:collops}. All four conserve mass, momentum and energy.

\begin{table}[!htb]
\caption{Collision operators.}\label{tab:collops}
\centering\small\renewcommand{\arraystretch}{1.3}\setlength{\tabcolsep}{4pt}
\begin{tabular}{@{}lll c@{}}
\toprule
Operator & $\mathcal{J}(f)$ & Collision integral or target distribution & $\mathit{Pr}$\\
\midrule
Full Boltzmann & $\kappa\,\mathcal{Q}(f,f)$ & $\mathcal{Q}(f,g)=\displaystyle\int_{\mathbb{R}^3}\!\int_{S^2}B\,\bigl[f(\boldsymbol v')g(\boldsymbol v_*')-f(\boldsymbol v)g(\boldsymbol v_*)\bigr]\,\mathrm{d}\boldsymbol\Omega\,\mathrm{d}\boldsymbol v_*$ & $0.661$\\
hard spheres \citep{Cercignani1988} & & bilinear, kernel $B=b\,|\boldsymbol v-\boldsymbol v_*|$, & \\
 & & $\kappa$ set by the discrete viscosity & \\
\addlinespace[4pt]
BGK \citep{BGK1954} & $\nu\,\bigl(\mathcal{M}[f]-f\bigr)$ & $\mathcal{M}[f]=\dfrac{n}{(2\pi T)^{3/2}}\exp\Bigl(-\dfrac{|\boldsymbol C|^2}{2T}\Bigr)$ & $1$\\
 & & local Maxwellian: same $n$, $\boldsymbol u$, $T$ as $f$ & \\
\addlinespace[4pt]
ES-BGK \citep{Holway1966,Andries2000} & $\nu\,\bigl(\mathcal{E}[f]-f\bigr)$ & $\mathcal{E}[f]=\dfrac{n}{\sqrt{\det(2\pi\boldsymbol\Lambda)}}\exp\bigl(-\tfrac12\,\boldsymbol C\cdot\boldsymbol\Lambda^{-1}\cdot\boldsymbol C\bigr)$ & $2/3$\\
 & & $\boldsymbol\Lambda=\dfrac{T}{\mathit{Pr}}\,\boldsymbol I+\Bigl(1-\dfrac1{\mathit{Pr}}\Bigr)\dfrac{\boldsymbol P}{n}$ & \\
 & & anisotropic Gaussian: same $n$, $\boldsymbol u$, $T$, & \\
 & & stress relaxes at $\nu/\mathit{Pr}$ & \\
\addlinespace[4pt]
Shakhov \citep{Shakhov1968} & $\nu\,\bigl(\mathcal{S}[f]-f\bigr)$ & $\mathcal{S}[f]=\mathcal{M}[f]\Bigl[1+\dfrac{1-\mathit{Pr}}{5}\,\dfrac{\boldsymbol q\cdot\boldsymbol C}{nT^2}\Bigl(\dfrac{|\boldsymbol C|^2}{T}-5\Bigr)\Bigr]$ & $2/3$\\
 & & Maxwellian with heat-flux correction: same $n$, $\boldsymbol u$, $T$, & \\
 & & heat flux relaxes at $\mathit{Pr}\,\nu$ & \\
\bottomrule
\end{tabular}
\end{table}

The macroscopic fluid quantities that enter the target distributions are the density, velocity, temperature, pressure tensor and heat flux of $f$,
\begin{equation}
 \begin{gathered}
 n=\int f\,\mathrm{d}\boldsymbol v,\qquad
 \boldsymbol u=\frac1n\int\boldsymbol v\,f\,\mathrm{d}\boldsymbol v,\qquad
 T=\frac1{3n}\int|\boldsymbol C|^2f\,\mathrm{d}\boldsymbol v,\\
 \boldsymbol P=\int\boldsymbol C\boldsymbol C\,f\,\mathrm{d}\boldsymbol v,\qquad
 \boldsymbol q=\int\tfrac12|\boldsymbol C|^2\boldsymbol C\,f\,\mathrm{d}\boldsymbol v,\qquad
 \boldsymbol C=\boldsymbol v-\boldsymbol u,
 \end{gathered}
 \label{eq:fmoments}
\end{equation}
and the collision frequency is
\begin{equation}
 \nu=\nu_\mu\,nT^{1-\eta}\quad\text{(BGK, Shakhov)},\qquad \nu=\mathit{Pr}\,\nu_\mu\,nT^{1-\eta}\quad\text{(ES-BGK)},
 \label{eq:nu}
\end{equation}
with $\eta=\tfrac12$, the exponent of the viscosity law $\mu\propto T^\eta$, for hard spheres. For $\mathit{Pr}=1$ the ES-BGK and Shakhov models reduce to BGK.

\subsection{Base flow}
The base flow is a parallel shear flow $\boldsymbol u=U(y)\,\boldsymbol e_x$ of uniform density and temperature, characterized by its Reynolds number $\mathit{Re}=\rho_0U\ell/\mu_0$ and Mach number $\Ma=U/\sqrt{\gamma RT_0}$, with $U$ and $\ell$ the reference speed and length of the profile. The Knudsen number is $\mathit{Kn}\propto\Ma/\mathit{Re}$. A profile that is not an exact solution of the Navier--Stokes equations is held steady, as in the classical parallel-flow approximation, by the body force that balances its viscous stress at each $y$,
\begin{equation}
 a(y)=-\frac{\mu_0}{\rho_0}\,U''(y)=-\frac{U''(y)}{\nu_\mu} ;
 \label{eq:force}
\end{equation}
it is of order $U/\mathit{Re}$, and for plane Poiseuille flow it is the constant pressure gradient. The profile may be given in closed form or, as for the Blasius layer, as nodal values.

The base state is steady and depends on $y$ only. With $\partial_t=\partial_x=\partial_z=0$, \eqref{eq:boltz} reduces to
\begin{equation}
 v_y\,\frac{\partial\bar f}{\partial y}+a\,\frac{\partial\bar f}{\partial v_x}=\mathcal{J}(\bar f).
 \label{eq:base}
\end{equation}
To first order in the Knudsen number its solution is the Maxwellian drifting with the local velocity $U(y)$, plus its Chapman--Enskog shear stress,
\begin{equation}
 \bar f(y,\boldsymbol v)=M(\boldsymbol c)+U'(y)\,\bar g(\boldsymbol c),\qquad \mathcal{L}_M\,\bar g=c_xc_y M,\qquad \boldsymbol c=\boldsymbol v-U(y)\,\boldsymbol e_x ,
 \label{eq:fbar}
\end{equation}
where $\boldsymbol c$ is the molecular velocity relative to the local flow and $\mathcal{L}_M$ is the linearization of $\mathcal{J}$ about $M$ (\S\ref{sec:linops}). The body force \eqref{eq:force} is the solvability condition of the next order.

The formulation restricts the base state, not the perturbation. The base state is near-continuum, $\mathit{Kn}\ll1$, so that terms of $\mathcal{O}(\mathit{Kn}^2)$ and the Knudsen layers of the base flow are dropped, a negligible approximation for the present flows, $10^{-7}\lesssim\mathit{Kn}\lesssim2\times10^{-3}$. It is also nearly incompressible, $\Ma^2\ll1$, and viscous heating is neglected, so that the base density and temperature are uniform, an $\mathcal{O}(\Ma^2)$ error, and the $\mathcal{O}(\Ma^2)$ dependence of the eigenvalues is not that of a fully consistent compressible base flow. The stability operator, by contrast, is the exact linearization about this base state, i.e., the perturbation is fully compressible, all moments of the distribution function are retained, and the collision operators and the boundary conditions hold at any $\mathit{Kn}$. The eigenvalues are thus correct to $\mathcal{O}(\Ma^2)$ at the Mach numbers computed and exact as $\Ma\to0$.

\subsection{Kinetic Linear Stability Theory (kLST)}
\subsubsection{Linearization, peculiar velocity and linearized collision operators}\label{sec:linops}
The distribution function is decomposed into the steady base state and a small perturbation, which is a normal mode in the homogeneous directions $x$ and $z$,
\begin{equation}
 f(\boldsymbol x,\boldsymbol v,t)=\bar f(y,\boldsymbol v)+\hat f(y,\boldsymbol v,t)\,\exp[\ii(\alpha x+\beta z)],\qquad |\hat f|\ll\bar f .
 \label{eq:ansatz}
\end{equation}
The perturbation is three-dimensional in physical space, with streamwise and spanwise wavenumbers $\alpha$ and $\beta$, $k=(\alpha^2+\beta^2)^{1/2}$, and its amplitude function $\hat f(y,\boldsymbol v,t)$ is \emph{four-dimensional}, resolved in $y$ and in all three components of the molecular velocity.

Base state and perturbation alike are written in the peculiar velocity $\boldsymbol c$ of \eqref{eq:fbar}, in which the base state has the same velocity dependence at every $y$, so that the collision operators linearized about it do not depend on $y$. With $f(\boldsymbol x,\boldsymbol v,t)=f(\boldsymbol x,\boldsymbol c,t)$, the derivative $\partial_y$ at fixed $\boldsymbol v$ becomes $\partial_y$ at fixed $\boldsymbol c$ plus the chain-rule term $-U'\partial_{c_x}$, $v_x=c_x+U$ and $\partial_{v_x}=\partial_{c_x}$, and the collision term is unchanged, since the collision operator is Galilean invariant. Equation \eqref{eq:boltz} becomes
\begin{equation}
 \frac{\partial f}{\partial t}+(c_x+U)\frac{\partial f}{\partial x}+c_y\frac{\partial f}{\partial y}+c_z\frac{\partial f}{\partial z}
 -\bigl(U'c_y-a\bigr)\frac{\partial f}{\partial c_x}=\mathcal{J}(f),
 \label{eq:boltz_c}
\end{equation}
in which the base flow enters through the Doppler term $U\partial_x$ and the shear term $U'c_y\partial_{c_x}$. The base-state equation \eqref{eq:base} is \eqref{eq:boltz_c} itself with $\partial_t=\partial_x=\partial_z=0$, i.e., base state and perturbation satisfy the \emph{same} kinetic equation, and any steady solution of \eqref{eq:boltz_c} is a consistent base state, to $\mathcal{O}(\Kn)$ for the present profiles and to within numerical errors for kinetic \citep{Zou2022} or DSMC \citep{Klothakis2022,Karpuzcu2026} base flows.

With $\bar f=M+U'\bar g$, the collision term of \eqref{eq:boltz_c} is expanded as
\begin{equation}
 \mathcal{J}(\bar f+\hat f)=\mathcal{J}(\bar f)+\mathcal{J}'(\bar f)\,\hat f+\mathcal{O}(\hat f^2),\qquad
 \mathcal{J}'(\bar f)=\mathcal{J}'(M)+U'\,\mathcal{J}''(M)[\bar g,\cdot\,]+\mathcal{O}(U'^2\bar g^2),
\end{equation}
which defines
\begin{equation}
 \mathcal{L}_M\hat f=\mathcal{J}'(M)\,\hat f,\qquad \mathcal{L}_{\bar g}\hat f=\mathcal{J}''(M)[\bar g,\hat f],
 \label{eq:Ldef}
\end{equation}
the derivative of $\mathcal{J}$ at the Maxwellian and the change of that derivative in the direction of the base non-equilibrium $\bar g$.
For the full operator the split is exact, because $\mathcal{Q}$ is bilinear, whereas for the relaxation models it holds to first order in the base non-equilibrium, the order to which $\bar f$ itself is defined. In every case $\mathcal{L}_{\bar g}$ is the kinetic form of the perturbation of the transport coefficients acting on the base shear, such as $\hat\mu U'$ in the shear stress. For the three relaxation models, with their base state $\bar g=-c_xc_yM/\nu_\mu$,
\begin{equation}
 \mathcal{L}_{\bar g}\hat f=\bigl[\hat\rho+(1-\eta)\hat T\bigr]c_xc_yM+(1-\mathit{Pr})\,M\,\phi(\hat{\boldsymbol u},\hat T;\boldsymbol c),
 \label{eq:Lgb}
\end{equation}
where $\hat\rho$, $\hat{\boldsymbol u}$ and $\hat T$ are moments of $\hat f$. The first term, common to the three models, is the perturbation $\nu_\mu[\hat\rho+(1-\eta)\hat T]$ of the collision frequency acting on the base non-equilibrium, and is the whole of $\mathcal{L}_{\bar g}$ for BGK, for which $\mathit{Pr}=1$. The second, a polynomial in $\boldsymbol c$ linear in $\hat{\boldsymbol u}$ and $\hat T$, comes from the anisotropic Gaussian of the stressed base state for ES-BGK and, for Shakhov, from the heat flux, which is defined about the perturbed mean velocity. The operators of the four models are derived in \ref{app:linops}.

\subsubsection{The equations of modal and non-modal kinetic LST}
Substituting \eqref{eq:ansatz}, written in $\boldsymbol c$, in \eqref{eq:boltz_c}, the terms of zeroth order cancel by \eqref{eq:base}, the derivatives $\partial_x$ and $\partial_z$ become $\ii\alpha$ and $\ii\beta$, and the collision term is linearized by \eqref{eq:Ldef}. The terms linear in $\hat f$ give
\begin{equation}
 \boxed{\;\frac{\partial\hat f}{\partial t}=-c_y\,\frac{\partial \hat f}{\partial y}-\ii\bigl[\alpha\,(c_x+U)+\beta c_z\bigr]\hat f+\bigl(U'c_y-a\bigr)\frac{\partial \hat f}{\partial c_x}+\mathcal{L}_M\hat f+U'\,\mathcal{L}_{\bar g}\hat f\equiv\mathcal{A}\hat f\;}
 \label{eq:ivp}
\end{equation}
The five terms are wall-normal streaming, the phase factor with the Doppler shift, the shear and body-force term, collisions about $M$, and collisions with the base non-equilibrium $U'\bar g$. The profile enters only through the Doppler and shear terms, and $\mathcal{L}_M$ and $\mathcal{L}_{\bar g}$ are formed once.

For modal solutions the perturbation is harmonic in time,
\begin{equation}
 \hat f(y,\boldsymbol c,t)=\hat f(y,\boldsymbol c)\,\mathrm{e}^{-\ii\omega t},
 \label{eq:harmonic}
\end{equation}
with complex frequency $\omega$ and complex phase speed $c=\omega/(\alpha U)$. The time derivative in \eqref{eq:ivp} is replaced by $-\ii\omega$, and the kLST eigenvalue problem, $\mathcal{A}\hat f=-\ii\omega\,\hat f$, is
\begin{equation}
 \boxed{\;-\ii\omega\,\hat f=-c_y\,\frac{\partial \hat f}{\partial y}-\ii\bigl[\alpha\,(c_x+U)+\beta c_z\bigr]\hat f+\bigl(U'c_y-a\bigr)\frac{\partial \hat f}{\partial c_x}+\mathcal{L}_M\hat f+U'\,\mathcal{L}_{\bar g}\hat f\;}
 \label{eq:evp}
\end{equation}

\subsubsection{Boundary conditions}\label{sec:bc}
The walls reflect diffusely with full accommodation at $T_0$. In $\boldsymbol c$ the walls are at rest and re-emit $M(\boldsymbol c)$ with a density fixed by impermeability. The condition is linear in $f$, so the perturbation satisfies it separately. At a lower wall, for the molecules that leave it, and at a lower or upper free-stream boundary $\mp y_\infty$, for the molecules that enter from the undisturbed stream,
\begin{equation}
 \begin{gathered}
 \hat f=\rho_w M,\quad \rho_w=\int_{c_y<0}|c_y|\,\hat f\,\mathrm{d}\boldsymbol c\Big/\!\int_{c_y>0}c_y\,M\,\mathrm{d}\boldsymbol c\quad(c_y>0,\ \text{lower wall});\\
 \hat f=0\quad(c_y\gtrless0,\ y=\mp y_\infty),
 \end{gathered}
 \label{eq:bc}
\end{equation}
with the mirror condition at an upper wall. At a free-stream boundary the molecules that enter come from the undisturbed stream, so that the perturbation of their distribution vanishes, while those that leave are unconstrained. The condition is non-reflecting at the kinetic level, and the macroscopic perturbations at $y_\infty$ are the moments of the outgoing half of $\hat f$, not prescribed values. Conditions are thus imposed only on the molecules that enter the domain, one per velocity, which is the correct count for the first-order operator $c_y\partial_y$. For all other molecules \eqref{eq:ivp} holds up to and including the boundary. No condition is imposed on any macroscopic perturbation, with no-slip, no-penetration and the isothermal wall emerging as $\mathit{Kn}\to0$, slip and temperature jump being of $\mathcal{O}(\mathit{Kn})$. The channels have two walls, the mixing layer two free boundaries, and the boundary layer a wall at $y=0$ and a free boundary at $y_\infty$.

\subsubsection{Kinetic Reynolds--Orr equation}\label{sec:ro}
In the energy norm $E=\tfrac12\langle\hat f,\hat f\rangle$, with $\langle\phi,\psi\rangle=\int(\phi,\psi)_c\,\mathrm{d}y$, which is the quadratic part of the relative entropy, the disturbance energy evolves as
\begin{equation}
 \begin{gathered}
 \frac{\mathrm{d}E}{\mathrm{d}t}=\mathcal{P}-\mathcal{D}_c-\mathcal{D}_w+\mathrm{Re}\,\langle\hat f,U'\mathcal{L}_{\bar g}\hat f\rangle,\\[4pt]
 \mathcal{P}=-\tfrac12\!\int\!\!\int\frac{(U'c_xc_y-a\,c_x)\,|\hat f|^2}{M}\,\mathrm{d}\boldsymbol c\,\mathrm{d}y,\\[4pt]
 \mathcal{D}_c=-\langle\hat f,\mathcal{L}_M\hat f\rangle\ge0,\qquad
 \mathcal{D}_w=\tfrac12\Bigl[\int\frac{c_y|\hat f|^2}{M}\,\mathrm{d}\boldsymbol c\Bigr]_{\rm lower}^{\rm upper}\ge0 .
 \end{gathered}
 \label{eq:RO}
\end{equation}
The shear production $\mathcal{P}$ is the only source. Collisions ($\mathcal{D}_c$, acting on the non-equilibrium part only) and walls or free boundaries ($\mathcal{D}_w$, by the conditions \eqref{eq:bc}) only remove energy. The last term, the exchange with the base non-equilibrium, is of relative order $\Ma\,\mathit{Kn}$.

\subsubsection{Quantification of transient energy growth}\label{sec:tg}
The non-modal quantities are defined in the energy norm $E$ of \S\ref{sec:ro}. The part of $E$ carried by the hydrodynamic moments, $\tfrac12\int\bigl(|\hat{\boldsymbol u}|^2+\hat\rho^2+\tfrac32\hat T^2\bigr)\,\mathrm{d}y$, is the disturbance energy of Chu, which as $\Ma\to0$ reduces to the kinetic energy $\tfrac12\int|\hat{\boldsymbol u}|^2\,\mathrm{d}y$ in which the Orr--Sommerfeld--Squire transient growth is computed. The adjoint of $\mathcal{A}$ in this norm is
\begin{equation}
 \mathcal{A}^{\dagger}=c_y\frac{\partial}{\partial y}+\ii\bigl[\alpha(c_x+U)+\beta c_z\bigr]
 -\bigl(U'c_y-a\bigr)\Bigl(\frac{\partial}{\partial c_x}+c_x\Bigr)+\mathcal{L}_M+U'\,\mathcal{L}_{\bar g}^{\dagger},
 \label{eq:adjoint}
\end{equation}
with $\mathcal{L}_M$ self-adjoint and the term $c_x$ coming from the weight $1/M$. The adjoint wall condition is diffuse reflection with the direction of streaming reversed, imposed on the molecules that arrive at the wall in the direct problem, and $\hat f^\dagger=0$ for the molecules that leave at a free boundary. The transient growth is the largest energy amplification attainable at time $t$ by any initial distribution,
\begin{equation}
 G(t)=\sup_{\hat f(0)\neq0}\frac{E\bigl(\hat f(t)\bigr)}{E\bigl(\hat f(0)\bigr)}=\bigl\|\mathrm{e}^{t\mathcal{A}}\bigr\|^{2},
 \qquad G_{\max}=\sup_{t\ge0}G(t),
 \label{eq:G}
\end{equation}
and the $\varepsilon$-pseudospectrum is the set of points at which the resolvent norm exceeds $1/\varepsilon$,
\begin{equation}
 \Lambda_\varepsilon(\mathcal{A})=\bigl\{z\in\mathbb{C}:\ \|(z\mathcal{I}-\mathcal{A})^{-1}\|>\varepsilon^{-1}\bigr\},
 \label{eq:pseudo}
\end{equation}
equivalently the set of eigenvalues of all perturbations of $\mathcal{A}$ of norm less than $\varepsilon$. On the real frequency axis, $R(\omega)=\|(-\ii\omega\mathcal{I}-\mathcal{A})^{-1}\|$ is the largest energy response to harmonic forcing of unit energy. The optimal growth $G(t)$ is the largest eigenvalue of $\mathrm{e}^{t\mathcal{A}^\dagger}\mathrm{e}^{t\mathcal{A}}$, the product of the adjoint and the direct propagators, and the corresponding eigenvector is the optimal initial distribution. All three require only shifted solves with $\mathcal{A}$ and $\mathcal{A}^\dagger$, without time stepping.

\subsubsection{Reconstruction of macroscopic perturbation quantities}\label{sec:recon_th}
The macroscopic perturbations are moments of $\hat f$ in the peculiar velocity,
\begin{equation}
 \begin{gathered}
 \hat\rho=\int \hat f\,\mathrm{d}\boldsymbol c,\qquad \hat{\boldsymbol u}=\int\boldsymbol c\,\hat f\,\mathrm{d}\boldsymbol c,\qquad
 \hat T=\tfrac13\!\int|\boldsymbol c|^2\hat f\,\mathrm{d}\boldsymbol c-\hat\rho,\\
 \hat P_{ij}=\int c_ic_j\,\hat f\,\mathrm{d}\boldsymbol c,\quad
 \hat{\boldsymbol q}=\int\tfrac12\boldsymbol c\,(|\boldsymbol c|^2-5)\,\hat f\,\mathrm{d}\boldsymbol c ,
 \end{gathered}
 \label{eq:moments}
\end{equation}
so that $\hat{\boldsymbol u}$ is the amplitude function of the velocity perturbation, to be compared with those of the Orr--Sommerfeld and Squire equations. $\hat P_{ij}$ is the perturbation of the pressure tensor and $\hat{\boldsymbol q}=(\hat q_x,\hat q_y,\hat q_z)$ that of the heat-flux vector. The heat flux is the flux of thermal energy relative to the local mean velocity, $\boldsymbol q=\int\tfrac12|\boldsymbol C|^2\boldsymbol C f\,\mathrm{d}\boldsymbol v$ with $\boldsymbol C=\boldsymbol v-\boldsymbol u$, \eqref{eq:fmoments}. Linearized about $M$, with $\boldsymbol C=\boldsymbol c-\hat{\boldsymbol u}$, it is $\int\tfrac12|\boldsymbol c|^2\boldsymbol c\,\hat f\,\mathrm{d}\boldsymbol c-\int\bigl[\tfrac12|\boldsymbol c|^2\hat{\boldsymbol u}+(\boldsymbol c\cdot\hat{\boldsymbol u})\,\boldsymbol c\bigr]M\,\mathrm{d}\boldsymbol c=\int\tfrac12|\boldsymbol c|^2\boldsymbol c\,\hat f\,\mathrm{d}\boldsymbol c-\tfrac52\hat{\boldsymbol u}$, which is the expression in \eqref{eq:moments}. The factor $5$ removes the enthalpy flux $\tfrac52\hat{\boldsymbol u}$, with enthalpy $c_pT=\tfrac52$ in the present units, carried by the perturbation velocity, so that $\hat{\boldsymbol q}$ is the conductive heat flux, orthogonal to the momentum moments. The mass, momentum and energy moments of \eqref{eq:ivp} are the linearized conservation laws, exactly and for any collision operator, since collisions have no such moments. With $\nabla=(\ii\alpha,\partial_y,\ii\beta)$ and $D=\partial_t+\ii\alpha U$, they read
\begin{equation}
 \begin{gathered}
 D\hat\rho+\nabla\!\cdot\hat{\boldsymbol u}=0 ,\qquad
 D \hat u_i+U'\hat u_y\,\delta_{ix}-a\hat\rho\,\delta_{ix}+\partial_j\hat P_{ij}=0 ,\\
 \tfrac32\,D(\hat\rho+\hat T)+\nabla\!\cdot\hat{\boldsymbol q}+\tfrac52\,\nabla\!\cdot\hat{\boldsymbol u}+U'\hat P_{xy}-a\,\hat u_x=0 .
 \end{gathered}
 \label{eq:cons}
\end{equation}
Since density and temperature are scaled by $\rho_0$ and $T_0$, $\hat\rho$ and $\hat T$ are relative perturbations, $\rho'/\rho_0$ and $T'/T_0$, and can be added. The third equation is the balance of internal energy per unit volume, $\tfrac32\rho T$ in the present units, whose linearization about $\rho=T=1$ is $\tfrac32(\hat\rho+\hat T)$, and $\tfrac52\hat{\boldsymbol u}$ is its enthalpy flux. The collision operator enters only through the closure of $\hat P_{ij}$ and $\hat{\boldsymbol q}$. As $\mathit{Kn}\to0$ the Chapman--Enskog expansion of $\hat f$ gives the Navier--Stokes closure, and \eqref{eq:cons} becomes the linearized compressible Navier--Stokes system of a monatomic gas with $\mu\propto T^\eta$ and the Prandtl number of the operator. As $\Ma\to0$ the acoustic modes, with $c=\mathcal{O}(1/\Ma)$, separate from the rest of the spectrum. For the remaining modes $\nabla\!\cdot\hat{\boldsymbol u}=\mathcal{O}(\Ma^2)$, the momentum equations reduce to the linearized incompressible Navier--Stokes equations, and the wall-normal velocity and vorticity satisfy the Orr--Sommerfeld and Squire equations, the limit in which the kinetic eigenvalues are compared with Orr--Sommerfeld values. The perturbations $\hat\rho$ and $\hat T$ are passive and form a thermal family of modes. The kinetic corrections to this limit are of $\mathcal{O}(\Ma^2)$ (compressibility, including the $\mathcal{L}_{\bar g}$ terms), of $\mathcal{O}(\mathit{Kn})$ (wall slip and temperature jump) and of $\mathcal{O}(\mathit{Kn}^2)$ (Burnett terms).

\section{Numerical methods}\label{sec:num}\label{sec:methods}
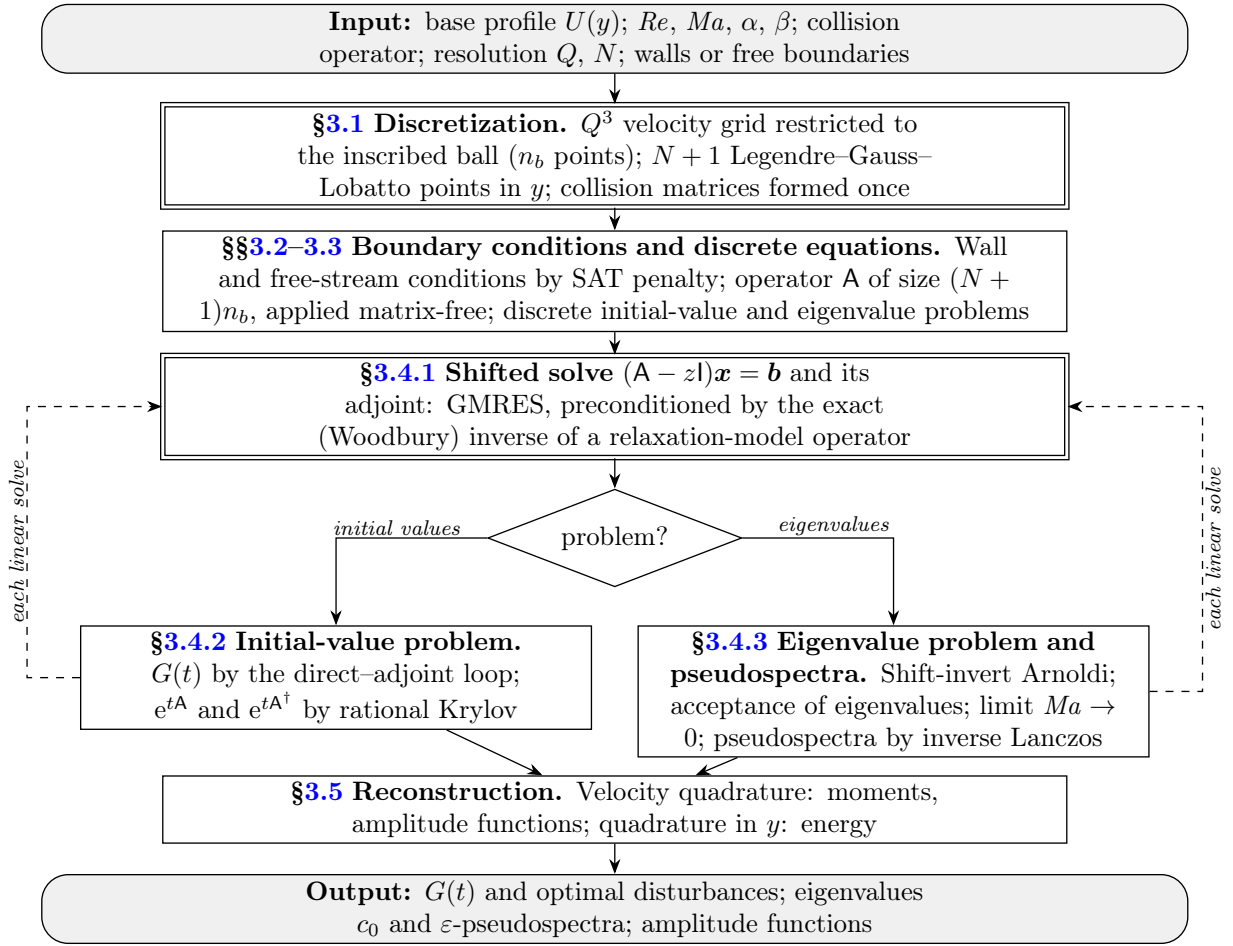
\begin{figure}[!htb]
\centering
\adjustbox{max width=\textwidth}{\begin{tikzpicture}[flow]
\node[term=150mm] (in) {\textbf{Input:} base profile $U(y)$; $\Rey$, $\Ma$, $\alpha$, $\beta$; collision
  operator; resolution $Q$, $N$; walls or free boundaries};
\node[pre=118mm, below=4mm of in] (D1) {\textbf{\S\ref{sec:disc} Discretization.} $Q^3$ velocity grid restricted to the
  inscribed ball ($n_b$ points); $N+1$ Legendre--Gauss--Lobatto points in $y$; collision matrices formed once};
\node[proc=118mm, below=of D1] (D2) {\textbf{\S\S\ref{sec:bcnum}--\ref{sec:discop} Boundary conditions and discrete equations.}
  Wall and free-stream conditions by SAT penalty; operator $\Amat$ of size $(N+1)n_b$, applied matrix-free;
  discrete initial-value and eigenvalue problems};
\node[pre=118mm, below=of D2] (B) {\textbf{\S\ref{sec:shsolve} Shifted solve} $(\Amat-z\mathsf{I})\bm{x}=\bm{b}$ and its adjoint:
  GMRES, preconditioned by the exact (Woodbury) inverse of a relaxation-model operator};
\node[dec=24mm, below=4mm of B] (q) {problem?};
\node[proc=66mm, below left=5mm and 3mm of q.south, anchor=north east] (I) {\textbf{\S\ref{sec:ivpgrowth} Initial-value problem.}
  $G(t)$ by the direct--adjoint loop; $\mathrm{e}^{t\Amat}$ and $\mathrm{e}^{t\Amat^\dagger}$ by rational Krylov};
\node[proc=66mm, below right=5mm and 3mm of q.south, anchor=north west] (E) {\textbf{\S\ref{sec:evpsol} Eigenvalue problem and
  pseudospectra.} Shift-invert Arnoldi; acceptance of eigenvalues; limit $\Ma\to0$; pseudospectra by inverse Lanczos};
\node[proc=118mm, below=6mm of I.south -| q.south, anchor=north] (R) {\textbf{\S\ref{sec:recon} Reconstruction.} Velocity
  quadrature: moments, amplitude functions; quadrature in $y$: energy};
\node[term=150mm, below=4mm of R] (out) {\textbf{Output:} $G(t)$ and optimal disturbances; eigenvalues $c_0$ and
  $\varepsilon$-pseudospectra; amplitude functions};
\foreach \a/\b in {in/D1, D1/D2, D2/B, B/q, I/R, E/R, R/out} \draw[arr] (\a) -- (\b);
\draw[arr] (q.west) -| node[lab, pos=0.3, above] {initial values} (I.north);
\draw[arr] (q.east) -| node[lab, pos=0.3, above] {eigenvalues} (E.north);
\draw[call] (I.west) -- ++(-7mm,0) coordinate (cl) -- node[lab, sloped, above] {each linear solve} (cl |- B.west) -- (B.west);
\draw[call] (E.east) -- ++(7mm,0) coordinate (cr) -- node[lab, sloped, below] {each linear solve} (cr |- B.east) -- (B.east);
\end{tikzpicture}}
\caption{Overview of the solver; each box carries the number of the subsection that describes it.}
\label{fig:flow_overview}
\end{figure}

The solver is summarized in figure~\ref{fig:flow_overview}, whose boxes are explained in the subsections below. Rounded boxes stand for data in and out, double-edged boxes for steps in which matrices are formed or factored, and dashed arrows for calls of the shifted solve.

\subsection{Discretization in velocity space and in the cross-stream direction}\label{sec:disc}
The distribution function in peculiar velocity, $f(y,\boldsymbol c)$, and with it the perturbation $\hat f(y,\boldsymbol c)$, is discretized with $Q$ points in each velocity direction and $N+1$ points in the cross-stream spatial direction, $y$. The peculiar velocity permits a Fourier representation in velocity space. In $\boldsymbol c$ the base state is the same Maxwellian, centered at $\boldsymbol c=0$, at every $y$, so that one fixed velocity domain, truncated where the Maxwellian is negligible and periodically extended, serves all cross-stream points. By contrast, in the laboratory velocity the Maxwellian would drift with $U(y)$, and no single centered domain could be used. On this single velocity domain the full hard-sphere collision integral, which acts in velocity space only and is invariant under translations of the velocity, can be evaluated by the fast Fourier-spectral method of \citet{GHHH2017}, a key enabler of the present work, in which the loss and gain terms become convolutions computed by three-dimensional fast Fourier transforms (FFTs) of size $Q^3$. Because the operator linearized about the base state is the same at every $y$, the FFTs are needed only once, to form the linearized operator as a dense matrix. The stability computations then apply this matrix and perform no FFTs. The relaxation models need no FFTs either, since they depend on $f$ only through its moments, evaluated by quadrature. They are nevertheless discretized on the same grid, so that the four operators share one discretization, and the differences between them in \S\ref{sec:res} are those of the collision models alone.

In velocity space the grid is uniform and periodic, as the Fourier-spectral method requires, with $Q$ points in each direction,
\begin{equation}
 c_j=-L+(j-1)\,\Delta v,\quad j=1,\dots,Q,\qquad \Delta v=\frac{2L}{Q},\qquad L=\frac{3+\sqrt2}{4}\,R_v ,
 \label{eq:vgrid}
\end{equation}
where $R_v$ is the truncation radius of the relative velocity, equal to $6$ in all runs, so that $L=6.62$. The unknowns are defined only on the ball $|\boldsymbol c|\le v_{\mathrm{cut}}=L-\Delta v$ inscribed in the cube, because in the corners of the cube the truncated collision integral vanishes and each node there would be an undamped ballistic mode. Since the ball has a radius of $Q/2-1$ grid spacings, it contains $n_b\approx\tfrac{\pi}{6}(Q-2)^3$ of the $Q^3$ nodes, between $30\%$ and $43\%$ of them for $12\le Q\le32$, with $n_b=515$ at $Q=12$, $3071$ at $Q=20$ and $14\,147$ at $Q=32$. The velocity grid also has a node at $c_y=0$. Velocity integrals, such as moments, wall fluxes and the energy norm, are evaluated by the rectangle rule, which is spectrally accurate for Gaussian-weighted integrands. Consequently, in $\boldsymbol c$ multiplication by $c_x$, $c_y$, $c_z$ becomes multiplication by diagonal matrices, $\partial_{c_x}$ a matrix $\mathsf D_{c_x}$, the collision operators dense $n_b\times n_b$ matrices, and moments weighted sums. The derivative $\partial_{c_x}$, the only velocity-space operation that is neither diagonal nor a collision matrix, comes from the shear and body-force term of \eqref{eq:boltz_c}. It is the Fourier derivative along each grid line of constant $(c_y,c_z)$, corrected by a rank-5 Maxwellian-weighted term so that its moments up to fourth order are exact. Without the correction, the plain Fourier derivative does not integrate by parts exactly against the non-periodic polynomials $c_x^p$ on the truncated ball, the discrete shear term then feeds the streamwise velocity back into the wall-normal one, and at $\alpha=0$ the lift-up turns this into a spurious unstable streamwise-invariant mode, as found in this work. The correction removes that mode, preserves the spectral accuracy for the Gaussian-weighted content of $\hat f$ and speeds up the velocity convergence by an order of magnitude.

\label{sec:discN}In the cross-stream direction the discretization is performed by Legendre--Gauss--Lobatto collocation on the points $\xi_j$, $j=0,\dots,N$, the zeros of $(1-\xi^2)P_N'(\xi)$, with quadrature weights $w_j$ and differentiation matrix $\mathsf D$. With $\mathsf W_y=\mathrm{diag}(w_j)$ they satisfy the summation-by-parts identity $\mathsf W_y\mathsf D+\mathsf D^{T}\mathsf W_y=\boldsymbol e_U\boldsymbol e_U^{T}-\boldsymbol e_L\boldsymbol e_L^{T}$ exactly, for every nodal vector, $\boldsymbol e_U$ and $\boldsymbol e_L$ being the unit vectors of the two boundary points. This identity is the discrete integration by parts on which the energy stability of the discrete operator rests. For this reason Legendre--Gauss--Lobatto points are preferred to Chebyshev points, since with Chebyshev points and Clenshaw--Curtis weights the identity holds only for resolved vectors, and the numerical abscissa of the gas at rest, zero here to round-off, grows like $N^2$. This would leave the eigenvalues unaffected but would spoil the non-modal quantities. The computational points are mapped onto the physical domain by
\begin{equation}
 y=\tfrac12\,\xi,\qquad
 y=y_\infty\frac{\sinh(b_y\xi)}{\sinh b_y},\qquad
 y=\frac{y_{1/2}\,y_\infty\,(1+\xi)}{y_\infty-\xi\,(y_\infty-2y_{1/2})},
 \label{eq:maps}
\end{equation}
for the channels, the mixing layer and the boundary layer, respectively, the second clustering the points about the center of the layer, the third placing half of the grid points between the wall and $y_{1/2}$ \citep{Malik1990,Theofilis1995}. The matrix $\mathsf D$ and the weights $w_j$ are transformed with $y'(\xi_j)$, which preserves the identity. The base profile is provided at the points $y_j$ in closed form or as nodal values, and $U'=\mathsf DU$, $U''=\mathsf DU'$. In $y$, therefore, $\partial_y$ becomes $\mathsf D$, and multiplication by $U$, $U'$ and $a$ becomes multiplication by diagonal matrices. The discrete distribution function is thus the matrix $\mathsf K\in\mathbb C^{(N+1)\times n_b}$, $\mathsf K_{jk}=\hat f(y_j,\boldsymbol c_k)$, with $n=(N+1)\,n_b$ unknowns. In the computations reported below, $n$ ranges from $65\times515=33\,475$ at the coarsest resolution, $N=64$ and $Q=12$, to $193\times14\,147=2\,730\,371$ at the finest, $N=192$ and $Q=32$. The number of unknowns thus grows with the fourth power of the resolution, as $NQ^3$, against the third power for a TriGlobal problem. An incompressible TriGlobal problem in primitive variables has four unknowns per grid point, so that the resolution of most results, $N=96$ and $Q=20$, with $n=97\times3071=297\,887$, is equivalent in size to a TriGlobal problem on about $42^3$ points, and the finest, $N=192$ and $Q=32$, to one on about $88^3$ points, both within the range of TriGlobal eigenvalue problems \citep{Theofilis2011}.

\label{sec:collnum}The full Boltzmann operator for hard spheres is evaluated by this method, in which the loss term is a convolution and the gain term a quadrature over the relative speed (Gauss--Legendre, $N_r=16$ nodes on $[0,R_v]$) and over the sphere (14-point Lebedev rule), each node costing a pair of three-dimensional FFTs. Since the base state linearizes about the same Maxwellian at every $y$, $\mathcal{L}_M$ and $\mathcal{L}_{\bar g}$ are formed once, column by column, as dense $n_b\times n_b$ matrices, which depend only on $(Q,N_r,R_v)$ and are cached for all flows. A projection makes them conserve mass, momentum and energy exactly. The operator is scaled so that its discrete viscosity is exactly $1/\nu_\mu$, which makes the Reynolds number exact at any $Q$. The relaxation models are built as exact discrete projections, each a gain of rank 5 (BGK), 8 (Shakhov) or 10 (ES-BGK) minus a multiple of the identity.

\subsection{Boundary conditions}\label{sec:bcnum}
The conditions \eqref{eq:bc} are imposed weakly, by a simultaneous approximation term (SAT) \cite{CarpenterGottliebAbarbanel1994}, a penalty on the molecules that enter the domain. All $(N+1)\,n_b$ nodal values are unknowns, \eqref{eq:ivp} is collocated at every point, and at a lower wall, $y_N$, the molecules with $c_y>0$ receive the additional term
\begin{equation}
 -\frac{|c_y|}{w_N}\Bigl(\hat f_N(\boldsymbol c)-\rho_w[\hat f_N]\,M(\boldsymbol c)\Bigr),\qquad c_y>0 ,
 \label{eq:sat}
\end{equation}
where $\rho_w[\hat f_N]$ is the re-emitted density of \eqref{eq:bc}, computed by the rectangle rule from the incident values at the same point. An upper wall is treated in the same way with the roles of $c_y>0$ and $c_y<0$ exchanged, and a free boundary is the same construction with $\rho_w=0$. These terms form the boundary penalty $\mathsf S_w$, which acts on the two boundary rows only. The penalty strength $|c_y|/w_N$ is chosen so that the discrete energy estimate reproduces the continuous one. With the summation-by-parts identity of \S\ref{sec:discN}, the streaming term contributes to $\mathrm{d}E/\mathrm{d}t$ only through the two boundary points, and at a lower wall streaming and penalty together contribute
\begin{equation}
 -\tfrac12\sum_{c_y<0}|c_y|\,\frac{|\hat f_N|^2}{M}\,\Delta v^3
 -\tfrac12\sum_{c_y>0}|c_y|\,\frac{|\hat f_N-\rho_wM|^2}{M}\,\Delta v^3
 +\tfrac12\sum_{c_y>0}|c_y|\,|\rho_w|^2M\,\Delta v^3 .
 \label{eq:satenergy}
\end{equation}
The first term is the energy carried into the wall by the incident molecules, the second the residual of the penalty, and the third the energy of the re-emitted Maxwellian. Because the velocity grid is symmetric in $c_y$, the Cauchy--Schwarz inequality applied to the discrete sum that defines $\rho_w$ bounds the third term by the first, so that \eqref{eq:satenergy} is non-positive for every nodal vector. At a free boundary $\rho_w=0$ and the bound is immediate. This is the discrete form of $\mathcal{D}_w\ge0$ in \eqref{eq:RO}, that is, of the linearized inequality of Darrozes and Guiraud \cite{DarrozesGuiraud1966}, according to which diffuse reflection at a wall at rest removes disturbance energy and never supplies it. The discretization is therefore energy-stable. The numerical abscissa of the gas at rest is zero to round-off and $G(t)\le1$ at rest, so that every transient energy growth computed in a shear flow is produced by the shear and not by the discretization. Imposing the conditions strongly, by eliminating the re-emitted wall values, gives the same eigenvalues in the limit but is not energy-stable and converges more slowly in $N$.

\subsection{Discrete equations}\label{sec:discop}
With the discretizations of \S\ref{sec:disc}, the operator $\mathcal{A}$ of the initial-value problem \eqref{eq:ivp} becomes a matrix $\mathsf A$. Each term of $\mathcal{A}$ acts on one index of $\mathsf K$ only, except the shear term, which is a product of one operation on each, and the action of $\mathsf A$ on $\mathsf K$ is
\begin{equation}
 \begin{aligned}
 \mathsf T\mathsf K={}&-\mathsf D\,\mathsf K\,\mathrm{diag}(c_y)
 -\ii\alpha\bigl[\mathsf K\,\mathrm{diag}(c_x)+\mathrm{diag}(U)\,\mathsf K\bigr]
 -\ii\beta\,\mathsf K\,\mathrm{diag}(c_z)\\
 &+\bigl[\mathrm{diag}(U')\,\mathsf K\,\mathrm{diag}(c_y)-\mathrm{diag}(a)\,\mathsf K\bigr]\mathsf D_{c_x}^{T}
 +\mathsf K\,\mathsf C^{T}+\mathrm{diag}(U')\,\mathsf K\,\mathsf G^{T}+\mathsf S_w(\mathsf K),
 \end{aligned}
 \label{eq:discrete}
\end{equation}
where multiplication from the left acts on the cross-stream index ($N$) and from the right on the velocity index ($Q$), $\mathsf C$ and $\mathsf G$ are the matrices of $\mathcal{L}_M$ and $\mathcal{L}_{\bar g}$, and $\mathsf S_w$ is the boundary penalty \eqref{eq:sat} of \S\ref{sec:bcnum}. With $\boldsymbol x=\mathrm{vec}(\mathsf K)$ and $\mathsf A\boldsymbol x=\mathrm{vec}(\mathsf T\mathsf K)$, the discrete counterparts of \eqref{eq:ivp} and \eqref{eq:evp} are
\begin{equation}
 \frac{\mathrm{d}\boldsymbol x}{\mathrm{d}t}=\mathsf A\boldsymbol x ,\qquad \mathsf A\boldsymbol x=-\ii\omega\,\boldsymbol x ,
 \label{eq:discrete_ivp_evp}
\end{equation}
the former an initial-value problem (IVP), the latter a standard eigenvalue problem (EVP) with no mass matrix, in which the discretized matrix $\mathsf A$ has dimensions $n\times n$, $n=(N+1)\,n_b$.

\subsection{Numerical solution of the IVP and EVP}\label{sec:ivpsol}
\subsubsection{Operator products and shifted solves}\label{sec:shsolve}
Due to its size, $\mathsf A$ is never formed. The collision matrices are applied as one dense matrix product over all $y$, and everything that couples $y$ and $\boldsymbol c$ matrix-free. A product costs $\mathcal{O}\bigl((N+1)\,n_b^2\bigr)$ for the full operator and $\mathcal{O}\bigl((N+1)\,n_b\,(N+n_m)\bigr)$ for a relaxation model of rank $n_m$. Every computation reduces to solving $(\mathsf{A}-z\mathsf{I})\boldsymbol x=\boldsymbol b$, or its adjoint, at a shift that changes rarely. These are solved by restarted GMRES, right-preconditioned with the exact inverse, by the Woodbury identity \cite{Hager1989}, of a relaxation-model operator, which consists of velocity-diagonal streaming, phase and loss terms plus a low-rank collision and wall part. For the models at rest this inverse is exact, and with shear GMRES needs three iterations. For the full operator the preconditioner uses Shakhov at the same viscosity, and GMRES resolves only the difference of the two operators. Round-off in the inner solve grows like $\Rey/\Ma^2$, so a self-check sets the inner tolerance just above it. The adjoint solve reuses the same factors. The construction is given in \ref{app:precond}.

\subsubsection{Initial-value problem and transient growth}\label{sec:ivpgrowth}
The transient growth \eqref{eq:G} is computed by a direct--adjoint loop \cite{LuchiniBottaro2014}. The discrete energy is $E=\tfrac12\boldsymbol x^{H}\mathsf W\boldsymbol x$, with the diagonal weight matrix $\mathsf W$ of entries $w_j\,\Delta v^3/M(\boldsymbol c_k)$, the Legendre--Gauss--Lobatto weights of \S\ref{sec:discN} in $y$ and the rectangle rule in $\boldsymbol c$. In the weighted variables $\mathsf W^{1/2}\boldsymbol x$, in which this norm is Euclidean and $\mathsf A^\dagger=\mathsf W^{-1}\mathsf A^{H}\mathsf W$ is the exact adjoint of $\mathsf A$, $G(t)$ is the largest eigenvalue of $\mathrm{e}^{t\mathsf A^\dagger}\mathrm{e}^{t\mathsf A}$, found by the power iteration
\begin{equation}
 \boldsymbol x_{k+1}=\frac{\mathrm{e}^{t\mathsf A^\dagger}\mathrm{e}^{t\mathsf A}\boldsymbol x_k}{\|\mathrm{e}^{t\mathsf A^\dagger}\mathrm{e}^{t\mathsf A}\boldsymbol x_k\|},\qquad
 G(t)=\lim_{k\to\infty}\frac{\|\mathrm{e}^{t\mathsf A}\boldsymbol x_k\|^2}{\|\boldsymbol x_k\|^2} ,
 \label{eq:loop}
\end{equation}
each step of which propagates the current initial distribution forward to time $t$ with the direct operator and the result back to $t=0$ with the adjoint. From a random start, as in the continuum analysis, the direct--adjoint loop converges in four to seven steps. The converged $\boldsymbol x$ is the optimal initial distribution and $\mathrm{e}^{t\mathsf A}\boldsymbol x$ the optimal response, and the loop is repeated for each time $t$. However, unlike usual direct--adjoint looping, no time stepping is involved. The matrix $\mathsf A$ is stiff, its collisional eigenvalues being of order $\nu_\mu\propto\mathit{Re}/\Ma$, $10^6$ to $10^8$ times the growth rates of interest, which would make explicit integration impossible and implicit integration costly. Instead, each propagation is evaluated directly at time $t$ by the shift-and-invert rational Krylov method \cite{vandenEshofHochbruck2006}. Arnoldi iteration in the energy inner product with $\mathsf R=(\mathsf I-\tau\mathsf A)^{-1}$ gives $\mathsf R\mathsf V_m=\mathsf V_{m+1}\underline{\mathsf H}_m$, and
\begin{equation}
 \mathrm{e}^{t\mathsf A}\boldsymbol v\approx\|\boldsymbol v\|\,\mathsf V_m\exp\!\Bigl[\frac{t}{\tau}\bigl(\mathsf I-\mathsf H_m^{-1}\bigr)\Bigr]\boldsymbol e_1 .
 \label{eq:sai}
\end{equation}
Each Arnoldi step is one shifted solve at the real shift $z=1/\tau$, with $\tau=t/4$, so that the preconditioner is factored once per time and serves the direct and the adjoint propagation. The transformation maps the stiff collisional and streaming eigenvalues close to zero, and \eqref{eq:sai} converges in a few tens of steps independently of the stiffness. At short times, $tU/\ell\lesssim150$ in the open flows, weakly damped acoustic waves excited by the random start vector have not yet decayed. The power iteration removes most of them, but the error estimate of \eqref{eq:sai} then stays between $10^{-6}$ and $7\times10^{-5}$ instead of falling below the tolerance of $10^{-7}$, which is still below the difference of $G$ from the Orr--Sommerfeld--Squire value. The initial slope of $G$, the numerical abscissa, is the largest eigenvalue of the Hermitian part $\tfrac12(\mathsf A+\mathsf A^\dagger)$ and is found by Lanczos iteration.

\subsubsection{Eigenvalue problem and pseudospectra}\label{sec:evpsol}
The eigenvalues are computed by shift-invert Arnoldi iteration on $(\mathsf{A}-z_0\mathsf{I})^{-1}$, with a shift $z_0$ near the target phase speed displaced by $3\times10^{-4}(1+\ii)$ from the reference value. From experimentation, the Krylov dimension is 6 for one eigenvalue and 24 to 64 for a group, from a random unit-norm start vector, and every Ritz pair is accepted on its true residual. Deeper in the spectrum a pair is kept only if its residual, times the condition number of the nearest reference eigenvalue, bounds the error below $0.1$. Jacobi--Davidson iteration with the same preconditioner was implemented as an independent check of single eigenvalues. Each case is run at three Mach numbers and extrapolated to $\Ma\to0$ by the fits of the results table. Each Arnoldi step is one shifted solve of \S\ref{sec:shsolve} at the fixed shift $z_0$. The pseudospectra \eqref{eq:pseudo} follow from the smallest singular value of $\mathsf W^{1/2}(z\mathsf I-\mathsf A)\mathsf W^{-1/2}$ on a grid of points $z$ in the complex plane, found by inverse Lanczos iteration \cite{Wright2001}, each step of which costs one direct and one adjoint shifted solve. The iteration converges in a few tens of steps, and the points of the grid are independent.

\subsection{Reconstruction of amplitude functions in physical space}\label{sec:recon}
The solvers return vectors $\boldsymbol x$ of the discrete distribution function, namely optimal initial distributions and their responses (\S\ref{sec:ivpgrowth}), and eigenvectors (\S\ref{sec:evpsol}). Every physical quantity is a moment \eqref{eq:moments} of $\hat f$, evaluated by quadrature in velocity space and, where it is integrated over the layer, in $y$. The vector is reshaped into $\mathsf K_{jk}=\hat f(y_j,\boldsymbol c_k)$, and the velocity integrals of the moments \eqref{eq:moments} are approximated at every collocation point by the rectangle rule over the $n_b$ nodes of the ball, which replaces $\int\cdots\,\mathrm{d}\boldsymbol c$ by $\sum_k\cdots\,\Delta v^3$,
\begin{subequations}\label{eq:moments_disc}
\begin{align}
 \hat\rho(y_j)&=\sum_{k=1}^{n_b}\hat f(y_j,\boldsymbol c_k)\,\Delta v^3,\\
 \hat{\boldsymbol u}(y_j)&=\sum_{k=1}^{n_b}\boldsymbol c_k\,\hat f(y_j,\boldsymbol c_k)\,\Delta v^3,\\
 \hat T(y_j)&=\sum_{k=1}^{n_b}\bigl(\tfrac13|\boldsymbol c_k|^2-1\bigr)\,\hat f(y_j,\boldsymbol c_k)\,\Delta v^3,\\
 \hat P_{il}(y_j)&=\sum_{k=1}^{n_b}c_{i,k}\,c_{l,k}\,\hat f(y_j,\boldsymbol c_k)\,\Delta v^3,\\
 \hat{\boldsymbol q}(y_j)&=\sum_{k=1}^{n_b}\tfrac12\boldsymbol c_k\bigl(|\boldsymbol c_k|^2-5\bigr)\,\hat f(y_j,\boldsymbol c_k)\,\Delta v^3,
 \qquad j=0,\dots,N.
\end{align}
\end{subequations}
The sums run over the velocity nodes only, at a fixed cross-stream index $j$. Velocity space is integrated out and the cross-stream coordinate is not, so that each moment is a function of the physical coordinate $y$, known at the $N+1$ collocation points. Multiplied by the normal-mode factor of \eqref{eq:ansatz}, and for a mode by $\mathrm{e}^{-\ii\omega t}$, it gives the perturbation of the macroscopic quantity in physical space, for example $u_x(x,y,z,t)=\mathrm{Re}\{\hat u_x(y)\,\mathrm{e}^{\ii(\alpha x+\beta z-\omega t)}\}$.
For all $y_j$ at once \eqref{eq:moments_disc} is a single product $\mathsf K\Psi$, with $\Psi_{km}=\psi_m(\boldsymbol c_k)\,\Delta v^3$ the weights of the thirteen moments. On the uniform grid the rectangle rule is the trapezoidal rule of an integrand that decays like the Maxwellian, which at the edge of the ball, $|\boldsymbol c|\approx6$, is of order $10^{-8}$ of its central value. The quadrature is therefore spectrally accurate, and its error is set by the truncation radius rather than by the spacing $\Delta v$. The node at $c_y=0$ makes the half-range sums of the wall condition \eqref{eq:bc} coincide with the trapezoidal rule as well. Because the discrete collision matrices conserve the five discrete moments exactly (\S\ref{sec:collnum}), the discrete moments of $\mathsf A\boldsymbol x$ satisfy the conservation laws \eqref{eq:cons} with the collision terms vanishing to round-off. The velocity moments $\hat u_x$, $\hat u_y$, $\hat u_z$ at the points $y_j$ are the amplitude functions of the streamwise, wall-normal and spanwise velocity perturbations ($\hat u$, $\hat v$, $\hat w$ in Orr--Sommerfeld notation), and are compared at the same points with the Orr--Sommerfeld and Squire solutions. No condition has been imposed on them at any boundary, so that no-slip at a wall and decay in the free stream are results of the computation.

Integrals across the layer use the Legendre--Gauss--Lobatto weights $w_j$ of \S\ref{sec:discN}. The discrete energy is
\begin{equation}
 E=\tfrac12\sum_j w_j\sum_k\frac{|\hat f_{jk}|^2}{M(\boldsymbol c_k)}\,\Delta v^3=\tfrac12\,\boldsymbol x^{H}\mathsf W\boldsymbol x ,
 \label{eq:Edisc}
\end{equation}
the norm with the weight matrix $\mathsf W$ of \S\ref{sec:ivpsol}, and the energy carried by each moment is $e_q=\sum_jw_j|\hat q(y_j)|^2$, $q\in\{\rho,u_x,u_y,u_z,T\}$. The hydrodynamic part of $\hat f$ is its projection onto the five discrete collision invariants $\psi_m\in\{1,c_x,c_y,c_z,|\boldsymbol c|^2\}$,
\begin{equation}
 \Pi_h\hat f=M\,\boldsymbol\psi^{T}\mathsf G^{-1}\Bigl(\sum_k\boldsymbol\psi(\boldsymbol c_k)\,\hat f_k\,\Delta v^3\Bigr),\qquad
 \mathsf G_{mn}=\sum_k\psi_m\psi_n\,M\,\Delta v^3 ,
 \label{eq:Pih}
\end{equation}
which reproduces the five discrete moments exactly. The energy of the remainder, $E\bigl((\mathsf I-\Pi_h)\hat f\bigr)$, measures how far the disturbance is from a local Maxwellian and is compared with its Chapman--Enskog estimate. Values between the points follow from the Legendre interpolant. Eigenfunctions are normalized by $\max_j|\hat u_x(y_j)|$, and optimal disturbances, as in the Orr--Sommerfeld--Squire computation with which they are compared, by their largest component. A mode is labeled in-plane if $e_{u_x}+e_{u_y}$ exceeds half of the total energy of the moments, spanwise if $e_{u_z}$ does, and thermodynamic otherwise. At $\beta=0$ these are the kinetic counterparts of the Orr--Sommerfeld, the Squire and the thermal modes.

\section{Verification}\label{sec:verification}
Before the eigenvalues are compared with those of the Orr--Sommerfeld equation, the discrete operator and the linear solver are tested on their own, against a solution known in closed form. The test separates two questions which a comparison of eigenvalues answers only together, namely whether the equations of \S\ref{sec:theory} have been implemented correctly, and how accurately a given grid represents the eigenfunctions of a shear flow.

\subsection{Manufactured solution}\label{sec:mms}
The manufactured amplitude function is
\begin{equation}
 \hat f_m(y,\boldsymbol c)=M(\boldsymbol c)\,\bigl[A(y)+w(y)\,p(\boldsymbol c)\bigr],\qquad -\tfrac12\le y\le\tfrac12 ,
 \label{eq:mms}
\end{equation}
with $w(y)=\cos\pi y$, which vanishes at both ends of the domain, and a cubic polynomial $p(\boldsymbol c)$ with ten complex coefficients and no symmetry, so that the density, velocity, temperature, stress and heat-flux moments of $\hat f_m$ are all non-zero. Two choices of $A$ cover the two kinds of boundary. With $A=(1+\tfrac12\ii)\exp(0.8\,y)$ the function reduces at each wall to $A\,M$, a re-emitted Maxwellian whose density satisfies the impermeability condition \eqref{eq:bc} identically. With $A=(1+\tfrac12\ii)\sin[\tfrac\pi2(\tfrac12-y)]\exp(0.8\,y)$ the function vanishes at $y=\tfrac12$, as the free-stream condition requires, and satisfies the wall condition at $y=-\tfrac12$. Neither $A$ nor $w$ is a polynomial, so that the collocation grid does not represent $\hat f_m$ exactly. Unlike an eigenfunction of the kinetic problem, $\hat f_m$ is smooth, since it is continuous across $c_y=0$ at the walls and has no Knudsen layer.

The base flow is $U(y)=U\tanh(3y)$ with $U=0.3$, held steady by the force \eqref{eq:force}, so that $U$, $U'$ and $a$ all vary across the domain, and the wavenumbers are $\alpha=1.3$ and $\beta=0.7$, so that the spanwise term is exercised as well. Two collision rates are used, namely $\nu_\mu=20$, at which streaming, shear and collisions are of comparable size, and $\nu_\mu=2000$, which is in the near-continuum regime of the stability computations. For the three relaxation models every term of the right-hand side of \eqref{eq:ivp} applied to $\hat f_m$ is a Maxwellian times a polynomial with coefficients that depend on $y$, and is evaluated in closed form, since the moments of $\hat f_m$ are Gaussian integrals, and $\mathcal{L}_M$ and $\mathcal{L}_{\bar g}$ are taken from Appendix~\ref{app:linops}. The closed form is coded independently of the stability code, which builds the same terms from discrete projections (\S\ref{sec:collnum}).

Two tests are made. In the first, the discrete operator is applied to the nodal values of $\hat f_m$ and the result is compared with the closed form at all points. With the weak wall conditions of \S\ref{sec:bcnum} every entry, the re-emitted wall values included, satisfies the discrete equation, and the penalty \eqref{eq:sat} vanishes on $\hat f_m$, which satisfies diffuse reflection exactly. In the second, the system $(\mathsf A-z_0\mathsf I)\,\boldsymbol x=\boldsymbol b$ is solved with $\boldsymbol b$ formed from the closed form, at the shift $z_0=0.5-\ii$, by the preconditioned GMRES iteration of \S\ref{sec:shsolve} with a tolerance of $10^{-12}$, and $\boldsymbol x$ is compared with $\hat f_m$. The first test involves the differentiation matrices, the velocity quadrature, the discrete projections and the base-stress terms. The second adds the penalty terms, the preconditioner and the inner iteration. Errors are relative, in the Euclidean norm.

\subsection{Operator and solver}\label{sec:mms_res}
\begin{figure}[!htbp]
\centering
\begin{tabular}{@{}c@{\hspace{0.02\textwidth}}c@{}}
\includegraphics[width=0.48\textwidth]{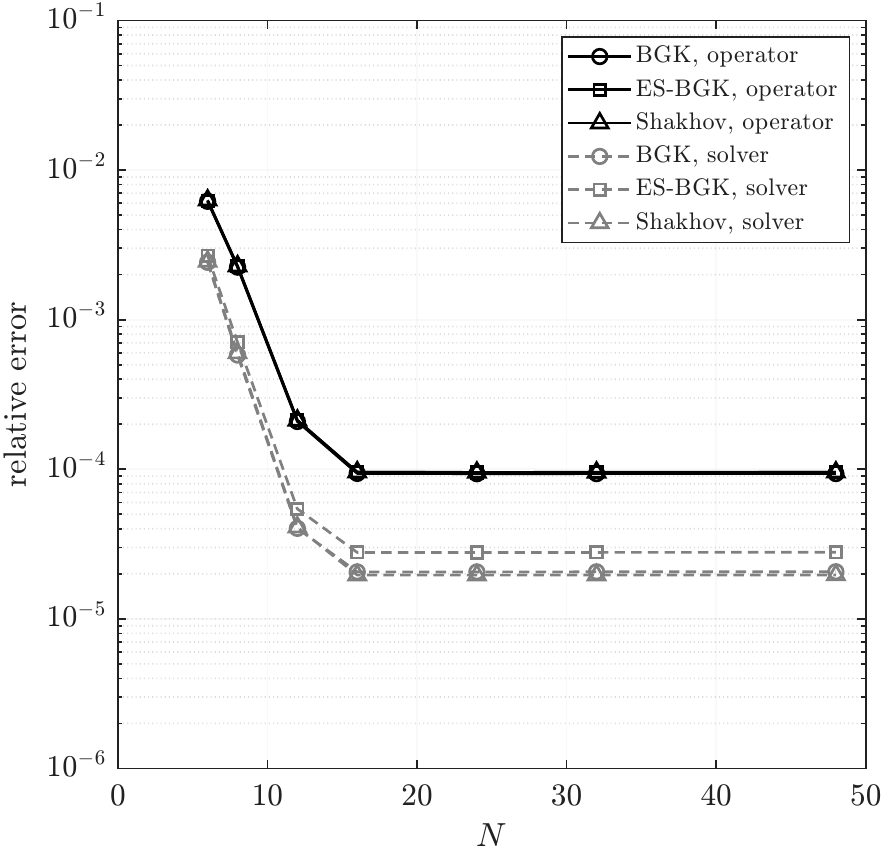} &
\includegraphics[width=0.48\textwidth]{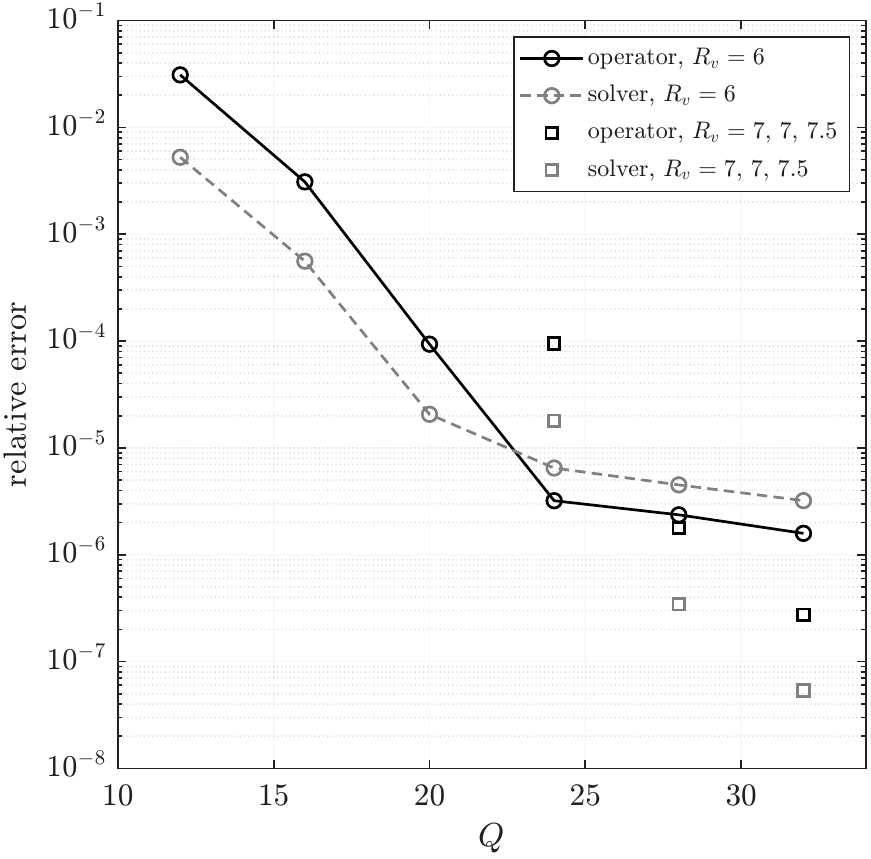}\\
(\textit{a}) & (\textit{b})
\end{tabular}
\caption{Manufactured solution, two walls, $\nu_\mu=20$: error of operator (black) and solver (gray). (\textit{a}) Against $N$, $Q=20$, relaxation models; (\textit{b}) against $Q$, $N=32$, BGK.}
\label{fig:mms}
\end{figure}

Table~\ref{tab:mms_N} gives the errors of the operator and of the solver for the three relaxation models and the full operator, with two walls and $\nu_\mu=20$, in the upper part against the cross-stream resolution $N$ at $Q=20$, $R_v=6$, and in the lower part against the velocity resolution $Q$ at $N=32$, for $R_v=6$ and for larger velocity balls, $v_{\mathrm{cut}}$ being the ball radius of \S\ref{sec:disc}. For the full operator the collision term of the right-hand side is evaluated with the discrete matrices and all other terms in closed form, so that its operator error is that of the streaming, shear, forcing and boundary terms alone. Figure~\ref{fig:mms} shows the same errors, of the operator in black and of the solver in gray, in (\textit{a}) against $N$ at $Q=20$, $R_v=6$, for the three relaxation models, and in (\textit{b}) against $Q$ at $N=32$ for BGK, with circles for $R_v=6$ and squares for $R_v=7$ ($Q=24$, $28$) and $R_v=7.5$ ($Q=32$).

\begin{table}[!htbp]
\centering\footnotesize\setlength{\tabcolsep}{4pt}
\begin{tabular}{@{}c@{}}
\adjustbox{max width=\textwidth}{\begin{tabular}{@{}rcccccccc@{}}
\toprule
 & \multicolumn{4}{c}{operator} & \multicolumn{4}{c}{solver}\\
\cmidrule(lr){2-5}\cmidrule(l){6-9}
$N$ & BGK & ES-BGK & Shakhov & full (HS)$^{\dagger}$ & BGK & ES-BGK & Shakhov & full (HS)$^{\dagger}$\\
\midrule
6 & $6.2\times10^{-3}$ & $6.2\times10^{-3}$ & $6.3\times10^{-3}$ & $6.8\times10^{-3}$ & $2.4\times10^{-3}$ & $2.7\times10^{-3}$ & $2.4\times10^{-3}$ & $2.4\times10^{-3}$\\
8 & $2.2\times10^{-3}$ & $2.3\times10^{-3}$ & $2.3\times10^{-3}$ & $2.4\times10^{-3}$ & $5.8\times10^{-4}$ & $7.1\times10^{-4}$ & $6.0\times10^{-4}$ & $5.9\times10^{-4}$\\
12 & $2.1\times10^{-4}$ & $2.1\times10^{-4}$ & $2.1\times10^{-4}$ & $2.3\times10^{-4}$ & $4.0\times10^{-5}$ & $5.4\times10^{-5}$ & $4.1\times10^{-5}$ & $3.9\times10^{-5}$\\
16 & $9.4\times10^{-5}$ & $9.6\times10^{-5}$ & $9.6\times10^{-5}$ & $1.0\times10^{-4}$ & $2.1\times10^{-5}$ & $2.8\times10^{-5}$ & $2.0\times10^{-5}$ & $1.8\times10^{-5}$\\
24 & $9.3\times10^{-5}$ & $9.5\times10^{-5}$ & $9.5\times10^{-5}$ & $1.0\times10^{-4}$ & $2.1\times10^{-5}$ & $2.8\times10^{-5}$ & $2.0\times10^{-5}$ & $1.7\times10^{-5}$\\
32 & $9.3\times10^{-5}$ & $9.5\times10^{-5}$ & $9.5\times10^{-5}$ & $1.0\times10^{-4}$ & $2.1\times10^{-5}$ & $2.8\times10^{-5}$ & $2.0\times10^{-5}$ & $1.8\times10^{-5}$\\
48 & $9.4\times10^{-5}$ & $9.5\times10^{-5}$ & $9.6\times10^{-5}$ & $1.0\times10^{-4}$ & $2.1\times10^{-5}$ & $2.8\times10^{-5}$ & $2.0\times10^{-5}$ & $1.8\times10^{-5}$\\
\bottomrule
\end{tabular}
}\\[10pt]
\adjustbox{max width=\textwidth}{\begin{tabular}{@{}rcccccccccc@{}}
\toprule
 & & & \multicolumn{4}{c}{operator} & \multicolumn{4}{c}{solver}\\
\cmidrule(lr){4-7}\cmidrule(l){8-11}
$Q$ & $R_v$ & $v_{\mathrm{cut}}$ & BGK & ES-BGK & Shakhov & full (HS)$^{\dagger}$ & BGK & ES-BGK & Shakhov & full (HS)$^{\dagger}$\\
\midrule
12 & 6 & 5.52 & $3.1\times10^{-2}$ & $3.1\times10^{-2}$ & $3.1\times10^{-2}$ & $3.4\times10^{-2}$ & $5.2\times10^{-3}$ & $7.4\times10^{-3}$ & $5.3\times10^{-3}$ & $1.1\times10^{-2}$\\
16 & 6 & 5.79 & $3.1\times10^{-3}$ & $3.1\times10^{-3}$ & $3.1\times10^{-3}$ & $3.3\times10^{-3}$ & $5.6\times10^{-4}$ & $8.2\times10^{-4}$ & $5.6\times10^{-4}$ & $5.6\times10^{-4}$\\
20 & 6 & 5.96 & $9.3\times10^{-5}$ & $9.5\times10^{-5}$ & $9.5\times10^{-5}$ & $1.0\times10^{-4}$ & $2.1\times10^{-5}$ & $2.8\times10^{-5}$ & $2.0\times10^{-5}$ & $1.8\times10^{-5}$\\
24 & 6 & 6.07 & $3.2\times10^{-6}$ & $2.9\times10^{-6}$ & $2.8\times10^{-6}$ & $2.7\times10^{-6}$ & $6.5\times10^{-6}$ & $4.9\times10^{-6}$ & $5.0\times10^{-6}$ & $3.1\times10^{-6}$\\
28 & 6 & 6.15 & $2.4\times10^{-6}$ & $2.1\times10^{-6}$ & $2.1\times10^{-6}$ & $2.0\times10^{-6}$ & $4.5\times10^{-6}$ & $3.4\times10^{-6}$ & $3.5\times10^{-6}$ & $2.2\times10^{-6}$\\
32 & 6 & 6.21 & $1.6\times10^{-6}$ & $1.4\times10^{-6}$ & $1.4\times10^{-6}$ & $1.3\times10^{-6}$ & $3.2\times10^{-6}$ & $2.4\times10^{-6}$ & $2.5\times10^{-6}$ & $1.4\times10^{-6}$\\
\midrule
24 & 7 & 7.08 & $9.5\times10^{-5}$ & $9.7\times10^{-5}$ & $9.7\times10^{-5}$ & $1.0\times10^{-4}$ & $1.8\times10^{-5}$ & $2.7\times10^{-5}$ & $1.8\times10^{-5}$ & $1.7\times10^{-5}$\\
28 & 7 & 7.17 & $1.8\times10^{-6}$ & $1.8\times10^{-6}$ & $1.8\times10^{-6}$ & $1.9\times10^{-6}$ & $3.4\times10^{-7}$ & $5.2\times10^{-7}$ & $3.4\times10^{-7}$ & $3.2\times10^{-7}$\\
32 & 7.5 & 7.76 & $2.7\times10^{-7}$ & $2.8\times10^{-7}$ & $2.8\times10^{-7}$ & $3.0\times10^{-7}$ & $5.4\times10^{-8}$ & $8.2\times10^{-8}$ & $5.4\times10^{-8}$ & $5.0\times10^{-8}$\\
\bottomrule
\end{tabular}
}
\end{tabular}
\caption{Manufactured solution, two walls, $\nu_\mu=20$: relative errors of operator and solver. Top: against $N$, $Q=20$, $R_v=6$. Bottom: against $Q$, $N=32$. $^{\dagger}$Collision term evaluated with the discrete matrices.}
\label{tab:mms_N}
\end{table}

Both errors decrease spectrally with $N$, by almost two orders of magnitude between $N=6$ and $16$, and then remain at the level set by the velocity grid. The three models behave identically, which verifies the projections and the base-stress terms of each. For a smooth solution the collocation in $y$ and the treatment of the boundaries are therefore spectrally accurate, and any slow residual drift of the eigenvalues of wall-bounded flows with $N$ is a property not of the implementation but of the eigenfunctions, which are discontinuous across $c_y=0$ at a wall and carry a Knudsen layer that the grid does not resolve.

At $R_v=6$ the errors fall by four orders of magnitude from $Q=12$ to $24$ and then level off near $10^{-6}$. The remaining error is that of the truncation of velocity space to the ball, whose radius grows only from $6.07$ to $6.21$ between $Q=24$ and $32$, and not of the grid spacing. With a larger ball the operator error of BGK is $1.8\times10^{-6}$ at $R_v=7$, $Q=28$ and $2.7\times10^{-7}$ at $R_v=7.5$, $Q=32$, the solver error $3.4\times10^{-7}$ and $5.4\times10^{-8}$, and the other three operators agree with these to within a factor of $1.7$. Two errors thus compete, that of the spacing $\Delta v$ and that of the radius $v_{\mathrm{cut}}$, and at $R_v=7$, $Q=24$, where the spacing is that of $R_v=6$, $Q=20$, the error is again of order $10^{-4}$. The cubic polynomial of \eqref{eq:mms} puts more weight on the tail of the Maxwellian than a hydrodynamic mode does, so that the test is a severe one for the truncation.

The moment-consistent velocity derivative of \S\ref{sec:disc} is included in these tests, and it raises the pointwise error on coarse velocity grids. Without the correction the operator error of BGK at $N=32$ is $8.2\times10^{-3}$, $5.8\times10^{-4}$, $1.6\times10^{-5}$ and $2.1\times10^{-6}$ at $Q=12$, $16$, $20$ and $24$, against $3.1\times10^{-2}$, $3.1\times10^{-3}$, $9.3\times10^{-5}$ and $3.2\times10^{-6}$ with it, the two coinciding to within a factor of $1.6$ from $Q=24$ on. The correction makes the moments $p\le4$ of the discrete derivative exact by moving the truncation error of each line from its tails, where the Maxwellian is small, into its core. The manufactured test, which weights all points alike and has a large tail, registers this as a larger pointwise error, while the eigenvalues, which depend on the low moments, converge faster with the correction than without it.

Table~\ref{tab:mms_cases} repeats the test, in its upper part, with a free upper boundary and at the larger collision rate, at $Q=20$, $R_v=6$, $N=32$, with the number of preconditioned GMRES iterations to a relative residual of $10^{-12}$. Its lower part gives the errors against the velocity grid for two walls at $\nu_\mu=2000$, $N=32$. The free-stream condition is verified to the same accuracy as the wall condition. The inner iteration takes twelve or thirteen steps at $\nu_\mu=20$, where the part of the shear term that the preconditioner omits is not small, and four or five at $\nu_\mu=2000$. At $\nu_\mu=2000$ the error of the solution is larger than that of the operator by a factor of about $200$, because in the near-continuum regime the inverse of the shifted operator amplifies a consistency error in proportion to the collision rate. This error, too, is that of the ball truncation and is removed by a larger ball. At $R_v=7$, $Q=28$ the solver error is $7\times10^{-7}$.

\begin{table}[!htbp]
\centering\footnotesize\setlength{\tabcolsep}{5pt}
\begin{tabular}{@{}c@{}}
\adjustbox{max width=\textwidth}{\begin{tabular}{@{}lclccc@{}}
\toprule
boundaries & $\nu_\mu$ & model & operator & solver & GMRES its.\\
\midrule
wall, wall & $20$ & BGK & $9.3\times10^{-5}$ & $2.1\times10^{-5}$ & 12\\
wall, wall & $20$ & ES-BGK & $9.5\times10^{-5}$ & $2.8\times10^{-5}$ & 13\\
wall, wall & $20$ & Shakhov & $9.5\times10^{-5}$ & $2.0\times10^{-5}$ & 12\\
wall, wall & $20$ & full (HS)$^{\dagger}$ & $1.0\times10^{-4}$ & $1.8\times10^{-5}$ & 29\\
wall, wall & $2000$ & BGK & $4.3\times10^{-6}$ & $7.4\times10^{-4}$ & 4\\
wall, wall & $2000$ & ES-BGK & $3.1\times10^{-6}$ & $5.0\times10^{-4}$ & 5\\
wall, wall & $2000$ & Shakhov & $2.5\times10^{-6}$ & $5.2\times10^{-4}$ & 4\\
wall, wall & $2000$ & full (HS)$^{\dagger}$ & $1.3\times10^{-6}$ & $1.3\times10^{-5}$ & 25\\
\midrule
wall, free & $20$ & BGK & $8.8\times10^{-5}$ & $3.0\times10^{-5}$ & 12\\
wall, free & $20$ & ES-BGK & $9.0\times10^{-5}$ & $4.0\times10^{-5}$ & 13\\
wall, free & $20$ & Shakhov & $9.0\times10^{-5}$ & $2.8\times10^{-5}$ & 12\\
wall, free & $20$ & full (HS)$^{\dagger}$ & $9.6\times10^{-5}$ & $2.5\times10^{-5}$ & 29\\
wall, free & $2000$ & BGK & $4.3\times10^{-6}$ & $1.1\times10^{-3}$ & 4\\
wall, free & $2000$ & ES-BGK & $3.1\times10^{-6}$ & $7.5\times10^{-4}$ & 5\\
wall, free & $2000$ & Shakhov & $2.5\times10^{-6}$ & $7.7\times10^{-4}$ & 4\\
wall, free & $2000$ & full (HS)$^{\dagger}$ & $1.3\times10^{-6}$ & $1.7\times10^{-5}$ & 25\\
\bottomrule
\end{tabular}
}\\[10pt]
\adjustbox{max width=\textwidth}{\begin{tabular}{@{}rcccccccccc@{}}
\toprule
 & & \multicolumn{4}{c}{operator} & \multicolumn{4}{c}{solver}\\
\cmidrule(lr){3-6}\cmidrule(l){7-10}
$Q$ & $R_v$ & BGK & ES-BGK & Shakhov & full (HS)$^{\dagger}$ & BGK & ES-BGK & Shakhov & full (HS)$^{\dagger}$\\
\midrule
20 & 6 & $4.3\times10^{-6}$ & $3.1\times10^{-6}$ & $2.5\times10^{-6}$ & $1.3\times10^{-6}$ & $7.4\times10^{-4}$ & $5.0\times10^{-4}$ & $5.2\times10^{-4}$ & $1.3\times10^{-5}$\\
24 & 6 & $2.5\times10^{-6}$ & $1.7\times10^{-6}$ & $1.6\times10^{-6}$ & $3.4\times10^{-8}$ & $4.3\times10^{-4}$ & $2.9\times10^{-4}$ & $2.9\times10^{-4}$ & $8.7\times10^{-6}$\\
28 & 6 & $1.7\times10^{-6}$ & $1.2\times10^{-6}$ & $1.1\times10^{-6}$ & $2.5\times10^{-8}$ & $2.9\times10^{-4}$ & $2.0\times10^{-4}$ & $2.0\times10^{-4}$ & $6.1\times10^{-6}$\\
28 & 7 & $2.1\times10^{-8}$ & $2.2\times10^{-8}$ & $2.2\times10^{-8}$ & $2.5\times10^{-8}$ & $7.1\times10^{-7}$ & $4.7\times10^{-7}$ & $4.8\times10^{-7}$ & $2.1\times10^{-8}$\\
\bottomrule
\end{tabular}
}
\end{tabular}
\caption{Manufactured solution. Top: two kinds of boundary and two collision rates, $Q=20$, $N=32$. Bottom: two walls, $\nu_\mu=2000$, $N=32$, against $Q$. $^{\dagger}$As in table~\ref{tab:mms_N}.}
\label{tab:mms_cases}
\end{table}

Since the manufactured function overstates the weight of the Maxwellian tail, the effect of the ball radius on the eigenvalues was measured directly. For the least-stable modes of the two channel flows (BGK, $\Ma=0.1$, $N=64$), enlarging the ball from $R_v=6$, $Q=24$ to $R_v=7$, $Q=28$, at the same spacing $\Delta v=0.55$, changes the eigenvalue by $5\times10^{-7}$ or less in the phase speed and $2\times10^{-7}$ or less in the growth parameter. At $R_v=6$ the truncation of velocity space does not limit the eigenvalues, whose dependence on $Q$ is that of the grid spacing. The path by which a profile given as nodal values enters (\S\ref{sec:discN}) was checked with plane Poiseuille flow supplied as a file, and it reproduces the eigenvalue obtained with the built-in profile to ten digits.

\subsection{The hard-sphere collision matrix}\label{sec:mms_full}
No closed form is available for the linearized hard-sphere operator acting on a function such as \eqref{eq:mms}. The test is therefore repeated for the full operator with the collision term of the right-hand side evaluated with the dense matrices, and the streaming, shear, forcing and boundary terms in closed form (columns full (HS) of tables~\ref{tab:mms_N} and~\ref{tab:mms_cases}). The operator error is then that of these terms alone, and the solver error measures the preconditioned solution of the system with the full operator. The errors against $N$ repeat those of the models, with spectral decay to the level of the velocity grid. Against $Q$ the solver error falls from $1.1\times10^{-2}$ at $Q=12$ to $3.1\times10^{-6}$ at $Q=24$ and, with the larger balls, to $3.2\times10^{-7}$ ($R_v=7$, $Q=28$) and $5.0\times10^{-8}$ ($R_v=7.5$, $Q=32$), as for the models. The inner iteration takes $29$ to $35$ steps to a residual of $10^{-12}$ at $\nu_\mu=20$ and $23$ to $27$ at $\nu_\mu=2000$, with either kind of boundary and ball. Only at $Q=12$, where the discrete operator is not resolved (table~\ref{tab:operator}), does it take $211$. At $\nu_\mu=2000$ the solver error is $9\times10^{-6}$ (two walls, $Q=24$) and $1.7\times10^{-5}$ (wall and free boundary, $Q=20$), against $5\times10^{-4}$ to $10^{-3}$ for the models. Here the collision term of the right-hand side is consistent with the discrete operator by construction, and the amplification of the consistency error by the collision rate is absent, which confirms that the amplified error of the models is that of their collision terms on the truncated ball.

The matrices themselves are verified by properties that the exact operator has. Table~\ref{tab:operator} characterizes the discrete hard-sphere operator on the ball, $R_v=6$, with $N_r=16$ radial points, namely the number $n_b$ of ball nodes, the ball radius, the viscosity and Prandtl number relative to the hard-sphere values $\mu_{\mathrm{HS}}=0.79588$ (mean-free-path units) and $\Pran_{\mathrm{HS}}=0.66069$, the defect $\max_m\|\mathcal{L}_M M\psi_m\|/\|M\psi_m\|$ of the Maxwellian null vectors, and the relative asymmetry $\|\mathsf B-\mathsf B^{T}\|_F/\|\mathsf B\|_F$ of the matrix
\begin{equation}
 \mathsf{B}_{ab}=\sum_{\boldsymbol c}\phi_a(\boldsymbol c)\,\bigl[\mathcal{L}_M(M\phi_b)\bigr](\boldsymbol c)\,\Delta v^3 ,
 \label{eq:Bsym}
\end{equation}
$\phi_a$ being the monomials in $\boldsymbol c$ up to the degree shown, which must be symmetric since the operator is self-adjoint in the inner product weighted by $1/M$. The five collision invariants are left null vectors to round-off, by the projection of \S\ref{sec:collnum}, and right null vectors to $3\times10^{-4}$ at $Q=20$. The Prandtl number is within $0.2\%$ of the hard-sphere value for $Q\ge20$. The viscosity is about $1.5\%$ above it for $Q\ge16$, which is immaterial for the stability results because the operator is scaled by its own discrete viscosity. On the quadratics, which carry the stress, the discrete operator is self-adjoint to $10^{-9}$ at $Q=20$, and on the cubics, which carry the heat flux, to $7\times10^{-5}$, without improvement from $Q=20$ to $24$, in line with the null-vector defect, which at fixed $N_r$ is presumably limited by the radial quadrature. The momentum transport of the discrete hard-sphere operator is thus verified to a higher accuracy than its energy transport.

\begin{table}[!htbp]
\centering\small
\adjustbox{max width=\textwidth}{\begin{tabular}{@{}rrccccccc@{}}
\toprule
 & & & & & & \multicolumn{3}{c}{asymmetry, degree $\le$}\\
\cmidrule(l){7-9}
$Q$ & $n_b$ & $v_{\mathrm{cut}}$ & $\mu_d/\mu_{\mathrm{HS}}$ & $\Pran_d/\Pran_{\mathrm{HS}}$ & null-vector defect & 2 & 3 & 4\\
\midrule
12 & 515 & 5.518 & 0.9823 & 0.8644 & $1.9\times10^{-1}$ & -- & -- & --\\
16 & 1419 & 5.794 & 1.0130 & 0.9638 & $3.4\times10^{-3}$ & $2.3\times10^{-7}$ & $3.1\times10^{-3}$ & $3.6\times10^{-3}$\\
20 & 3071 & 5.959 & 1.0148 & 0.9995 & $2.9\times10^{-4}$ & $1.0\times10^{-9}$ & $7.1\times10^{-5}$ & $5.1\times10^{-4}$\\
24 & 5575 & 6.070 & 1.0149 & 1.0021 & $3.7\times10^{-4}$ & $6.4\times10^{-11}$ & $8.1\times10^{-5}$ & $3.3\times10^{-4}$\\
\bottomrule
\end{tabular}
}
\caption{The discrete hard-sphere operator on the ball, $R_v=6$, $N_r=16$: nodes, radius, viscosity and Prandtl number, null-vector defect and asymmetry of \eqref{eq:Bsym}.}
\label{tab:operator}
\end{table}

\FloatBarrier
\subsection{Energy stability, adjoint and transient growth}\label{sec:verif_nonmodal}
The non-modal quantities of \S\ref{sec:tg} rest on four properties of the discrete operator, each of which is tested separately here, with the BGK model at $Q=16$, $R_v=6$ unless stated otherwise. For random vectors the adjoint identity $\langle\mathsf A\boldsymbol x,\boldsymbol y\rangle=\langle\boldsymbol x,\mathsf A^{H}\boldsymbol y\rangle$ holds to $10^{-17}$ relative to $\|\mathsf A\boldsymbol x\|\|\boldsymbol y\|$, for the complete operator and for each of its parts (streaming with the wall penalty, collisions, shear and force, $U'\mathcal{L}_{\bar g}$), at rest and in Couette flow, and the adjoint preconditioner of \S\ref{sec:ivpgrowth} satisfies the corresponding identity to $10^{-13}$. For smooth distributions $\hat f=M\,p(\boldsymbol c)\,s(y)$, with $p$ a random cubic polynomial and $s$ a smooth function, the four terms of the discrete energy rate $\mathrm{Re}\langle\hat f,\mathsf A\hat f\rangle$ in the norm \eqref{eq:Edisc} were evaluated separately (Couette flow, $\Rey=20$, $\Ma=0.1$, $N=24$). They add up to the total to $10^{-14}$, the collision term is negative and the discrete collision operator is self-adjoint to $10^{-14}$, and the shear and force term agrees with the production of \eqref{eq:RO}, evaluated by quadrature of its continuous expression, to $4\times10^{-6}$ ($6\times10^{-5}$ with the uncorrected velocity derivative).

The largest eigenvalue $\omega_{\mathrm{num}}$ of the Hermitian part of $\mathsf A$ in the energy norm, the numerical abscissa, bounds the initial growth, $G(t)\le\mathrm{e}^{2\omega_{\mathrm{num}}t}$, and $\omega_{\mathrm{num}}\le0$ is equivalent to $G(t)\le1$ for all $t$. It is computed by Lanczos iteration. For the gas at rest the continuous operator has $\omega_{\mathrm{num}}=0$ by \eqref{eq:RO}. Table~\ref{tab:abscissa} gives, in its upper part, the discrete numerical abscissa of the gas at rest, in units of $\sqrt{RT_0}/H$ (BGK, $Q=12$, $\nu_\mu=\sqrt2$, $\alpha=0.5$, $\beta=1$), for three discretizations of the cross-stream direction and of the walls, namely Chebyshev points with strongly imposed wall values, Legendre points with strongly imposed wall values, and the present Legendre points with weak wall conditions. Only the last inherits the property, while the positive values of the other two grow like $N^2$. In shear flow the kinetic numerical abscissa is the inviscid value $\max|U'|/2$ at every Reynolds number, with $\omega_{\mathrm{num}}=0.5006$ and $0.5003$ (in units $U/h$) for Couette flow at $\Rey=20$ and $1000$, and $0.988$ for Poiseuille flow at $\Rey=1000$, which tends to $\max|U'|/2=1$ as the wall region is resolved. This is not a discretization effect but a property of \eqref{eq:RO}. A hydrodynamic distribution has no non-equilibrium part, so its collisional dissipation is zero and its instantaneous energy rate is the production alone, viscous dissipation appearing only after one collision time $1/\nu_\mu$. The kinetic description therefore has no counterpart of the energy-stability Reynolds number of the Navier--Stokes equations ($20.7$ for Couette flow), below which no growth whatever is possible. The additional growth it admits is confined to times of order $1/\nu_\mu$, that is to an energy gain of relative order $\Ma\,\Kn$.

\begin{table}[!htbp]
\centering\small
\begin{tabular}{@{}c@{}}
\begin{tabular}{@{}rrrr@{}}
\toprule
$N$ & Chebyshev, strong walls & Legendre, strong walls & Legendre, weak walls\\
\midrule
16 & $+279$ & $+20.0$ & $5\times10^{-14}$\\
24 & $+641$ & -- & -- \\
32 & $+1148$ & $+77.0$ & $3\times10^{-12}$\\
48 & $+2598$ & -- & -- \\
\bottomrule
\end{tabular}\\[10pt]
\begin{tabular}{@{}rll@{}}
\toprule
$N$ & Chebyshev, strong walls & Legendre, weak walls\\
\midrule
32 & $0.6057838-0.1193627\,\ii$ & $0.6056978-0.1193647\,\ii$\\
48 & $0.6057312-0.1193644\,\ii$ & $0.6056979-0.1193648\,\ii$\\
64 & $0.6057148-0.1193646\,\ii$ & $0.6056980-0.1193648\,\ii$\\
96 & $0.6057035-0.1193647\,\ii$ & $0.6056982-0.1193648\,\ii$\\
\bottomrule
\end{tabular}
\end{tabular}
\caption{Discretizations of the cross-stream direction and of the walls. Top: numerical abscissa of the gas at rest (exact value zero). Bottom: Couette eigenvalue, $(\Rey,\Ma,\alpha,\beta)=(1000,0.1,1,0)$.}
\label{tab:abscissa}
\end{table}

The lower part of table~\ref{tab:abscissa} compares the strongly and the weakly imposed wall conditions for the least-stable mode of Couette flow at $\Rey=1000$, $\Ma=0.1$, $\alpha=1$, BGK, $Q=16$. The weak conditions converge an order of magnitude faster in $N$, and the two treatments approach the same limit. The moment-consistent velocity derivative was found necessary by the transient-growth computations. For streamwise-invariant disturbances of Couette flow ($\Rey=1000$, $\Ma=0.1$, $\alpha=0$, $\beta h=1.66$) the least-damped eigenvalue $-\ii\omega$ of the uncorrected discretization is real and \emph{positive}, $+0.074$ at $Q=12$ and $+0.0134$ at $Q=16$ (in units of $U/h$), while Couette flow is stable and the Orr--Sommerfeld--Squire value is $-0.005223$. The spurious growth rate is independent of the Mach number and decreases with $Q$ and with the truncation radius. It is a coupling of the streamwise velocity back into the wall-normal one through the discrete moments of the shear term, which the lift-up mechanism amplifies. With the correction the same eigenvalue is $-0.005217$ at $Q=16$ and $Q=20$ alike, and the change of the Couette eigenvalue of table~\ref{tab:abscissa} from $Q=16$ to $Q=24$ is $8\times10^{-5}$ without the correction and $5\times10^{-6}$ with it.

The term $U'\mathcal{L}_{\bar g}$ changes the eigenvalues at $\mathcal{O}(\Ma^2)$. Switching it off changes $c$ of the least-stable Couette mode ($\Rey=1000$, $\alpha=1$, BGK, $Q=16$, $N=48$) by $2.10\times10^{-5}$, $5.31\times10^{-6}$ and $1.34\times10^{-6}$ at $\Ma=0.2$, $0.1$ and $0.05$, and that of the Poiseuille mode ($\Rey=10^4$) by $3.07\times10^{-6}$, $7.7\times10^{-7}$ and $1.9\times10^{-7}$, that is ratios of $4.0$ for each halving of $\Ma$, with a coefficient that decreases with $\Rey$. Finally, for the gas at rest $G(t)=0.983$, $0.911$, $0.114$ and $2.9\times10^{-7}$ at $t=0.3$, $1$, $3$ and $10$, below one as the $H$ theorem requires.

\FloatBarrier

\section{Results}\label{sec:res}\label{sec:results}
\subsection{Accuracy and computational cost}\label{sec:res_acc}
Table~\ref{tab:modal} presents the phase speed $c=c_r+\ii c_i$ of the least-stable two-dimensional mode of the four flows with each collision operator, on a velocity grid of $Q=20$ points per direction in a ball of radius $R_v=6$. The cross-stream grids are $N=128$ and $192$, extrapolated to $N\to\infty$, for the channel flows, $N=96$ with free boundaries at $y=\pm40$ for the mixing layer, and $N=128$ with a free boundary at $y=60$, about $35\,\delta^*$, for the Blasius boundary layer. Each value at $\Ma\to0$ is a fit through three Mach numbers, of which the two common to all flows are shown, namely $c(\Ma)=c_0+c_1\Ma+c_2\Ma^2$ through $\Ma=0.2$, $0.1$ and $0.05$ for the channel flows and through $\Ma=0.1$, $0.05$ and $0.025$ for the Blasius boundary layer, and $c(\Ma)=c_0+c_2\Ma^2+c_4\Ma^4$ through $\Ma=0.1$, $0.05$ and $0.025$ for the mixing layer, which has no wall and therefore no linear term from wall slip. Reynolds numbers and wavenumbers are based on the half-gap $h$ for the channel flows and on the length scale $\ell$ of the profile for the open flows, the half-thickness of the mixing layer and $\ell=(\mu_0x/\rho_0U_\infty)^{1/2}$ for the Blasius boundary layer. Times are in units of $h/U$ or $\ell/U$. The reference values come from an independent Chebyshev collocation Orr--Sommerfeld solver, converged to $10^{-8}$ for the channels, where it reproduces Orszag's $0.23752649+0.00373967\,\ii$, and run on the kinetic collocation points for the open flows.

\begin{table}[!htb]
\centering\footnotesize\setlength{\tabcolsep}{3.5pt}
\adjustbox{max width=\textwidth}{%
\begin{tabular}{@{}llccc@{}}
\toprule
Case & Model & $\Ma=0.1$ & $\Ma=0.05$ & $\Ma\to0$\\
\midrule
Couette, $\Rey=10^3$ & full Boltzmann (HS) & $0.6056900-0.1193633\,\ii$ & $0.6054798-0.1192596\,\ii$ & $0.6053480-0.1192249\,\ii$\\
 & BGK & $0.6056950-0.1193653\,\ii$ & $0.6054785-0.1192637\,\ii$ & $0.6053429-0.1192298\,\ii$\\
 & ES-BGK & $0.6056888-0.1193686\,\ii$ & $0.6054769-0.1192645\,\ii$ & $0.6053429-0.1192297\,\ii$\\
 & Shakhov & $0.6056888-0.1193683\,\ii$ & $0.6054770-0.1192644\,\ii$ & $0.6053429-0.1192297\,\ii$\\
 & Orr--Sommerfeld & & & $0.6053430-0.1192302\,\ii$\\
\midrule
Poiseuille, $\Rey=10^4$ & full Boltzmann (HS) & $0.2376171+0.0037202\,\ii$ & $0.2375406+0.0037349\,\ii$ & $0.2375248+0.0037395\,\ii$\\
 & BGK & $0.2376165+0.0037191\,\ii$ & $0.2375415+0.0037345\,\ii$ & $0.2375264+0.0037393\,\ii$\\
 & ES-BGK & $0.2376182+0.0037202\,\ii$ & $0.2375420+0.0037348\,\ii$ & $0.2375264+0.0037393\,\ii$\\
 & Shakhov & $0.2376182+0.0037202\,\ii$ & $0.2375420+0.0037348\,\ii$ & $0.2375264+0.0037393\,\ii$\\
 & Orr--Sommerfeld & & & $0.2375265+0.0037397\,\ii$\\
\midrule
Mixing layer, $\Rey=100$ & full Boltzmann (HS) & $0.0000000+0.3884693\,\ii$ & $0.0000000+0.3920766\,\ii$ & $0.0000000+0.3932823\,\ii$\\
 & BGK & $0.0000000+0.3884718\,\ii$ & $0.0000000+0.3920708\,\ii$ & $0.0000000+0.3932735\,\ii$\\
 & ES-BGK & $0.0000000+0.3884569\,\ii$ & $0.0000000+0.3920670\,\ii$ & $0.0000000+0.3932736\,\ii$\\
 & Shakhov & $0.0000000+0.3884567\,\ii$ & $0.0000000+0.3920670\,\ii$ & $0.0000000+0.3932735\,\ii$\\
 & Orr--Sommerfeld & & & $0.0000000+0.3932734\,\ii$\\
\midrule
Blasius, $\Rey=580$ & full Boltzmann (HS) & $0.3644583+0.0078034\,\ii$ & $0.3641881+0.0079221\,\ii$ & $0.3641207+0.0079602\,\ii$\\
 & BGK & $0.3644537+0.0078005\,\ii$ & $0.3641880+0.0079209\,\ii$ & $0.3641226+0.0079595\,\ii$\\
 & ES-BGK & $0.3644588+0.0078031\,\ii$ & $0.3641893+0.0079216\,\ii$ & $0.3641226+0.0079595\,\ii$\\
 & Shakhov & $0.3644587+0.0078031\,\ii$ & $0.3641892+0.0079216\,\ii$ & $0.3641226+0.0079595\,\ii$\\
 & Orr--Sommerfeld & & & $0.3641229+0.0079598\,\ii$\\
\bottomrule
\end{tabular}}
\caption{Phase speed of the least-stable mode, $\beta=0$, $Q=20$: kinetic values at $\Ma=0.1$ and $0.05$, their limit $\Ma\to0$, and the Orr--Sommerfeld value of each flow.}
\label{tab:modal}
\end{table}

\begin{figure}[!htbp]
\centering
\begin{tabular}{@{}c@{\hspace{0.02\textwidth}}c@{}}
\includegraphics[width=0.48\textwidth]{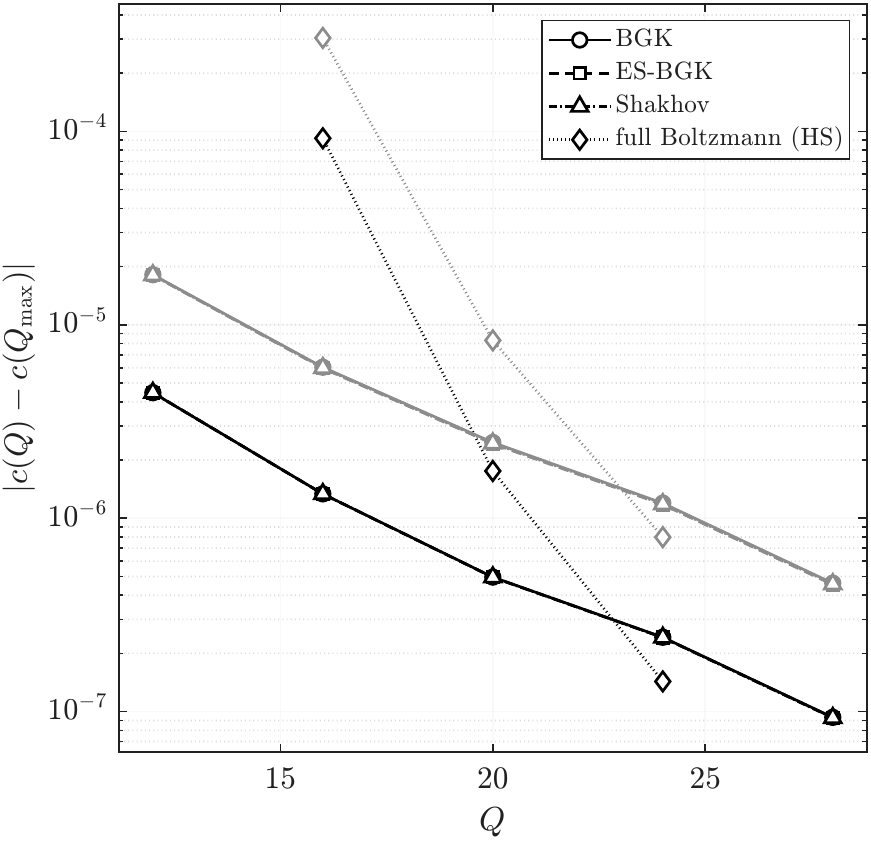} &
\includegraphics[width=0.48\textwidth]{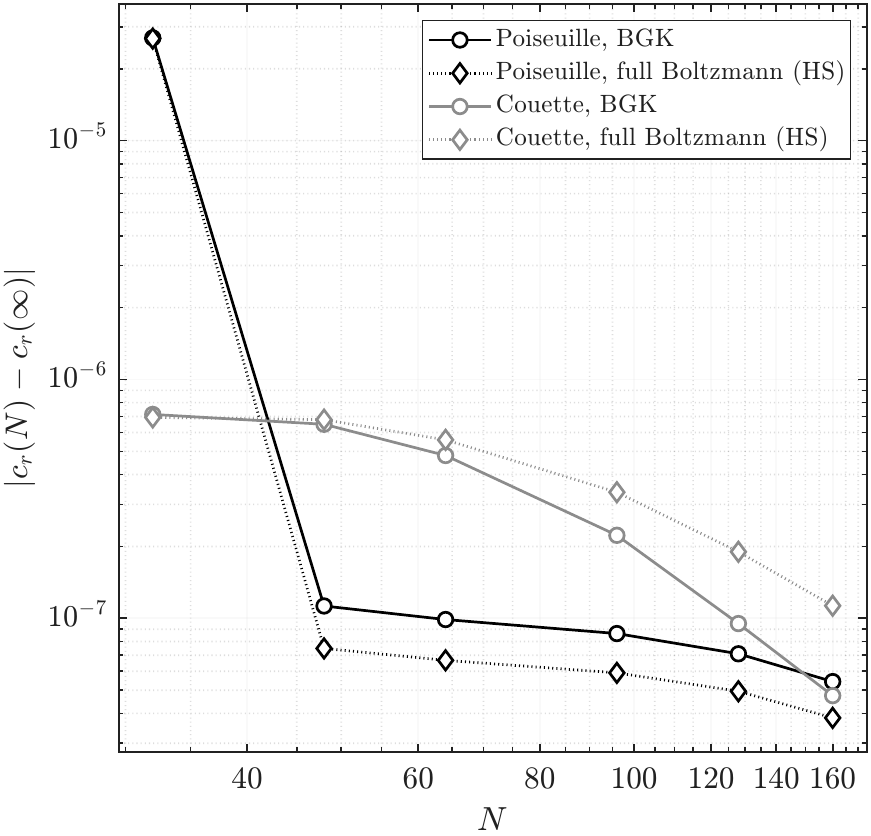}\\
(\textit{a}) & (\textit{b})
\end{tabular}
\caption{Discretization errors of the channel-flow eigenvalues, $\Ma=0.1$; black, Poiseuille; gray, Couette. (\textit{a}) Velocity grid, $N=64$; (\textit{b}) cross-stream grid, $Q=20$.}
\label{fig:conv}
\end{figure}

\begin{figure}[!htbp]
\centering
\begin{tabular}{@{}c@{\hspace{0.02\textwidth}}c@{}}
\includegraphics[width=0.48\textwidth]{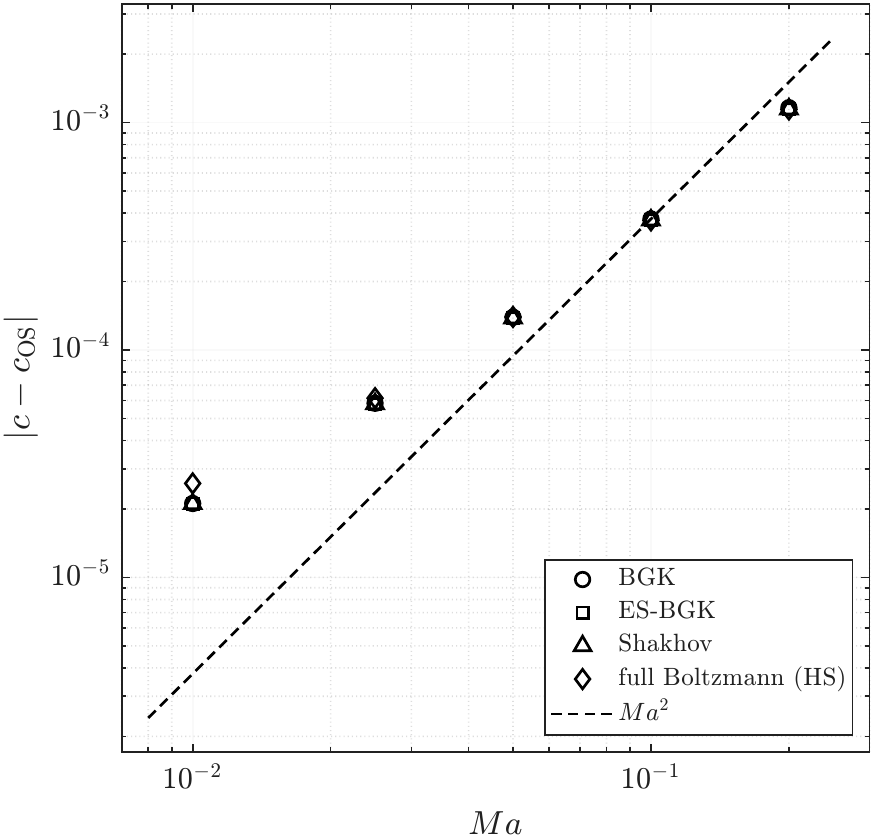} &
\includegraphics[width=0.48\textwidth]{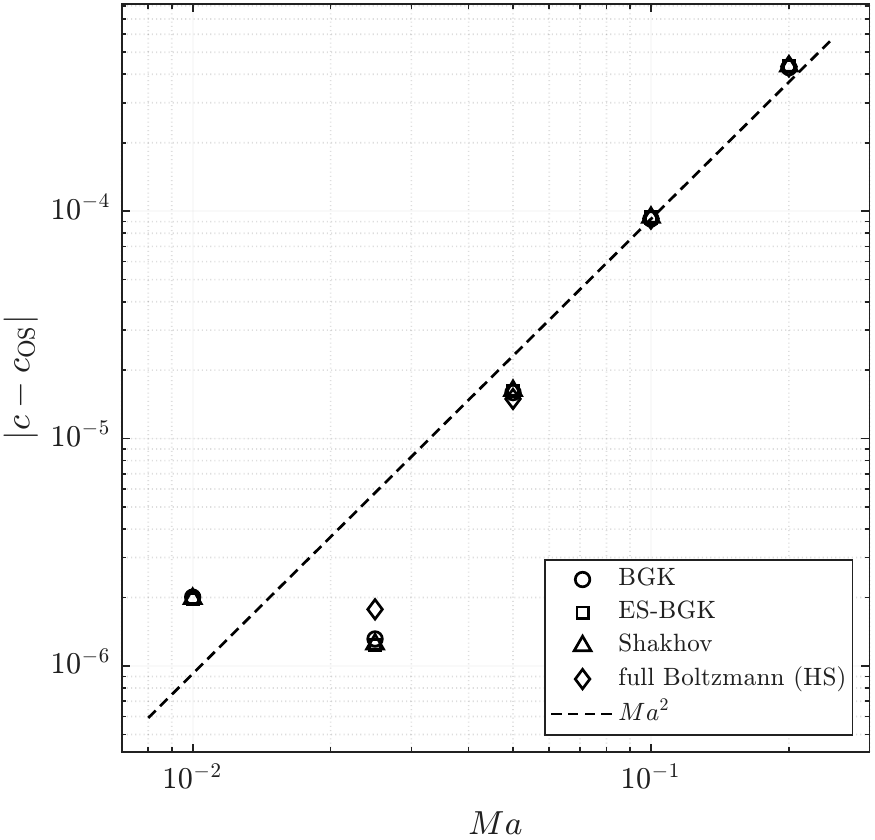}\\
(\textit{a}) & (\textit{b})\\[4pt]
\includegraphics[width=0.48\textwidth]{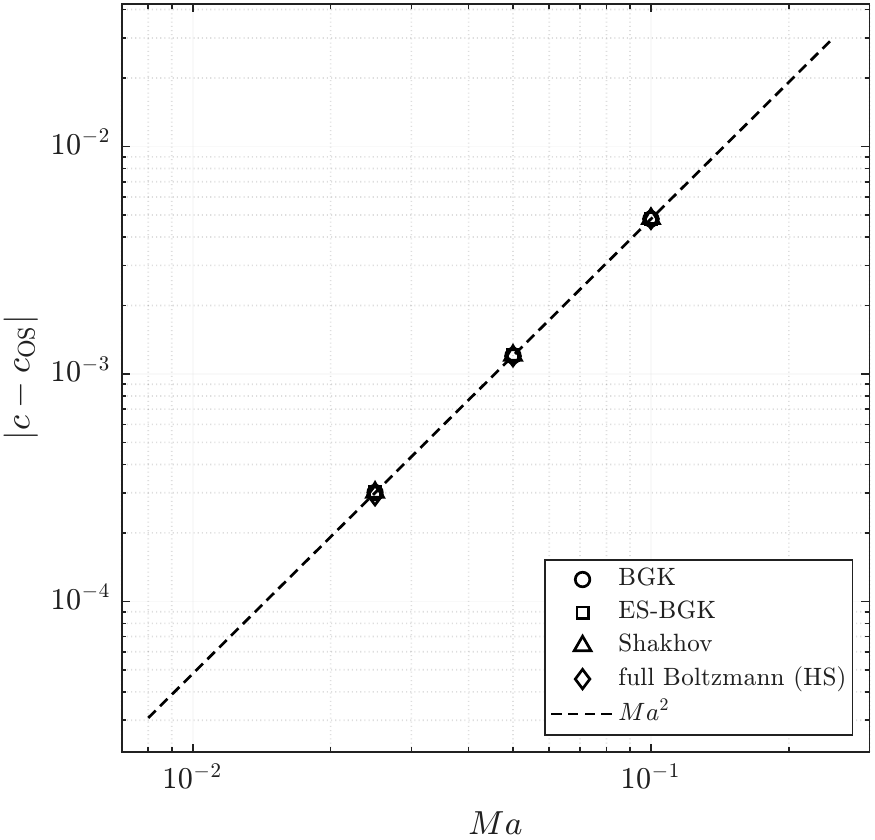} &
\includegraphics[width=0.48\textwidth]{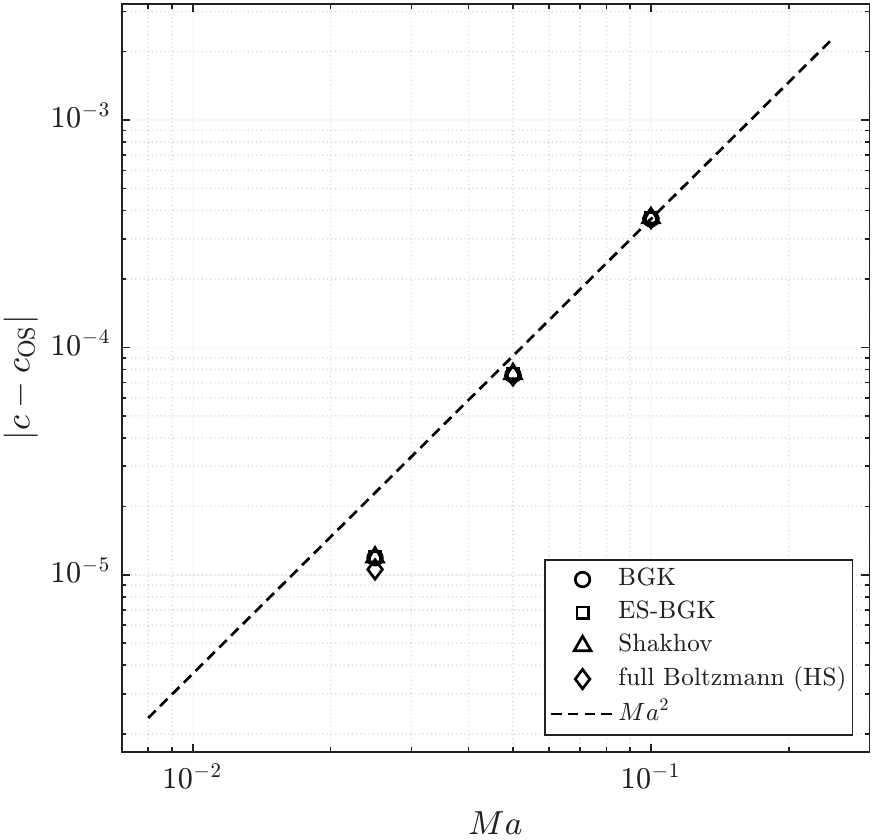}\\
(\textit{c}) & (\textit{d})
\end{tabular}
\caption{Distance of the kinetic eigenvalue from the Orr--Sommerfeld value against $\Ma$, $Q=20$, $\beta=0$: (\textit{a}) Couette, $(\Rey,\alpha)=(10^3,1)$; (\textit{b}) Poiseuille, $(10^4,1)$; (\textit{c}) mixing layer, $(100,0.4446)$; (\textit{d}) Blasius, $(580,0.179)$.}
\label{fig:mach}
\end{figure}

Figure~\ref{fig:conv} shows the discretization errors of the channel-flow eigenvalues at $\Ma=0.1$, for Poiseuille (black) and Couette (gray). Panel (\textit{a}) shows the velocity-grid error $|c(Q)-c(Q_{\max})|$ at $N=64$ for the four collision operators, the three relaxation models coinciding to the scale of the plot, and panel (\textit{b}) the cross-stream error $|c_r(N)-c_r(\infty)|$ at $Q=20$ for BGK and the full operator, with $c_r(\infty)$ extrapolated from the two finest grids of the same operator. In figure~\ref{fig:mach} the distance of the kinetic eigenvalue from its Orr--Sommerfeld counterpart is shown against the Mach number for the four operators at $Q=20$, $\beta=0$, for plane Couette flow at $(\Rey,\alpha)=(10^3,1)$ in (\textit{a}), plane Poiseuille flow at $(10^4,1)$ in (\textit{b}), the mixing layer at $(100,0.4446)$ in (\textit{c}) and the Blasius boundary layer at $(580,0.179)$ in (\textit{d}). The dashed line is the $\Ma^2$ law through the BGK value at $\Ma=0.1$, and for the channel flows the BGK sequence extends to $\Ma=0.01$.

For the relaxation models the change of the eigenvalue is reduced by a factor of two to three with every refinement of the velocity grid, and at $Q=20$ the eigenvalue is within $5\times10^{-7}$ (Poiseuille) and $2.5\times10^{-6}$ (Couette) of its value on the finest grid. This rapid convergence is due to the moment-consistent velocity derivative of \S\ref{sec:disc}, since the perturbation is close to a Maxwellian times a low-order polynomial, whose moments the corrected derivative reproduces on any grid. The full operator, whose discrete matrix itself changes with the grid, starts from a larger error and converges faster, to below $10^{-5}$ at $Q=20$ and $10^{-6}$ at $Q=24$. In the cross-stream direction every grid from $N=48$ on is within $7\times10^{-7}$ of the extrapolated value, for both flows and both operators. At $N=32$ the critical layer of Poiseuille flow at $\Rey=10^4$ is not resolved and the error is $2.7\times10^{-5}$. The remaining slow decrease, of order $10^{-7}$, is attributed to the Knudsen layers at the walls, of thickness $\mathcal{O}(\Kn)$, which no grid of this kind resolves.

The approach to the incompressible limit in figure~\ref{fig:mach} takes two forms. In the mixing layer, which has no wall, the distance from the Orr--Sommerfeld value is reduced by a factor of four with every halving of the Mach number, for all four operators, which is the $\Ma^2$ law of compressibility. When a wall is present, a term linear in $\Ma$ appears, indicating the first effect of rarefaction at fixed $\Rey$, since $\Kn\propto\Ma/\Rey$. Its sign and size are consistent with first-order velocity slip of the perturbation at the walls, although this identification has not been confirmed independently. In Couette flow the linear term dominates below $\Ma\approx0.05$, where the distance falls much more slowly than the $\Ma^2$ law. In Poiseuille flow and in the Blasius boundary layer the linear and quadratic terms have opposite signs in $c_r$ (table~\ref{tab:fitcoef}) and partly cancel, and in Poiseuille flow the distance passes through a minimum before the linear term takes over at $\Ma=0.01$. At $\Ma=0.1$ the kinetic eigenvalues are $9\times10^{-5}$ (Poiseuille) to $5\times10^{-3}$ (mixing layer) from the Orr--Sommerfeld values, so that the extrapolation is what makes the comparison at the accuracy of continuum stability theory possible.

The limits $c_0$ of table~\ref{tab:modal} agree with Orr--Sommerfeld to $1\times10^{-7}$ to $5\times10^{-7}$ for the three relaxation models in all four flows, and to $1.7\times10^{-6}$ to $8.9\times10^{-6}$ for the full operator. The latter difference is the velocity-grid error of the full operator at $Q=20$. In Couette flow it equals, to within $5\times10^{-7}$, the change of the full-operator eigenvalue from $Q=20$ to $Q=24$, and at $Q=24$ the full operator agrees with Orr--Sommerfeld to $4.6\times10^{-7}$ to $1.1\times10^{-6}$ in the four flows.

Table~\ref{tab:fitcoef} lists the coefficients of the Mach-number fits at $Q=20$. For the wall-bounded flows these are the real part of the linear coefficient $c_1$, the corresponding derivative of the phase speed with respect to the Knudsen number, $\Kn=\sqrt{\pi\gamma/2}\,\Ma/(2\Rey)$ for the channels and $\sqrt{\pi\gamma/2}\,\Ma/\Rey$ for the Blasius boundary layer, and the coefficient $c_2$ of $\Ma^2$, and for the mixing layer $c_2$ alone. The linear coefficient is the same for the three relaxation models to within $0.5\%$ in each flow, although their Prandtl numbers differ, and that of the full operator is smaller in magnitude by $2\%$ (Couette) to $2.4\%$ (Poiseuille, Blasius). The three models share the viscosity and the diffuse-reflection condition, and a difference of this sign and size is what the dependence of the slip coefficient on the collision operator would produce, although it has not been analyzed further. The quadratic coefficient $c_2$ instead groups the operators by their Prandtl number. ES-BGK, Shakhov and the full operator, all with $\Pran\approx\tfrac23$, agree with one another to about $1\%$ in every flow, and BGK, with $\Pran=1$, differs from them by up to $4\%$. The agreement of two relaxation models with the full operator, in which the base-stress terms of $\mathcal{L}_{\bar g}$ are contained automatically, is an independent check of the model forms of these terms in \ref{app:linops}. The coefficient $c_2$ is not that of a fully consistent compressible flow, since the base state has uniform temperature.

\begin{table}[!htbp]
\centering\footnotesize\setlength{\tabcolsep}{4pt}
\adjustbox{max width=\textwidth}{\begin{tabular}{@{}llcccc@{}}
\toprule
flow & coefficient & BGK & ES-BGK & Shakhov & full Boltzmann (HS)\\
\midrule
Couette & $\mathrm{Re}(c_1)$ & $1.89\times10^{-3}$ & $1.89\times10^{-3}$ & $1.90\times10^{-3}$ & $1.85\times10^{-3}$\\
 & $\partial c_r/\partial\Kn$ & $+1.171$ & $+1.171$ & $+1.172$ & $+1.142$\\
 & $c_2$ & $+0.0162-0.0135\,\ii$ & $+0.0156-0.0139\,\ii$ & $+0.0156-0.0138\,\ii$ & $+0.0157-0.0138\,\ii$\\
\midrule
Poiseuille & $\mathrm{Re}(c_1)$ & $-2.95\times10^{-4}$ & $-2.95\times10^{-4}$ & $-2.95\times10^{-4}$ & $-2.88\times10^{-4}$\\
 & $\partial c_r/\partial\Kn$ & $-1.823$ & $-1.823$ & $-1.823$ & $-1.781$\\
 & $c_2$ & $+0.0120-0.0021\,\ii$ & $+0.0121-0.0020\,\ii$ & $+0.0121-0.0020\,\ii$ & $+0.0121-0.0020\,\ii$\\
\midrule
Blasius & $\mathrm{Re}(c_1)$ & $-6.96\times10^{-4}$ & $-6.97\times10^{-4}$ & $-6.97\times10^{-4}$ & $-6.79\times10^{-4}$\\
 & $\partial c_r/\partial\Kn$ & $-0.250$ & $-0.250$ & $-0.250$ & $-0.243$\\
 & $c_2$ & $+0.0401-0.0164\,\ii$ & $+0.0406-0.0161\,\ii$ & $+0.0406-0.0161\,\ii$ & $+0.0405-0.0161\,\ii$\\
\midrule
mixing layer, $\Rey=100$ & $c_2$ & $-0.4814\,\ii$ & $-0.4829\,\ii$ & $-0.4829\,\ii$ & $-0.4826\,\ii$\\
\bottomrule
\end{tabular}
}
\caption{Coefficients of the Mach-number fits of table~\ref{tab:modal}: real part of the linear coefficient $c_1$, the corresponding $\partial c_r/\partial\Kn$, and the coefficient $c_2$ of $\Ma^2$.}
\label{tab:fitcoef}
\end{table}

The cost of one eigenvalue on a single core is presented in table~\ref{tab:cost}, excluding the assembly of the hard-sphere matrices, namely the number of preconditioned GMRES iterations per linear solve, the peak memory and the CPU time. In the upper part, results are shown at $Q=20$ for the channel flows at $\Ma=0.1$, $N=96$, the mixing layer at $\Rey=100$, $\Ma=0.05$, $N=96$, and the Blasius boundary layer at $\Ma=0.05$, $N=128$, while in the lower part results are presented for plane Couette flow at $\Ma=0.1$, at the coarsest and finest $N$ and on finer velocity grids. For the relaxation models, for which the preconditioner is the exact inverse of the operator, three inner iterations suffice in every case, while the computation of an eigenvalue takes $7$ to $15$\,s and requires $170$ to $320$\,MB. The full operator takes $20$ to $38$ inner iterations and $8$ to $24$ times the time of a relaxation model, and requires two to four times its memory. Its iteration count hardly depends on the resolution, so that its cost follows the size of the problem, moderately in $N$ and steeply in $Q$, where the dense collision matrices of size $n_b^2$ dominate. The hard-sphere matrices are assembled once for all flows, in about two and a half minutes per matrix at $Q=20$ on two cores and about an hour at $Q=28$ on a single core.

\begin{table}[!htbp]
\centering\scriptsize\setlength{\tabcolsep}{4pt}
\adjustbox{max width=\textwidth}{\begin{tabular}{@{}lrrrrrrrrrrrr@{}}
\toprule
 & \multicolumn{3}{c}{Couette} & \multicolumn{3}{c}{Poiseuille} & \multicolumn{3}{c}{mixing layer} & \multicolumn{3}{c}{Blasius}\\
\cmidrule(lr){2-4}\cmidrule(lr){5-7}\cmidrule(lr){8-10}\cmidrule(lr){11-13}
operator & its. & MB & s & its. & MB & s & its. & MB & s & its. & MB & s\\
\midrule
BGK & 3 & 172 & 7.2 & 3 & 172 & 6.6 & 3 & 187 & 8.5 & 3 & 265 & 14\\
ES-BGK & 3 & 204 & 7.2 & 3 & 204 & 6.8 & 3 & 222 & 8.1 & 3 & 322 & 15\\
Shakhov & 3 & 179 & 7.0 & 3 & 179 & 6.8 & 3 & 189 & 8.0 & 3 & 276 & 15\\
full Boltzmann (HS) & 23 & 429 & 76 & 38 & 611 & 159 & 20 & 424 & 67 & 25 & 805 & 299\\
\midrule
\multicolumn{13}{@{}l}{\textit{Plane Couette flow, $\Ma=0.1$: dependence on $N$ ($Q=20$) and on $Q$ ($N=64$)}}\\
 & \multicolumn{3}{c}{$N=32$} & \multicolumn{3}{c}{$N=192$} & \multicolumn{3}{c}{$Q=24$, $N=64$} & \multicolumn{3}{c}{$Q=28$, $N=64$}\\
\cmidrule(lr){2-4}\cmidrule(lr){5-7}\cmidrule(lr){8-10}\cmidrule(lr){11-13}
BGK & 3 & 69 & 0.9 & 3 & 451 & 25 & 3 & 178 & 7.3 & 3 & 275 & 13\\
full Boltzmann (HS) & 24 & 241 & 56 & 26 & 829 & 231 & 21 & 777 & 144 & 21 & 1762 & 489\\
\bottomrule
\end{tabular}
}
\caption{Cost of one eigenvalue on one core: GMRES iterations per linear solve, peak memory (MB) and CPU time (s). Top: the four flows; bottom: plane Couette flow at other $N$ and $Q$.}
\label{tab:cost}
\end{table}

\paragraph{Resolution in the cross-stream direction and the free-stream boundary}
Table~\ref{tab:conv_N} gives, in its upper part, the distance of the least-stable eigenvalue of the four flows from its value on the reference grid, the finest one run for that flow, for the four operators at $Q=20$. For the channel flows the reference is the extrapolation $N\to\infty$ in $N^{-2}$ through $N=160$ and $192$, the channel flows are at $\Ma=0.1$, and the mixing layer ($\alpha=0.4446$, $\Rey=100$, $y_\infty=40$) and the Blasius boundary layer ($\Rey=580$, $\alpha=0.179$, $y_\infty=60$) at $\Ma=0.05$. Its lower part gives the dependence on the position $y_\infty$ of the free boundaries, for the mixing layer at $N=96$ and the Blasius boundary layer at $N=128$. In Couette flow the growth parameter is converged to $10^{-7}$ already at $N=32$ for every operator, and the phase speed is within $10^{-6}$ of its extrapolated value at $N=32$ and within $3.4\times10^{-7}$ at $N=96$. Its slow residual drift, of a few $10^{-7}$ between $N=32$ and $192$, is the effect of the Knudsen layers. In Poiseuille flow $N=32$ does not resolve the critical layer, so that the growth parameter is in error by $3.4\times10^{-5}$, about one percent of its value, and the phase speed by $2.7\times10^{-5}$, for every operator. From $N=48$ on every operator is within $1.6\times10^{-7}$ of its extrapolated value, and the growth parameter does not change in the seventh digit from $N=48$ to $192$. In the mixing layer the eigenvalue is converged to nine digits in $N$ and in $y_\infty$ for every operator. Without a wall there is no Knudsen-layer drift, convergence in $N$ is spectral, with differences from the finest grid of $3\times10^{-5}$ at $N=32$, $1.4\times10^{-7}$ at $N=48$ and $2\times10^{-9}$ at $N=64$, the same for the four operators to two digits, and the free-stream condition leaves no trace once $\exp(-\alpha y_\infty)$ is negligible. In the Blasius boundary layer the eigenvalue changes by at most $1.2\times10^{-7}$ between $N=64$ and $160$, and the limits $\Ma\to0$ formed on these grids change by less than $3\times10^{-8}$. Moving the free boundary from $y_\infty=60$ to $100$ changes the eigenvalue by $3\times10^{-9}$ for every operator, while $y_\infty=30$ is too close, with a change of $1.3\times10^{-5}$.

\begin{table}[!htbp]
\centering\scriptsize\setlength{\tabcolsep}{3.5pt}
\begin{tabular}{@{}c@{}}
\adjustbox{max width=\textwidth}{\begin{tabular}{@{}llrrrrrrr@{}}
\toprule
flow & operator & $N=32$ & $N=48$ & $N=64$ & $N=96$ & $N=128$ & $N=160$ & $N=192$\\
\midrule
Couette & BGK & $7.2\times10^{-7}$ & $6.5\times10^{-7}$ & $4.8\times10^{-7}$ & $2.2\times10^{-7}$ & $9.5\times10^{-8}$ & $4.7\times10^{-8}$ & $3.3\times10^{-8}$\\
 & ES-BGK & $6.3\times10^{-7}$ & $5.1\times10^{-7}$ & $3.3\times10^{-7}$ & $1.0\times10^{-7}$ & $2.9\times10^{-8}$ & $1.1\times10^{-8}$ & $7.5\times10^{-9}$\\
 & Shakhov & $9.6\times10^{-7}$ & $8.2\times10^{-7}$ & $5.9\times10^{-7}$ & $2.5\times10^{-7}$ & $1.0\times10^{-7}$ & $4.8\times10^{-8}$ & $3.4\times10^{-8}$\\
 & full Boltzmann (HS) & $7.0\times10^{-7}$ & $6.8\times10^{-7}$ & $5.6\times10^{-7}$ & $3.4\times10^{-7}$ & $1.9\times10^{-7}$ & $1.1\times10^{-7}$ & $7.8\times10^{-8}$\\
\midrule
Poiseuille & BGK & $4.3\times10^{-5}$ & $1.1\times10^{-7}$ & $9.9\times10^{-8}$ & $8.6\times10^{-8}$ & $7.1\times10^{-8}$ & $5.4\times10^{-8}$ & $3.8\times10^{-8}$\\
 & ES-BGK & $4.3\times10^{-5}$ & $1.3\times10^{-7}$ & $1.2\times10^{-7}$ & $1.0\times10^{-7}$ & $8.1\times10^{-8}$ & $6.0\times10^{-8}$ & $4.2\times10^{-8}$\\
 & Shakhov & $4.3\times10^{-5}$ & $1.6\times10^{-7}$ & $1.4\times10^{-7}$ & $1.2\times10^{-7}$ & $9.9\times10^{-8}$ & $7.6\times10^{-8}$ & $5.3\times10^{-8}$\\
 & full Boltzmann (HS) & $4.3\times10^{-5}$ & $7.5\times10^{-8}$ & $6.7\times10^{-8}$ & $5.9\times10^{-8}$ & $4.9\times10^{-8}$ & $3.8\times10^{-8}$ & $2.7\times10^{-8}$\\
\midrule
mixing layer & BGK & $3.2\times10^{-5}$ & $1.4\times10^{-7}$ & $1.6\times10^{-9}$ & $0$ & -- &  & \\
 & ES-BGK & $3.1\times10^{-5}$ & $1.4\times10^{-7}$ & $1.7\times10^{-9}$ & $0$ & -- &  & \\
 & Shakhov & $3.1\times10^{-5}$ & $1.4\times10^{-7}$ & $1.7\times10^{-9}$ & $0$ & -- &  & \\
 & full Boltzmann (HS) & $3.2\times10^{-5}$ & $1.4\times10^{-7}$ & $1.7\times10^{-9}$ & $0$ & -- &  & \\
\midrule
Blasius & BGK &  &  & $9.0\times10^{-8}$ & $5.9\times10^{-8}$ & $2.7\times10^{-8}$ & -- & \\
 & ES-BGK &  &  & $9.9\times10^{-8}$ & $5.9\times10^{-8}$ & $2.5\times10^{-8}$ & -- & \\
 & Shakhov &  &  & $1.2\times10^{-7}$ & $7.9\times10^{-8}$ & $3.6\times10^{-8}$ & -- & \\
 & full Boltzmann (HS) &  &  & $6.5\times10^{-8}$ & $4.5\times10^{-8}$ & $2.2\times10^{-8}$ & -- & \\
\bottomrule
\end{tabular}
}\\[10pt]
\adjustbox{max width=\textwidth}{\begin{tabular}{@{}llrrrr@{}}
\toprule
flow & operator & \multicolumn{4}{c}{$y_\infty$}\\
\midrule
 mixing layer & & 10 & 20 & 40 & 80\\
\midrule
 & BGK & $2.1\times10^{-4}$ & $2.9\times10^{-8}$ & $0$ & --\\
 & ES-BGK & $2.1\times10^{-4}$ & $2.9\times10^{-8}$ & $0$ & --\\
 & Shakhov & $2.1\times10^{-4}$ & $2.9\times10^{-8}$ & $0$ & --\\
 & full Boltzmann (HS) & $2.1\times10^{-4}$ & $2.9\times10^{-8}$ & $0$ & --\\
\midrule
 Blasius & & 30 & 60 & 100 & \\
\midrule
 & BGK & $1.3\times10^{-5}$ & $2.9\times10^{-9}$ & -- & \\
 & ES-BGK & $1.3\times10^{-5}$ & $2.8\times10^{-9}$ & -- & \\
 & Shakhov & $1.3\times10^{-5}$ & $3.7\times10^{-9}$ & -- & \\
 & full Boltzmann (HS) & $1.3\times10^{-5}$ & $2.3\times10^{-9}$ & -- & \\
\bottomrule
\end{tabular}
}
\end{tabular}
\caption{Cross-stream convergence of the least-stable eigenvalue, $Q=20$: distance from the finest grid. Top: against $N$; bottom: against the position $y_\infty$ of the free boundaries.}
\label{tab:conv_N}
\end{table}

\paragraph{Resolution in velocity space}
Table~\ref{tab:conv_Q} gives the corresponding distances against the velocity grid, $R_v=6$, from the finest grid run for each flow and operator, namely the channel flows at $\Ma=0.1$, $N=64$, the mixing layer at $\Ma=0.05$, $N=96$, and the Blasius boundary layer at $\Ma=0.05$, $N=128$, and n.c.\ marks runs in which the inner iteration of the full operator does not converge. For the three relaxation models the behavior is the same in all four flows and nearly independent of the model. The change of $c$ from one grid to the next decreases by a factor of two to three per step, and at $Q=12$, the coarsest grid, $c$ is already within $2\times10^{-5}$ (Couette), $5\times10^{-6}$ (Poiseuille, Blasius) and $10^{-6}$ (mixing layer) of the finest value. With the uncorrected Fourier derivative the error at $Q=12$ is $4\times10^{-3}$ for Poiseuille flow. In the channel flows, where finer grids were run, the growth parameter is converged to $10^{-8}$ while the phase speed continues to drift by a few $10^{-7}$ per step up to $Q=32$, an effect attributed to the half-range wall integrals, whose integrands are discontinuous in $c_y$. The full operator converges more slowly, because its discrete matrix itself depends on the grid (table~\ref{tab:operator}). Its inner iteration does not converge at $Q=12$, consistently with the $20\%$ null-vector defect of the matrix at that resolution, $Q=16$ is in error by $1$ to $3\times10^{-4}$, and $Q=20$ by $2\times10^{-6}$ to $10^{-5}$. From $Q=24$ to $Q=28$ ($n_b=9171$) the full-operator eigenvalue changes by $7\times10^{-7}$ in Couette and $10^{-7}$ in Poiseuille flow, against $7\times10^{-7}$ for BGK.

\begin{table}[!htbp]
\centering\scriptsize\setlength{\tabcolsep}{4pt}
\adjustbox{max width=\textwidth}{\begin{tabular}{@{}llrrrrrr@{}}
\toprule
flow & operator & $Q=12$ & $Q=16$ & $Q=20$ & $Q=24$ & $Q=28$ & $Q=32$\\
\midrule
Couette & BGK & $1.8\times10^{-5}$ & $6.0\times10^{-6}$ & $2.5\times10^{-6}$ & $1.2\times10^{-6}$ & $4.6\times10^{-7}$ & --\\
 & ES-BGK & $1.8\times10^{-5}$ & $6.0\times10^{-6}$ & $2.4\times10^{-6}$ & $1.2\times10^{-6}$ & $4.5\times10^{-7}$ & --\\
 & Shakhov & $1.8\times10^{-5}$ & $6.0\times10^{-6}$ & $2.4\times10^{-6}$ & $1.2\times10^{-6}$ & $4.6\times10^{-7}$ & --\\
 & full Boltzmann (HS) & n.c. & $3.0\times10^{-4}$ & $8.3\times10^{-6}$ & $8.0\times10^{-7}$ & -- & \\
\midrule
Poiseuille & BGK & $4.5\times10^{-6}$ & $1.3\times10^{-6}$ & $5.0\times10^{-7}$ & $2.4\times10^{-7}$ & $9.3\times10^{-8}$ & --\\
 & ES-BGK & $4.5\times10^{-6}$ & $1.3\times10^{-6}$ & $4.9\times10^{-7}$ & $2.4\times10^{-7}$ & $9.3\times10^{-8}$ & --\\
 & Shakhov & $4.5\times10^{-6}$ & $1.3\times10^{-6}$ & $5.0\times10^{-7}$ & $2.4\times10^{-7}$ & $9.3\times10^{-8}$ & --\\
 & full Boltzmann (HS) & n.c. & $9.2\times10^{-5}$ & $1.8\times10^{-6}$ & $1.4\times10^{-7}$ & -- & \\
\midrule
mixing layer & BGK & $1.2\times10^{-6}$ & $3.6\times10^{-7}$ & $6.9\times10^{-8}$ & -- &  & \\
 & ES-BGK & $1.2\times10^{-6}$ & $3.6\times10^{-7}$ & $7.0\times10^{-8}$ & -- &  & \\
 & Shakhov & $1.2\times10^{-6}$ & $3.6\times10^{-7}$ & $6.9\times10^{-8}$ & -- &  & \\
 & full Boltzmann (HS) & n.c. & $1.1\times10^{-4}$ & $9.7\times10^{-6}$ & -- &  & \\
\midrule
Blasius & BGK & $4.8\times10^{-6}$ & $1.3\times10^{-6}$ & $2.9\times10^{-7}$ & -- &  & \\
 & ES-BGK & $4.8\times10^{-6}$ & $1.3\times10^{-6}$ & $2.9\times10^{-7}$ & -- &  & \\
 & Shakhov & $4.8\times10^{-6}$ & $1.3\times10^{-6}$ & $2.9\times10^{-7}$ & -- &  & \\
 & full Boltzmann (HS) & n.c. & $1.1\times10^{-4}$ & $2.0\times10^{-6}$ & -- &  & \\
\bottomrule
\end{tabular}
}
\caption{Velocity-space convergence of the least-stable eigenvalue, $R_v=6$: distance from the finest grid of each flow and operator; n.c., not converged.}
\label{tab:conv_Q}
\end{table}

Table~\ref{tab:fullQ} gives the distance of the limit $c_0$ from the Orr--Sommerfeld value for BGK and the full operator at $Q=20$, $24$ and $28$, the latter for the channel flows only. For BGK it does not change with $Q$, $1.2\times10^{-7}$ to $6.5\times10^{-7}$, whereas for the full operator it falls from $1.7\times10^{-6}$ to $8.9\times10^{-6}$ at $Q=20$ to $4.6\times10^{-7}$ to $1.1\times10^{-6}$ at $Q=24$ and, in the channel flows, to the BGK level, $3.7\times10^{-7}$ and $4.5\times10^{-7}$, at $Q=28$. The full operator thus needs a velocity grid finer by about four nodes per direction than the models for the same accuracy.

\begin{table}[!htbp]
\centering\small\setlength{\tabcolsep}{5pt}
\adjustbox{max width=\textwidth}{\begin{tabular}{@{}llccc@{}}
\toprule
 & & \multicolumn{3}{c}{$|c_0-c_{\mathrm{OS}}|$}\\
\cmidrule(l){3-5}
flow & operator & $Q=20$ & $Q=24$ & $Q=28$\\
\midrule
Couette & BGK & $4.6\times10^{-7}$ & $5.3\times10^{-7}$ & $6.5\times10^{-7}$\\
 & full Boltzmann (HS) & $7.3\times10^{-6}$ & $5.2\times10^{-7}$ & $3.7\times10^{-7}$\\
\midrule
Poiseuille & BGK & $3.7\times10^{-7}$ & $3.3\times10^{-7}$ & $4.1\times10^{-7}$\\
 & full Boltzmann (HS) & $1.7\times10^{-6}$ & $4.6\times10^{-7}$ & $4.5\times10^{-7}$\\
\midrule
mixing layer & BGK & $2.0\times10^{-7}$ & $1.2\times10^{-7}$ & \\
 & full Boltzmann (HS) & $8.9\times10^{-6}$ & $1.1\times10^{-6}$ & \\
\midrule
Blasius & BGK & $4.0\times10^{-7}$ & $3.0\times10^{-7}$ & \\
 & full Boltzmann (HS) & $2.2\times10^{-6}$ & $6.5\times10^{-7}$ & \\
\bottomrule
\end{tabular}
}
\caption{Distance of the limit $\Ma\to0$ from the Orr--Sommerfeld value against $Q$, BGK and full operator.}
\label{tab:fullQ}
\end{table}

\paragraph{Mach-number sequences and small Mach numbers}
Table~\ref{tab:mach_app} lists the Mach-number sequences behind the limits of table~\ref{tab:modal}, at $Q=20$, namely for the channel flows (and BGK at $Q=32$) the extrapolations $N\to\infty$ in $N^{-2}$ of the runs at $N=128$ and $192$, which differ from them by at most $3\times10^{-7}$, and for the mixing layer ($\Rey=100$, $N=96$, $c_i$ only) and the Blasius boundary layer ($\Rey=580$, $N=128$) the standard grids. With the eigenvalues extrapolated in $N$ the limits are unchanged to $10^{-7}$. In Couette flow the three relaxation models give $c_0=0.6053431-0.1192298\,\ii$ (BGK) and $0.6053432-0.1192297\,\ii$ (ES-BGK, Shakhov), within $2\times10^{-7}$ in $c_r$ and $5\times10^{-7}$ in $c_i$ of Orr--Sommerfeld, and BGK at $Q=32$ gives a difference of $-4.7\times10^{-7}+2.5\times10^{-7}\,\ii$. In Poiseuille flow the three models give one and the same limit to seven digits, $c_0=0.2375264+0.0037393\,\ii$, which differs from the value of \citet{Orszag1971} by $-0.7\times10^{-7}-3.6\times10^{-7}\,\ii$. For the full operator the difference is $-1.7\times10^{-6}-0.2\times10^{-6}\,\ii$, so that the growth parameter of the unstable mode is reproduced with the full Boltzmann operator to $2\times10^{-7}$, and the offset of the phase speed is its velocity-grid error at $Q=20$, which the change from $Q=20$ to $Q=24$ ($+1.6\times10^{-6}$) cancels to $10^{-7}$. In Couette flow, likewise, the full-operator offset $5.1\times10^{-6}+5.3\times10^{-6}\,\ii$ equals the change of its eigenvalue from $Q=20$ to $24$ to within $5\times10^{-7}$ in both parts.

\begin{table}[!htbp]
\centering\scriptsize\setlength{\tabcolsep}{3.5pt}
\adjustbox{max width=\textwidth}{\begin{tabular}{@{}llcccc@{}}
\toprule
flow & operator & $\Ma=0.2$ & $0.1$ & $0.05$ & $0.025$\\
\midrule
Couette & BGK & $0.6063714-0.1197716\,\ii$ & $0.6056950-0.1193653\,\ii$ & $0.6054785-0.1192637\,\ii$ & \\
 & ES-BGK & $0.6063469-0.1197846\,\ii$ & $0.6056888-0.1193686\,\ii$ & $0.6054769-0.1192645\,\ii$ & \\
 & Shakhov & $0.6063463-0.1197835\,\ii$ & $0.6056888-0.1193683\,\ii$ & $0.6054770-0.1192644\,\ii$ & \\
 & full Boltzmann (HS) & $0.6063455-0.1197779\,\ii$ & $0.6056900-0.1193633\,\ii$ & $0.6054798-0.1192596\,\ii$ & \\
 & BGK, $Q=32$ & $0.6063676-0.1197718\,\ii$ & $0.6056928-0.1193655\,\ii$ & $0.6054771-0.1192639\,\ii$ & \\
\midrule
Poiseuille & BGK & $0.2379460+0.0036562\,\ii$ & $0.2376165+0.0037191\,\ii$ & $0.2375415+0.0037345\,\ii$ & \\
 & ES-BGK & $0.2379527+0.0036605\,\ii$ & $0.2376182+0.0037202\,\ii$ & $0.2375420+0.0037348\,\ii$ & \\
 & Shakhov & $0.2379526+0.0036606\,\ii$ & $0.2376182+0.0037202\,\ii$ & $0.2375420+0.0037348\,\ii$ & \\
 & full Boltzmann (HS) & $0.2379520+0.0036602\,\ii$ & $0.2376171+0.0037202\,\ii$ & $0.2375406+0.0037349\,\ii$ & \\
 & BGK, $Q=32$ & $0.2379467+0.0036562\,\ii$ & $0.2376170+0.0037191\,\ii$ & $0.2375419+0.0037345\,\ii$ & \\
\midrule
mixing layer & BGK &  & $0.3884718\,\ii$ & $0.3920708\,\ii$ & $0.3929727\,\ii$\\
 & ES-BGK &  & $0.3884569\,\ii$ & $0.3920670\,\ii$ & $0.3929718\,\ii$\\
 & Shakhov &  & $0.3884567\,\ii$ & $0.3920670\,\ii$ & $0.3929718\,\ii$\\
 & full Boltzmann (HS) &  & $0.3884693\,\ii$ & $0.3920766\,\ii$ & $0.3929807\,\ii$\\
\midrule
Blasius & BGK &  & $0.3644537+0.0078005\,\ii$ & $0.3641880+0.0079209\,\ii$ & $0.3641303+0.0079504\,\ii$\\
 & ES-BGK &  & $0.3644588+0.0078031\,\ii$ & $0.3641893+0.0079216\,\ii$ & $0.3641306+0.0079506\,\ii$\\
 & Shakhov &  & $0.3644587+0.0078031\,\ii$ & $0.3641892+0.0079216\,\ii$ & $0.3641306+0.0079506\,\ii$\\
 & full Boltzmann (HS) &  & $0.3644583+0.0078034\,\ii$ & $0.3641881+0.0079221\,\ii$ & $0.3641291+0.0079512\,\ii$\\
\bottomrule
\end{tabular}
}
\caption{Mach-number sequences of the least-stable eigenvalue, $Q=20$, behind the limits of table~\ref{tab:modal}.}
\label{tab:mach_app}
\end{table}

Tables~\ref{tab:lowMa} extend the sequences of the channel flows to $\Ma=0.01$ at $N=96$ for the four operators, with the distance from the Orr--Sommerfeld value, the number of inner iterations, the memory and the CPU time. In Couette flow the growth parameter of the relaxation models at $\Ma=0.01$ is $1.1\times10^{-6}$ from the Orr--Sommerfeld value, as the quadratic term of the fit predicts, and the phase speed $2.1\times10^{-5}$, the slip term $c_1\Ma$. For the full operator the distances carry, in addition, the velocity-grid offset found above. In Poiseuille flow the growth parameter of the three models at $\Ma=0.01$ is $5\times10^{-7}$ from the value of \citet{Orszag1971}, and the phase speed $2\times10^{-6}$, the sum of the linear and quadratic terms at this Mach number. The full operator takes $23$ to $25$ inner iterations down to $\Ma=0.025$ and $35$ at $\Ma=0.01$ in Couette flow. In Poiseuille flow it follows the models down to $\Ma=0.025$, with about $40$ inner iterations, while at $\Ma=0.01$ its inner iteration reaches the limit of 150 steps and the eigenpair, with a residual of $2\times10^{-7}$, is not accepted.

\begin{table}[!htbp]
\centering\scriptsize\setlength{\tabcolsep}{3pt}
\begin{tabular}{@{}c@{}}
\adjustbox{max width=\textwidth}{\begin{tabular}{@{}llrrrrr@{}}
\toprule
operator & $\Ma$ & 0.2 & 0.1 & 0.05 & 0.025 & 0.01\\
\midrule
BGK & $c_r$ & $0.6063713$ & $0.6056947$ & $0.6054782$ & $0.6054007$ & $0.6053641$\\
 & $c_i$ & $-0.1197716$ & $-0.1193653$ & $-0.1192637$ & $-0.1192383$ & $-0.1192313$\\
 & $|c-c_{\mathrm{OS}}|$ & $1.2\times10^{-3}$ & $3.8\times10^{-4}$ & $1.4\times10^{-4}$ & $5.8\times10^{-5}$ & $2.1\times10^{-5}$\\
 & its. & 4 & 3 & 3 & 3 & 2\\
 & mem.~(MB) & 177 & 172 & 172 & 172 & 168\\
 & CPU~(s) & 8.3 & 7.2 & 6.8 & 6.5 & 5.8\\
\midrule
ES-BGK & $c_r$ & $0.6063468$ & $0.6056886$ & $0.6054767$ & $0.6054003$ & $0.6053640$\\
 & $c_i$ & $-0.1197846$ & $-0.1193686$ & $-0.1192645$ & $-0.1192385$ & $-0.1192313$\\
 & $|c-c_{\mathrm{OS}}|$ & $1.1\times10^{-3}$ & $3.7\times10^{-4}$ & $1.4\times10^{-4}$ & $5.8\times10^{-5}$ & $2.1\times10^{-5}$\\
 & its. & 4 & 3 & 3 & 3 & 2\\
 & mem.~(MB) & 204 & 204 & 204 & 204 & 204\\
 & CPU~(s) & 8.1 & 7.2 & 6.8 & 6.8 & 6.0\\
\midrule
Shakhov & $c_r$ & $0.6063461$ & $0.6056885$ & $0.6054767$ & $0.6054003$ & $0.6053640$\\
 & $c_i$ & $-0.1197835$ & $-0.1193683$ & $-0.1192644$ & $-0.1192385$ & $-0.1192313$\\
 & $|c-c_{\mathrm{OS}}|$ & $1.1\times10^{-3}$ & $3.7\times10^{-4}$ & $1.4\times10^{-4}$ & $5.8\times10^{-5}$ & $2.1\times10^{-5}$\\
 & its. & 4 & 3 & 3 & 3 & 2\\
 & mem.~(MB) & 183 & 179 & 179 & 179 & 174\\
 & CPU~(s) & 7.7 & 7.0 & 6.5 & 6.6 & 6.1\\
\midrule
full Boltzmann (HS) & $c_r$ & $0.6063452$ & $0.6056896$ & $0.6054796$ & $0.6054043$ & $0.6053686$\\
 & $c_i$ & $-0.1197779$ & $-0.1193633$ & $-0.1192596$ & $-0.1192337$ & $-0.1192265$\\
 & $|c-c_{\mathrm{OS}}|$ & $1.1\times10^{-3}$ & $3.7\times10^{-4}$ & $1.4\times10^{-4}$ & $6.1\times10^{-5}$ & $2.6\times10^{-5}$\\
 & its. & 25 & 23 & 24 & 25 & 35\\
 & mem.~(MB) & 438 & 429 & 520 & 601 & 592\\
 & CPU~(s) & 109 & 76 & 94 & 157 & 190\\
\bottomrule
\end{tabular}
}\\[8pt]
\adjustbox{max width=\textwidth}{\begin{tabular}{@{}llrrrrr@{}}
\toprule
operator & $\Ma$ & 0.2 & 0.1 & 0.05 & 0.025 & 0.01\\
\midrule
BGK & $c_r$ & $0.2379461$ & $0.2376166$ & $0.2375416$ & $0.2375264$ & $0.2375245$\\
 & $c_i$ & $0.0036562$ & $0.0037191$ & $0.0037345$ & $0.0037383$ & $0.0037392$\\
 & $|c-c_{\mathrm{OS}}|$ & $4.3\times10^{-4}$ & $9.2\times10^{-5}$ & $1.6\times10^{-5}$ & $1.3\times10^{-6}$ & $2.0\times10^{-6}$\\
 & its. & 4 & 3 & 3 & 2 & 2\\
 & mem.~(MB) & 177 & 172 & 172 & 172 & 168\\
 & CPU~(s) & 11 & 6.6 & 7.0 & 6.7 & 5.9\\
\midrule
ES-BGK & $c_r$ & $0.2379529$ & $0.2376183$ & $0.2375420$ & $0.2375265$ & $0.2375245$\\
 & $c_i$ & $0.0036605$ & $0.0037202$ & $0.0037348$ & $0.0037384$ & $0.0037393$\\
 & $|c-c_{\mathrm{OS}}|$ & $4.3\times10^{-4}$ & $9.4\times10^{-5}$ & $1.6\times10^{-5}$ & $1.2\times10^{-6}$ & $2.0\times10^{-6}$\\
 & its. & 4 & 3 & 3 & 2 & 2\\
 & mem.~(MB) & 204 & 204 & 204 & 204 & 204\\
 & CPU~(s) & 8.7 & 6.8 & 6.8 & 6.5 & 5.6\\
\midrule
Shakhov & $c_r$ & $0.2379529$ & $0.2376183$ & $0.2375420$ & $0.2375265$ & $0.2375245$\\
 & $c_i$ & $0.0036606$ & $0.0037202$ & $0.0037348$ & $0.0037384$ & $0.0037392$\\
 & $|c-c_{\mathrm{OS}}|$ & $4.3\times10^{-4}$ & $9.4\times10^{-5}$ & $1.6\times10^{-5}$ & $1.2\times10^{-6}$ & $2.0\times10^{-6}$\\
 & its. & 4 & 3 & 3 & 2 & 2\\
 & mem.~(MB) & 183 & 179 & 179 & 179 & 174\\
 & CPU~(s) & 11 & 6.8 & 6.8 & 6.4 & 5.5\\
\midrule
full Boltzmann (HS) & $c_r$ & $0.2379521$ & $0.2376172$ & $0.2375407$ & $0.2375251$ & n.c.\\
 & $c_i$ & $0.0036602$ & $0.0037202$ & $0.0037349$ & $0.0037386$ & n.c.\\
 & $|c-c_{\mathrm{OS}}|$ & $4.3\times10^{-4}$ & $9.3\times10^{-5}$ & $1.5\times10^{-5}$ & $1.8\times10^{-6}$ & \\
 & its. & 37 & 38 & 41 & 39 & 163$^\dagger$\\
 & mem.~(MB) & 638 & 611 & 606 & 615 & 729\\
 & CPU~(s) & 218 & 159 & 243 & 267 & 508\\
\bottomrule
\end{tabular}
}
\end{tabular}
\caption{Channel flows to small Mach number, $Q=20$, $N=96$. Top: plane Couette flow; bottom: plane Poiseuille flow.}
\label{tab:lowMa}
\end{table}

Table~\ref{tab:smallMa} gives, for BGK, the collision rate, the conditioning ratio $\Rey/\Ma^2$, the accuracy floor of the preconditioned solve, the eigenvalue and the true residual of the eigenpair along the same sequences. The accuracy of the solve degrades roughly in proportion to $\Ma^{-2}$, to $3\times10^{-5}$ at $\Ma=0.01$ in Couette flow and $10^{-5}$ in Poiseuille flow, whose conditioning ratio is ten times larger, but the eigenpair residual remains below $10^{-9}$ for the models ($2.2\times10^{-9}$ for the full operator). Running at very small Mach number is therefore possible, but not needed, since the fit through moderate Mach numbers gives the same limit at a better-conditioned operating point.

\begin{table}[!htbp]
\centering\scriptsize\setlength{\tabcolsep}{4pt}
\begin{tabular}{@{}c@{}}
\adjustbox{max width=\textwidth}{\begin{tabular}{@{}lccccc@{}}
\toprule
$\Ma$ & $\nu_\mu$ & $\Rey/\Ma^2$ & solve floor & $c$ & residual\\
\midrule
0.2 & $7.7\times10^{3}$ & $2\times10^{4}$ & $4.3\times10^{-10}$ & $0.6063713-0.1197716\,\ii$ & $9.9\times10^{-10}$\\
0.1 & $1.5\times10^{4}$ & $1\times10^{5}$ & $1.2\times10^{-7}$ & $0.6056947-0.1193653\,\ii$ & $1.9\times10^{-11}$\\
0.05 & $3.1\times10^{4}$ & $4\times10^{5}$ & $2.2\times10^{-8}$ & $0.6054782-0.1192637\,\ii$ & $2.2\times10^{-11}$\\
0.025 & $6.2\times10^{4}$ & $2\times10^{6}$ & $7.4\times10^{-8}$ & $0.6054007-0.1192383\,\ii$ & $4.8\times10^{-11}$\\
0.01 & $1.5\times10^{5}$ & $1\times10^{7}$ & $3.0\times10^{-5}$ & $0.6053641-0.1192313\,\ii$ & $1.4\times10^{-10}$\\
\bottomrule
\end{tabular}
}\\[8pt]
\adjustbox{max width=\textwidth}{\begin{tabular}{@{}lccccc@{}}
\toprule
$\Ma$ & $\nu_\mu$ & $\Rey/\Ma^2$ & solve floor & $c$ & residual\\
\midrule
0.2 & $7.7\times10^{4}$ & $2\times10^{5}$ & $1.9\times10^{-8}$ & $0.2379461+0.0036562\,\ii$ & $7.1\times10^{-11}$\\
0.1 & $1.5\times10^{5}$ & $1\times10^{6}$ & $7.8\times10^{-8}$ & $0.2376166+0.0037191\,\ii$ & $1.3\times10^{-10}$\\
0.05 & $3.1\times10^{5}$ & $4\times10^{6}$ & $2.8\times10^{-7}$ & $0.2375416+0.0037345\,\ii$ & $3.6\times10^{-10}$\\
0.025 & $6.2\times10^{5}$ & $2\times10^{7}$ & $1.1\times10^{-5}$ & $0.2375264+0.0037383\,\ii$ & $7.9\times10^{-10}$\\
0.01 & $1.5\times10^{6}$ & $1\times10^{8}$ & $7.1\times10^{-6}$ & $0.2375245+0.0037392\,\ii$ & $2.0\times10^{-9}$\\
\bottomrule
\end{tabular}
}
\end{tabular}
\caption{Conditioning at small Mach number, BGK, $Q=20$, $N=96$. Top: plane Couette flow; bottom: plane Poiseuille flow.}
\label{tab:smallMa}
\end{table}

\paragraph{Cross-stream resolution and the Reynolds number}
In the Orr--Sommerfeld equation the viscous term carries the factor $1/\Rey$, and the resolution needed grows with the Reynolds number, whereas the kinetic equation has no such term. Table~\ref{tab:cross_Re} gives, for the least-stable wall mode of plane Poiseuille flow at $\alpha=1$ and $\Rey=10^4$, $10^5$ and $10^6$, the error of the eigenvalue on $N$ intervals relative to the same method on its finest grid, $N=256$ for Orr--Sommerfeld and $N=192$ for kLST with the BGK model at $\Ma=0.1$, $Q=20$, and n.f.\ marks a mode not found. The number of intervals at which the mode is first resolved grows with the Reynolds number for both methods and by the same amount, from about $32$ at $\Rey=10^4$ to $64$ at $10^5$ and $96$ at $10^6$, because the kinetic eigenfunction carries the same wall and critical layers, of thickness $(\alpha\Rey)^{-1/3}$, as the Orr--Sommerfeld one, and the grid has to resolve them however the equation is written. Once they are resolved, both errors fall spectrally and at the same rate. On the coarse grids the kinetic error is the smaller of the two, by a factor of two to three ($4.3\times10^{-5}$ against $1.1\times10^{-4}$ at $\Rey=10^4$, $N=32$, and $3.8\times10^{-5}$ against $7.8\times10^{-5}$ at $\Rey=10^6$, $N=96$). On fine grids the kinetic error levels off at a few $10^{-8}$ at $\Rey=10^4$, the slow Knudsen-layer drift, and below $10^{-7}$ at the other two Reynolds numbers. At fixed resolution the accuracy of the kinetic eigenvalues therefore does not deteriorate with the Reynolds number once the viscous layers of the eigenfunction are resolved. The Reynolds number enters through the resolution of those layers, as for the Orr--Sommerfeld equation, and through the conditioning of the inner solve, which degrades like $\Rey/\Ma^2$. In these runs the relative error of the probe solve rises from $10^{-8}$ at $\Rey=10^4$ to $10^{-5}$ at $10^6$, with the residual of the eigenpair below $5\times10^{-7}$ in every case.

\begin{table}[!htbp]
\centering\small\setlength{\tabcolsep}{5pt}
\adjustbox{max width=\textwidth}{\begin{tabular}{@{}rcccccc@{}}
\toprule
 & \multicolumn{2}{c}{$\Rey=10^4$} & \multicolumn{2}{c}{$\Rey=10^5$} & \multicolumn{2}{c}{$\Rey=10^6$}\\
\cmidrule(lr){2-3}\cmidrule(lr){4-5}\cmidrule(l){6-7}
$N$ & OS & kLST & OS & kLST & OS & kLST\\
\midrule
24 & $1.4\times10^{-3}$ & $8.7\times10^{-4}$ & $6.9\times10^{-3}$ & n.f. & $1.0\times10^{-2}$ & n.f.\\
32 & $1.1\times10^{-4}$ & $4.3\times10^{-5}$ & $6.2\times10^{-3}$ & n.f. & $1.9\times10^{-2}$ & n.f.\\
48 & $1.0\times10^{-6}$ & $7.5\times10^{-8}$ & $1.3\times10^{-3}$ & $1.1\times10^{-3}$ & $6.0\times10^{-3}$ & n.f.\\
64 & $1.8\times10^{-8}$ & $6.1\times10^{-8}$ & $4.3\times10^{-5}$ & $1.8\times10^{-5}$ & $3.7\times10^{-3}$ & $1.4\times10^{-3}$\\
96 & $2.0\times10^{-8}$ & $4.9\times10^{-8}$ & $4.3\times10^{-8}$ & $8.9\times10^{-10}$ & $7.8\times10^{-5}$ & $3.8\times10^{-5}$\\
128 & $2.0\times10^{-8}$ & $3.3\times10^{-8}$ & $1.5\times10^{-10}$ & $1.9\times10^{-9}$ & $9.6\times10^{-7}$ & $8.3\times10^{-8}$\\
\bottomrule
\end{tabular}
}
\caption{Plane Poiseuille flow, $\alpha=1$: error of the least-stable eigenvalue against $N$ at three Reynolds numbers, Orr--Sommerfeld and kLST (BGK, $\Ma=0.1$, $Q=20$); n.f., not found.}
\label{tab:cross_Re}
\end{table}

\FloatBarrier

\subsection{Modal analysis}\label{sec:res_modal}
\begin{figure}[!htbp]
\centering
\includegraphics[width=\textwidth]{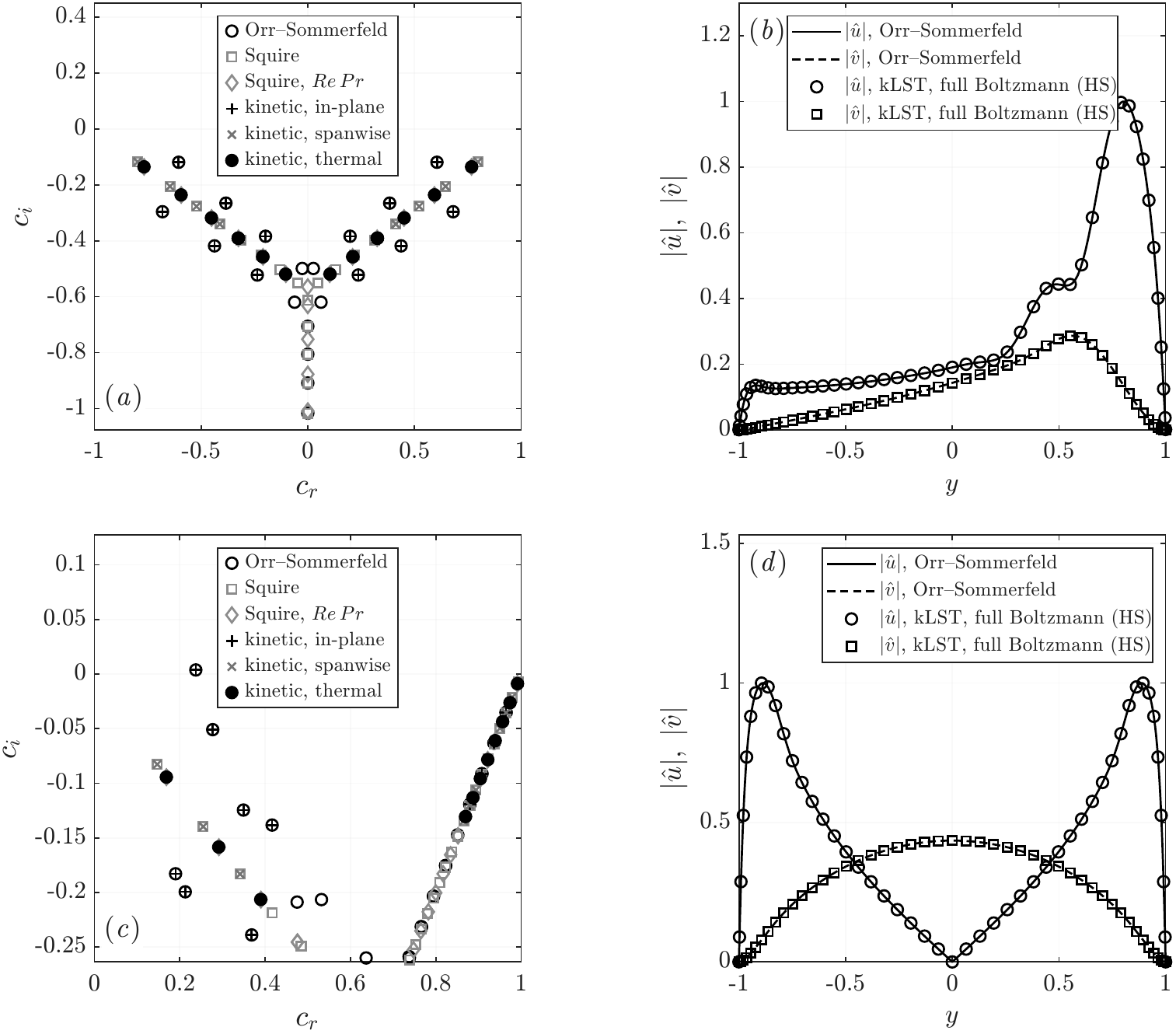}
\caption{Modal kLST of the channel flows, full operator, $\Ma=0.05$. (\textit{a}) Spectrum and (\textit{b}) amplitude functions of plane Couette flow at $(\Rey,\alpha,\beta)=(10^3,1,0)$; (\textit{c}, \textit{d}) the same for plane Poiseuille flow at $(10^4,1,0)$.}
\label{fig:modal_ch}
\end{figure}

The modal results are for two-dimensional disturbances, $\beta=0$, the most amplified in the limit $\Ma\to0$ by Squire's theorem. Figure~\ref{fig:modal_ch} shows the modal results of the channel flows, computed with the full Boltzmann operator for hard spheres at $\Ma=0.05$, $Q=20$, $N=96$. In figure~\ref{fig:modal_ch}(\textit{a}) the eigenvalues of plane Couette flow at $(\Rey,\alpha,\beta)=(10^3,1,0)$ that pass the acceptance test of \S\ref{sec:evpsol} are shown, with the Orr--Sommerfeld and Squire spectra and the Squire spectrum at $\Rey\,\Pran$, $\Pran=0.6604$. Only the neighborhoods of the shifts have been searched, so a reference mode without a kinetic partner has not necessarily been looked for. Figure~\ref{fig:modal_ch}(\textit{b}) shows the amplitude functions $|\hat u|$, $|\hat v|$ of the least-stable mode, kinetic (symbols) against Orr--Sommerfeld (lines), normalized by $\max|\hat u|$, with a largest difference of $1.3\times10^{-3}$ in $|\hat u|$. Figures~\ref{fig:modal_ch}(\textit{c}) and (\textit{d}) show the same for plane Poiseuille flow at $(\Rey,\alpha,\beta)=(10^4,1,0)$, whose least-stable mode is unstable. The amplitude functions are shown at every second collocation point, and the largest difference is $1.6\times10^{-3}$ in $|\hat u|$.

\begin{figure}[!htbp]
\centering
\includegraphics[width=\textwidth]{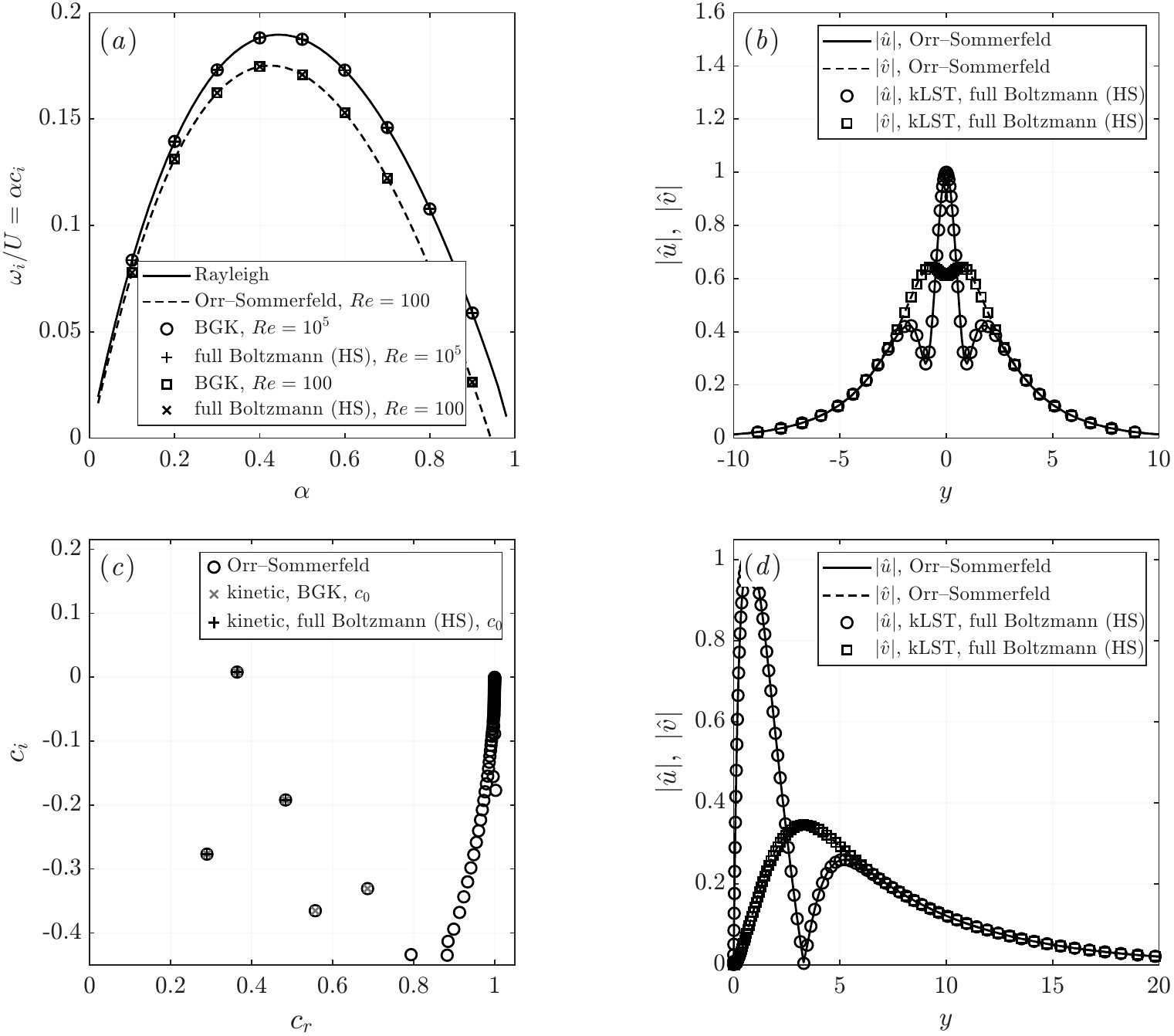}
\caption{Modal kLST of the open flows. (\textit{a}) Growth rate of the mixing layer against $\alpha$; (\textit{b}) its amplitude functions at $(\Rey,\Ma,\alpha,\beta)=(100,0.05,0.4446,0)$. (\textit{c}) Spectrum of the Blasius boundary layer at $(\Rey,\alpha,\beta)=(580,0.179,0)$; (\textit{d}) its Tollmien--Schlichting amplitude functions, $\Ma=0.05$.}
\label{fig:modal_open}
\end{figure}

In figure~\ref{fig:modal_open}(\textit{a}) the temporal growth rate $\alpha c_i$ of the mixing layer is shown against the wavenumber, for the Rayleigh and Orr--Sommerfeld equations at $\Rey=100$ (lines) and for kinetic eigenvalues with the BGK model and with the full Boltzmann operator for hard spheres at $\Rey=10^5$ and $\Rey=100$, extrapolated to $\Ma=0$ from $\Ma=0.1$ and $0.05$ (symbols). Figure~\ref{fig:modal_open}(\textit{b}) shows the amplitude functions of the velocity perturbation at $(\Rey,\Ma,\alpha,\beta)=(100,0.05,0.4446,0)$, full operator at the collocation points (symbols) against Orr--Sommerfeld (lines), normalized by the maximum of $|\hat u|$. The BGK amplitudes differ from those of the full operator by less than $1.2\times10^{-5}$. In figure~\ref{fig:modal_open}(\textit{c}) the Orr--Sommerfeld spectrum of the Blasius boundary layer at $(\Rey,\alpha,\beta)=(580,0.179,0)$ is shown for $c_i>-0.45$, with the limits $c_0$ of the kinetic eigenvalues of its five discrete modes, BGK and full operator (the latter for the first three). The Tollmien--Schlichting wave, mode~1, is in table~\ref{tab:modal}. The damped modes 2--5 are listed in table~\ref{tab:blasius_modes} as limits $\Ma\to0$ at $Q=20$, $N=128$, with their distance from the Orr--Sommerfeld values, and n.c.\ marks the modes for which the full operator did not converge. Figure~\ref{fig:modal_open}(\textit{d}) shows the amplitude functions of the Tollmien--Schlichting wave, full operator at $\Ma=0.05$, $N=128$ (symbols, at the collocation points), against Orr--Sommerfeld (lines). The BGK amplitudes differ from those of the full operator by at most $2\times10^{-5}$.

\begin{table}[!htbp]
\centering\footnotesize\setlength{\tabcolsep}{4pt}
\adjustbox{max width=\textwidth}{\begin{tabular}{@{}llrrrr@{}}
\toprule
operator & mode & 2 & 3 & 4 & 5\\
\midrule
BGK & $c_r$ & $0.4839286$ & $0.2897138$ & $0.6862610$ & $0.5571953$\\
 & $c_i$ & $-0.1920669$ & $-0.2768659$ & $-0.3307711$ & $-0.3653420$\\
 & $|c-c_{\mathrm{OS}}|$ & $1.1\times10^{-6}$ & $8.3\times10^{-7}$ & $5.9\times10^{-7}$ & $1.4\times10^{-6}$\\
\midrule
ES-BGK & $c_r$ & $0.4839286$ & $0.2897138$ & $0.6862588$ & $0.5571953$\\
 & $c_i$ & $-0.1920669$ & $-0.2768659$ & $-0.3307695$ & $-0.3653418$\\
 & $|c-c_{\mathrm{OS}}|$ & $1.1\times10^{-6}$ & $8.3\times10^{-7}$ & $2.8\times10^{-6}$ & $1.5\times10^{-6}$\\
\midrule
Shakhov & $c_r$ & $0.4839286$ & $0.2897138$ & $0.6862616$ & $0.5571953$\\
 & $c_i$ & $-0.1920669$ & $-0.2768659$ & $-0.3307701$ & $-0.3653419$\\
 & $|c-c_{\mathrm{OS}}|$ & $1.1\times10^{-6}$ & $8.2\times10^{-7}$ & $6.5\times10^{-7}$ & $1.4\times10^{-6}$\\
\midrule
full Boltzmann (HS) & $c_r$ & $0.4839236$ & $0.2897099$ & n.c. & n.c.\\
 & $c_i$ & $-0.1920624$ & $-0.2768619$ & n.c. & n.c.\\
 & $|c-c_{\mathrm{OS}}|$ & $7.8\times10^{-6}$ & $6.4\times10^{-6}$ &  & \\
\midrule
Orr--Sommerfeld & $c_r$ & $0.4839295$ & $0.2897144$ & $0.6862613$ & $0.5571966$\\
 & $c_i$ & $-0.1920675$ & $-0.2768665$ & $-0.3307706$ & $-0.3653424$\\
\bottomrule
\end{tabular}
}
\caption{Damped discrete modes 2--5 of the Blasius boundary layer, $(\Rey,\alpha,\beta)=(580,0.179,0)$: limits $\Ma\to0$ and distance from the Orr--Sommerfeld values.}
\label{tab:blasius_modes}
\end{table}

In both channel flows the kinetic eigenvalues lie on the Orr--Sommerfeld spectrum. With the full operator twelve of the first twenty Couette modes, 1 to 10, 13 and 14 on the arms of the Y-shaped spectrum, and fourteen of the first twenty Poiseuille modes, the unstable mode, the A-branch modes and the leading P-branch pairs, are recovered at each Mach number. Their limits $c_0$ lie within $7\times10^{-6}$ to $3.5\times10^{-5}$ (Couette modes 1 to 10) and $1.4\times10^{-6}$ to $6.8\times10^{-6}$ (Poiseuille) of the Orr--Sommerfeld values, the level of the velocity-grid error of the full operator at $Q=20$. The modes not recovered lie at the branch point and on the stem of the Couette spectrum and, in Poiseuille flow, are the P-branch pairs with condition numbers above $2.7\times10^3$. Next to them the relaxation-model operator is itself ill-conditioned, its inverse loses accuracy, and the preconditioned inner iteration of the full operator does not converge. With BGK, for which the preconditioner is the solver, eighteen of the twenty Couette modes and all twenty Poiseuille modes are recovered. The same runs return two further families. The spanwise modes lie on the Squire spectrum. The thermodynamic modes carry density and temperature in equal parts, as an isobaric entropy disturbance must, and lie on the Squire spectrum evaluated at $\Rey\,\Pran$, that is on the spectrum of the linearized energy equation. For $\Pran\neq1$ they therefore separate from the spanwise family, by $7\times10^{-3}$ (Poiseuille) to $3\times10^{-2}$ (Couette) for the full operator. The velocity moments of the kinetic eigenvectors coincide with the Orr--Sommerfeld amplitude functions to $1.3\times10^{-3}$ and $1.6\times10^{-3}$, and satisfy no-slip and no-penetration to plotting accuracy, although neither is imposed, since the wall condition acts on the distribution function alone.

In the mixing layer the kinetic growth rates follow the Orr--Sommerfeld curve at $\Rey=100$ and the Rayleigh curve at $\Rey=10^5$ across the unstable band, for both operators. The relaxation models differ from Orr--Sommerfeld by at most $2\times10^{-6}$ in $c_i$ for $\alpha\ge0.5$. The difference grows towards long waves, to $2\times10^{-5}$ at $\alpha=0.1$, where the two-point fit through $\Ma=0.1$ and $0.05$ leaves the $\Ma^4$ term. The full operator differs by at most $1.7\times10^{-5}$, its velocity-grid error included. At $\Rey=10^5$ the kinetic growth parameter is within $4\times10^{-5}$ of the Rayleigh value $0.4266802$ for every operator, the difference being the viscous correction which the Orr--Sommerfeld equation shares. The kinetic operator, in which nothing is inviscid, thus recovers the inviscid instability as its high-Reynolds-number limit. The amplitude functions decay in the free stream with no condition imposed on the velocity moments at the free boundaries.

In the Blasius boundary layer the relaxation models recover the Tollmien--Schlichting wave and the four damped discrete modes of \citet{Mack1976} to $3\times10^{-6}$ or better, the strongly damped fourth and fifth modes included (tables~\ref{tab:modal} and~\ref{tab:blasius_modes}). Four-digit values reported for these modes with the BGK model differ from the Orr--Sommerfeld values by up to $1.3\%$ \cite{Ma2024}. The full operator recovers the first three modes to $2.2\times10^{-6}$, $7.8\times10^{-6}$ and $6.4\times10^{-6}$, whereas for the fourth and fifth its inner iteration does not converge. The no-slip behavior of the amplitude functions at the wall and their decay in the free stream both follow from the conditions on the distribution function.

\paragraph{The leading twenty modes of the channel flows}
The spectra of figure~\ref{fig:modal_ch} were computed by shift-invert Arnoldi iteration with several shifts. In Couette flow nine shifts with Krylov dimension $m=40$ were placed along the two arms and the stem of the Y-shaped spectrum, at $\Ma=0.2$, $0.1$ and $0.05$, followed by a second pass with one shift ($m=24$) at each Orr--Sommerfeld value of modes 11 to 20 and a third ($m=64$) at modes 17 and 19. In Poiseuille flow eight shifts with $m=40$ were placed along the A and P branches of \citet{Mack1976}. Tables~\ref{tab:spec_full} and~\ref{tab:spec_bgk} list the twenty least-stable Orr--Sommerfeld modes of the two flows with their condition numbers $\varkappa$ and the kinetic eigenvalues at $Q=20$, $N=96$, for the full operator and for BGK, at $\Ma=0.05$ and as the limit $c_0$ of the fit through the three Mach numbers. A dash marks a mode not captured at that Mach number or not at all three, and for Poiseuille flow the parity is that of the wall-normal velocity. The condition numbers of the first twenty modes range from $19$ to $3.6\times10^{6}$ in Couette and from $8$ to $6\times10^{4}$ in Poiseuille flow.

\begin{table}[!htbp]
\centering\scriptsize\setlength{\tabcolsep}{3pt}
\begin{tabular}{@{}c@{}}
\adjustbox{max width=\textwidth}{\begin{tabular}{@{}rlllll@{}}
\toprule
 & Orr--Sommerfeld & $\varkappa$ & kinetic, $\Ma=0.05$ & kinetic, $c_0$ & $|c_0-c_{\mathrm{OS}}|$\\
\midrule
1 & $-0.605343-0.119230\,\ii$ & $3.1\times10^{1}$ & $-0.605480-0.119260\,\ii$ & $-0.605348-0.119225\,\ii$ & $7.3\times10^{-6}$\\
2 & $0.605343-0.119230\,\ii$ & $3.1\times10^{1}$ & $0.605480-0.119260\,\ii$ & $0.605348-0.119225\,\ii$ & $7.3\times10^{-6}$\\
3 & $-0.383757-0.265338\,\ii$ & $5.1\times10^{2}$ & $-0.383898-0.265377\,\ii$ & $-0.383764-0.265330\,\ii$ & $1.0\times10^{-5}$\\
4 & $0.383757-0.265338\,\ii$ & $5.1\times10^{2}$ & $0.383898-0.265377\,\ii$ & $0.383764-0.265330\,\ii$ & $1.1\times10^{-5}$\\
5 & $0.681133-0.296194\,\ii$ & $1.9\times10^{1}$ & $0.681183-0.296157\,\ii$ & $0.681140-0.296185\,\ii$ & $1.1\times10^{-5}$\\
6 & $-0.681133-0.296194\,\ii$ & $1.9\times10^{1}$ & $-0.681183-0.296157\,\ii$ & $-0.681140-0.296185\,\ii$ & $1.1\times10^{-5}$\\
7 & $0.198124-0.383884\,\ii$ & $1.1\times10^{4}$ & $0.198261-0.383935\,\ii$ & $0.198114-0.383888\,\ii$ & $1.1\times10^{-5}$\\
8 & $-0.198123-0.383884\,\ii$ & $1.1\times10^{4}$ & $-0.198261-0.383920\,\ii$ & $-0.198112-0.383850\,\ii$ & $3.5\times10^{-5}$\\
9 & $0.437283-0.418287\,\ii$ & $5.9\times10^{2}$ & $0.437335-0.418238\,\ii$ & $0.437294-0.418275\,\ii$ & $1.7\times10^{-5}$\\
10 & $-0.437283-0.418288\,\ii$ & $5.9\times10^{2}$ & $-0.437335-0.418240\,\ii$ & $-0.437295-0.418281\,\ii$ & $1.3\times10^{-5}$\\
11 & $0.025998-0.499120\,\ii$ & $1.1\times10^{6}$ & -- & -- & --\\
12 & $-0.026007-0.499125\,\ii$ & $1.1\times10^{6}$ & -- & -- & --\\
13 & $-0.236610-0.522604\,\ii$ & $3.7\times10^{4}$ & $-0.236718-0.522674\,\ii$ & $-0.236771-0.522909\,\ii$ & $3.4\times10^{-4}$\\
14 & $0.236610-0.522605\,\ii$ & $3.7\times10^{4}$ & $0.236759-0.522462\,\ii$ & $0.236901-0.522368\,\ii$ & $3.8\times10^{-4}$\\
15 & $0.061895-0.619723\,\ii$ & $2.3\times10^{6}$ & -- & -- & --\\
16 & $-0.061898-0.619740\,\ii$ & $2.3\times10^{6}$ & -- & -- & --\\
17 & $0.000017-0.705831\,\ii$ & $3.6\times10^{6}$ & -- & -- & --\\
18 & $-0.000010-0.805657\,\ii$ & $1.7\times10^{6}$ & -- & -- & --\\
19 & $0.000004-0.908702\,\ii$ & $7.8\times10^{5}$ & -- & -- & --\\
20 & $-0.000001-1.016280\,\ii$ & $3.9\times10^{5}$ & -- & -- & --\\
\bottomrule
\end{tabular}
}\\[8pt]
\adjustbox{max width=\textwidth}{\begin{tabular}{@{}rclllll@{}}
\toprule
 & parity & Orr--Sommerfeld & $\varkappa$ & kinetic, $\Ma=0.05$ & kinetic, $c_0$ & $|c_0-c_{\mathrm{OS}}|$\\
\midrule
1 & even & $0.237526+0.003740\,\ii$ & $5.6\times10^{1}$ & $0.237541+0.003735\,\ii$ & $0.237525+0.003739\,\ii$ & $1.7\times10^{-6}$\\
2 & odd & $0.964631-0.035167\,\ii$ & $7.6\times10^{0}$ & $0.964632-0.035166\,\ii$ & $0.964632-0.035166\,\ii$ & $1.4\times10^{-6}$\\
3 & even & $0.964643-0.035187\,\ii$ & $8.1\times10^{0}$ & $0.964643-0.035185\,\ii$ & $0.964643-0.035185\,\ii$ & $1.4\times10^{-6}$\\
4 & odd & $0.277204-0.050899\,\ii$ & $4.5\times10^{1}$ & $0.277181-0.050939\,\ii$ & $0.277200-0.050897\,\ii$ & $4.5\times10^{-6}$\\
5 & odd & $0.936317-0.063201\,\ii$ & $4.0\times10^{1}$ & $0.936318-0.063199\,\ii$ & $0.936318-0.063200\,\ii$ & $2.3\times10^{-6}$\\
6 & even & $0.936352-0.063252\,\ii$ & $4.3\times10^{1}$ & $0.936353-0.063250\,\ii$ & $0.936353-0.063250\,\ii$ & $2.3\times10^{-6}$\\
7 & odd & $0.907983-0.091223\,\ii$ & $1.7\times10^{2}$ & $0.907985-0.091219\,\ii$ & $0.907985-0.091220\,\ii$ & $3.2\times10^{-6}$\\
8 & even & $0.908056-0.091313\,\ii$ & $1.8\times10^{2}$ & $0.908058-0.091310\,\ii$ & $0.908058-0.091310\,\ii$ & $3.2\times10^{-6}$\\
9 & odd & $0.879627-0.119233\,\ii$ & $6.5\times10^{2}$ & $0.879629-0.119227\,\ii$ & $0.879630-0.119230\,\ii$ & $4.3\times10^{-6}$\\
10 & even & $0.879756-0.119371\,\ii$ & $6.9\times10^{2}$ & $0.879759-0.119368\,\ii$ & $0.879759-0.119367\,\ii$ & $4.6\times10^{-6}$\\
11 & even & $0.349107-0.124502\,\ii$ & $6.8\times10^{2}$ & $0.349094-0.124488\,\ii$ & $0.349103-0.124499\,\ii$ & $4.9\times10^{-6}$\\
12 & odd & $0.416351-0.138227\,\ii$ & $6.4\times10^{2}$ & $0.416355-0.138259\,\ii$ & $0.416344-0.138225\,\ii$ & $6.8\times10^{-6}$\\
13 & odd & $0.851246-0.147234\,\ii$ & $2.7\times10^{3}$ & -- & -- & --\\
14 & even & $0.851449-0.147426\,\ii$ & $2.7\times10^{3}$ & -- & -- & --\\
15 & odd & $0.822835-0.175229\,\ii$ & $1.2\times10^{4}$ & -- & -- & --\\
16 & even & $0.823137-0.175478\,\ii$ & $1.2\times10^{4}$ & -- & -- & --\\
17 & even & $0.190059-0.182822\,\ii$ & $5.0\times10^{1}$ & $0.190052-0.182828\,\ii$ & $0.190057-0.182819\,\ii$ & $3.7\times10^{-6}$\\
18 & odd & $0.212726-0.199361\,\ii$ & $2.2\times10^{1}$ & $0.212733-0.199351\,\ii$ & $0.212722-0.199357\,\ii$ & $5.1\times10^{-6}$\\
19 & odd & $0.794388-0.203221\,\ii$ & $6.2\times10^{4}$ & -- & -- & --\\
20 & even & $0.794818-0.203529\,\ii$ & $5.9\times10^{4}$ & -- & -- & --\\
\bottomrule
\end{tabular}
}
\end{tabular}
\caption{The twenty least-stable Orr--Sommerfeld modes and the eigenvalues of the full operator, $Q=20$, $N=96$, $\beta=0$. Top: Couette, $(\Rey,\alpha)=(10^3,1)$; bottom: Poiseuille, $(10^4,1)$.}
\label{tab:spec_full}
\end{table}

\begin{table}[!htbp]
\centering\scriptsize\setlength{\tabcolsep}{3pt}
\begin{tabular}{@{}c@{}}
\adjustbox{max width=\textwidth}{\begin{tabular}{@{}rlllll@{}}
\toprule
 & Orr--Sommerfeld & $\varkappa$ & kinetic, $\Ma=0.05$ & kinetic, $c_0$ & $|c_0-c_{\mathrm{OS}}|$\\
\midrule
1 & $-0.605343-0.119230\,\ii$ & $3.1\times10^{1}$ & $-0.605478-0.119264\,\ii$ & $-0.605343-0.119230\,\ii$ & $5.0\times10^{-7}$\\
2 & $0.605343-0.119230\,\ii$ & $3.1\times10^{1}$ & $0.605478-0.119264\,\ii$ & $0.605343-0.119230\,\ii$ & $6.3\times10^{-7}$\\
3 & $-0.383757-0.265338\,\ii$ & $5.1\times10^{2}$ & $-0.383892-0.265379\,\ii$ & $-0.383756-0.265336\,\ii$ & $1.6\times10^{-6}$\\
4 & $0.383757-0.265338\,\ii$ & $5.1\times10^{2}$ & $0.383892-0.265379\,\ii$ & $0.383756-0.265336\,\ii$ & $1.4\times10^{-6}$\\
5 & $0.681133-0.296194\,\ii$ & $1.9\times10^{1}$ & $0.681179-0.296163\,\ii$ & $0.681135-0.296192\,\ii$ & $2.6\times10^{-6}$\\
6 & $-0.681133-0.296194\,\ii$ & $1.9\times10^{1}$ & $-0.681179-0.296163\,\ii$ & $-0.681135-0.296192\,\ii$ & $2.8\times10^{-6}$\\
7 & $0.198124-0.383884\,\ii$ & $1.1\times10^{4}$ & $0.198253-0.383931\,\ii$ & $0.198123-0.383878\,\ii$ & $6.1\times10^{-6}$\\
8 & $-0.198123-0.383884\,\ii$ & $1.1\times10^{4}$ & $-0.198254-0.383932\,\ii$ & $-0.198125-0.383882\,\ii$ & $2.6\times10^{-6}$\\
9 & $0.437283-0.418287\,\ii$ & $5.9\times10^{2}$ & $0.437331-0.418242\,\ii$ & $0.437288-0.418287\,\ii$ & $4.4\times10^{-6}$\\
10 & $-0.437283-0.418288\,\ii$ & $5.9\times10^{2}$ & $-0.437331-0.418242\,\ii$ & $-0.437288-0.418287\,\ii$ & $4.3\times10^{-6}$\\
11 & $0.025998-0.499120\,\ii$ & $1.1\times10^{6}$ & $0.026006-0.499124\,\ii$ & $0.026586-0.499335\,\ii$ & $6.3\times10^{-4}$\\
12 & $-0.026007-0.499125\,\ii$ & $1.1\times10^{6}$ & $-0.025990-0.499096\,\ii$ & $-0.026180-0.499145\,\ii$ & $1.7\times10^{-4}$\\
13 & $-0.236610-0.522604\,\ii$ & $3.7\times10^{4}$ & $-0.236665-0.522547\,\ii$ & $-0.236613-0.522604\,\ii$ & $2.9\times10^{-6}$\\
14 & $0.236610-0.522605\,\ii$ & $3.7\times10^{4}$ & $0.236667-0.522547\,\ii$ & $0.236618-0.522605\,\ii$ & $7.9\times10^{-6}$\\
15 & $0.061895-0.619723\,\ii$ & $2.3\times10^{6}$ & $0.061903-0.619692\,\ii$ & $0.061904-0.619654\,\ii$ & $7.0\times10^{-5}$\\
16 & $-0.061898-0.619740\,\ii$ & $2.3\times10^{6}$ & $-0.061897-0.619701\,\ii$ & $-0.061911-0.619692\,\ii$ & $5.0\times10^{-5}$\\
17 & $0.000017-0.705831\,\ii$ & $3.6\times10^{6}$ & -- & -- & --\\
18 & $-0.000010-0.805657\,\ii$ & $1.7\times10^{6}$ & -- & -- & --\\
19 & $0.000004-0.908702\,\ii$ & $7.8\times10^{5}$ & $-0.000042-0.908670\,\ii$ & $-0.000384-0.909555\,\ii$ & $9.4\times10^{-4}$\\
20 & $-0.000001-1.016280\,\ii$ & $3.9\times10^{5}$ & $-0.000001-1.016065\,\ii$ & $-0.000019-1.016438\,\ii$ & $1.6\times10^{-4}$\\
\bottomrule
\end{tabular}
}\\[8pt]
\adjustbox{max width=\textwidth}{\begin{tabular}{@{}rclllll@{}}
\toprule
 & parity & Orr--Sommerfeld & $\varkappa$ & kinetic, $\Ma=0.05$ & kinetic, $c_0$ & $|c_0-c_{\mathrm{OS}}|$\\
\midrule
1 & even & $0.237526+0.003740\,\ii$ & $5.6\times10^{1}$ & $0.237542+0.003735\,\ii$ & $0.237526+0.003739\,\ii$ & $3.8\times10^{-7}$\\
2 & odd & $0.964631-0.035167\,\ii$ & $7.6\times10^{0}$ & $0.964631-0.035166\,\ii$ & $0.964631-0.035167\,\ii$ & $1.0\times10^{-7}$\\
3 & even & $0.964643-0.035187\,\ii$ & $8.1\times10^{0}$ & $0.964643-0.035186\,\ii$ & $0.964643-0.035186\,\ii$ & $1.4\times10^{-7}$\\
4 & odd & $0.277204-0.050899\,\ii$ & $4.5\times10^{1}$ & $0.277183-0.050940\,\ii$ & $0.277203-0.050898\,\ii$ & $1.5\times10^{-6}$\\
5 & odd & $0.936317-0.063201\,\ii$ & $4.0\times10^{1}$ & $0.936316-0.063200\,\ii$ & $0.936317-0.063201\,\ii$ & $2.3\times10^{-7}$\\
6 & even & $0.936352-0.063252\,\ii$ & $4.3\times10^{1}$ & $0.936352-0.063251\,\ii$ & $0.936352-0.063251\,\ii$ & $2.6\times10^{-7}$\\
7 & odd & $0.907983-0.091223\,\ii$ & $1.7\times10^{2}$ & $0.907983-0.091221\,\ii$ & $0.907983-0.091222\,\ii$ & $3.4\times10^{-7}$\\
8 & even & $0.908056-0.091313\,\ii$ & $1.8\times10^{2}$ & $0.908057-0.091313\,\ii$ & $0.908057-0.091313\,\ii$ & $4.7\times10^{-7}$\\
9 & odd & $0.879627-0.119233\,\ii$ & $6.5\times10^{2}$ & $0.879626-0.119230\,\ii$ & $0.879627-0.119233\,\ii$ & $2.0\times10^{-7}$\\
10 & even & $0.879756-0.119371\,\ii$ & $6.9\times10^{2}$ & $0.879756-0.119370\,\ii$ & $0.879757-0.119370\,\ii$ & $1.5\times10^{-6}$\\
11 & even & $0.349107-0.124502\,\ii$ & $6.8\times10^{2}$ & $0.349097-0.124489\,\ii$ & $0.349106-0.124502\,\ii$ & $7.1\times10^{-7}$\\
12 & odd & $0.416351-0.138227\,\ii$ & $6.4\times10^{2}$ & $0.416360-0.138261\,\ii$ & $0.416353-0.138229\,\ii$ & $2.9\times10^{-6}$\\
13 & odd & $0.851246-0.147234\,\ii$ & $2.7\times10^{3}$ & $0.851245-0.147229\,\ii$ & $0.851246-0.147232\,\ii$ & $1.6\times10^{-6}$\\
14 & even & $0.851449-0.147426\,\ii$ & $2.7\times10^{3}$ & $0.851449-0.147426\,\ii$ & $0.851448-0.147427\,\ii$ & $2.3\times10^{-6}$\\
15 & odd & $0.822835-0.175229\,\ii$ & $1.2\times10^{4}$ & $0.822830-0.175217\,\ii$ & $0.822823-0.175210\,\ii$ & $2.3\times10^{-5}$\\
16 & even & $0.823137-0.175478\,\ii$ & $1.2\times10^{4}$ & $0.823129-0.175487\,\ii$ & $0.823110-0.175513\,\ii$ & $4.4\times10^{-5}$\\
17 & even & $0.190059-0.182822\,\ii$ & $5.0\times10^{1}$ & $0.190055-0.182830\,\ii$ & $0.190059-0.182822\,\ii$ & $4.8\times10^{-7}$\\
18 & odd & $0.212726-0.199361\,\ii$ & $2.2\times10^{1}$ & $0.212736-0.199353\,\ii$ & $0.212725-0.199360\,\ii$ & $1.3\times10^{-6}$\\
19 & odd & $0.794388-0.203221\,\ii$ & $6.2\times10^{4}$ & $0.794372-0.203215\,\ii$ & $0.794349-0.203220\,\ii$ & $3.9\times10^{-5}$\\
20 & even & $0.794818-0.203529\,\ii$ & $5.9\times10^{4}$ & $0.794823-0.203524\,\ii$ & $0.794832-0.203516\,\ii$ & $1.9\times10^{-5}$\\
\bottomrule
\end{tabular}
}
\end{tabular}
\caption{As table~\ref{tab:spec_full}, for BGK.}
\label{tab:spec_bgk}
\end{table}

With the full operator, Couette modes 1 to 10 and 13, 14, all on the arms of the Y, are recovered at each Mach number, with residuals of $10^{-9}$ to $10^{-7}$. Their limits differ from Orr--Sommerfeld by $7\times10^{-6}$ to $3.5\times10^{-5}$ for modes 1 to 10, the level of the velocity-grid error of the full operator at $Q=20$, and by $3.4$ to $3.8\times10^{-4}$ for the pair 13, 14 ($\varkappa=3.7\times10^4$). Plane Couette flow has the symmetry $c\to-c^{*}$, which the discretization preserves but the iteration does not impose, so that the distance between a computed eigenvalue and the mirror image of its partner measures the iteration error directly. At $\Ma=0.05$ it is $10^{-8}$ or less for modes 1, 2, 5, 6, $2\times10^{-7}$ and $2\times10^{-6}$ for modes 3, 4 and 9, 10, $1.5\times10^{-5}$ for modes 7, 8 ($\varkappa=10^4$) and $2\times10^{-4}$ for the pair 13, 14. The remaining eight modes, at the branch point and on the stem, are not reached by the full operator, since at every shift with $c_i\le-0.5$ its inner iteration does not converge. The cause is the preconditioner, whose self-check $(\mathsf A_{\mathrm{pc}}-z_0\mathsf I)\mathsf P^{-1}\boldsymbol b-\boldsymbol b$ is $5\times10^{-10}$ at the shifts next to the leading modes, $10^{-6}$ to $2\times10^{-6}$ at $c_i\approx-0.42$, where the iteration still converges, and $7\times10^{-5}$ to $5\times10^{-2}$ at $c_i\le-0.5$. The relaxation-model operator is itself ill-conditioned there, and a preconditioner that is not an accurate inverse of the model cannot be an effective one for the full operator. With BGK, for which the preconditioner is the solver, eighteen of the twenty Couette modes are recovered, namely modes 1 to 10 with residuals of $10^{-11}$ to $10^{-8}$ and limits within $5\times10^{-7}$ to $6\times10^{-6}$ of Orr--Sommerfeld, modes 13 and 14 to $3\times10^{-6}$ and $8\times10^{-6}$, modes 15, 16 to $5$ to $7\times10^{-5}$, the pair at the branch point, modes 11 and 12 ($\varkappa=10^6$), to $2\times10^{-4}$ and $6\times10^{-4}$, and modes 19 and 20 on the stem to $9\times10^{-4}$. Mode 18 passes the acceptance test at $\Ma=0.1$ only, and mode 17, the worst conditioned of the set ($\varkappa=3.6\times10^6$), at no Mach number. Its Ritz pairs have residuals of $3$ to $6\times10^{-8}$, which with this condition number do not determine the eigenvalue, also with $m=64$ and the shift at the Orr--Sommerfeld value.

In Poiseuille flow the full operator recovers fourteen of the twenty modes at each Mach number, with residuals of $10^{-9}$ to $10^{-6}$, namely the unstable mode, the A-branch modes 4, 11, 12, 17 and 18, and the P-branch pairs 2, 3 and 5 to 10, which are nearly degenerate, differ by $10^{-5}$ to $2\times10^{-4}$, and are separated by the parity of $\hat v$. Their limits lie within $1.4\times10^{-6}$ to $6.8\times10^{-6}$ of Orr--Sommerfeld, with a median of $3.4\times10^{-6}$. The six missing modes are the P-branch pairs 13 to 16, 19 and 20, with $\varkappa\ge2.7\times10^{3}$, next to which the preconditioner self-check rises to $5\times10^{-6}$ to $2\times10^{-4}$ and the inner iteration does not converge. With BGK all twenty are recovered at each Mach number, with limits within $10^{-7}$ to $4.4\times10^{-5}$ of Orr--Sommerfeld (median $10^{-6}$). The P-branch modes with $\varkappa<10^3$ and the A-branch modes agree to $3\times10^{-6}$ or better, and the pairs 15, 16 and 19, 20 ($\varkappa\approx10^4$ and $6\times10^4$) to $2$ to $4\times10^{-5}$. No acoustic mode appears in the windows searched, as expected for phase speeds $|c|\le1$ at these Mach numbers.

\paragraph{Spanwise and thermodynamic families}
Table~\ref{tab:spec_models} compares the leading part of the spectrum for the four operators at $Q=20$, $N=96$, namely for the in-plane family the number $n$ of the twenty least-stable Orr--Sommerfeld modes recovered at $\Ma=0.05$ and at all three Mach numbers, with the largest and median distance of the limit $c_0$ from Orr--Sommerfeld, and for the spanwise and thermodynamic families at $\Ma=0.05$ the number of accepted modes and their median distance from the Squire spectrum at $\Rey$ ($d_{\mathrm{Sq}}$) and, for the thermodynamic family, at $\Rey\,\Pran$ ($d_{\Pran}$). In Poiseuille flow the Shakhov and ES-BGK models recover all twenty in-plane modes, like BGK, with median distances of $6\times10^{-7}$ to $10^{-6}$, whereas in Couette flow the two models, run without the second pass, recover 14 and 15 modes. The spanwise modes lie on the Squire spectrum, with a median distance of $5\times10^{-7}$ (Poiseuille) to $10^{-4}$ (Couette) for BGK, comparable to the Mach-number offset of the in-plane modes. The thermodynamic modes carry equal density and temperature fractions, and for BGK, with $\Pran=1$, nearly coincide with the spanwise ones. For $\Pran\ne1$ they lie on the Squire spectrum at $\Rey\,\Pran$, with a median distance of $1.1\times10^{-5}$ for the Shakhov and ES-BGK models in Poiseuille flow. For the full operator, with the discrete Prandtl number $0.6604$ of table~\ref{tab:operator}, the median distance is $2.0\times10^{-4}$, sixty to seventy times larger than for its in-plane and spanwise modes. This would correspond to a Prandtl number different by a few tenths of a percent and has not been explained.

\begin{table}[!htbp]
\centering\scriptsize\setlength{\tabcolsep}{3pt}
\adjustbox{max width=\textwidth}{\begin{tabular}{@{}llcllclcll@{}}
\toprule
 & & \multicolumn{3}{c}{in-plane} & \multicolumn{2}{c}{spanwise} & \multicolumn{3}{c}{thermodynamic}\\
flow & operator & $n$ & $\max|c_0-c_{\mathrm{OS}}|$ & median & $n$ & $d_{\mathrm{Sq}}$ & $n$ & $d_{\Pran}$ & $d_{\mathrm{Sq}}$\\
\midrule
Poiseuille & BGK & 20, 20 & $4.4\times10^{-5}$ & $1.0\times10^{-6}$ & 15 & $4.7\times10^{-7}$ & 27 & $2.1\times10^{-5}$ & $2.1\times10^{-5}$\\
 & Shakhov & 20, 20 & $8.3\times10^{-5}$ & $6.3\times10^{-7}$ & 24 & $1.5\times10^{-6}$ & 20 & $1.1\times10^{-5}$ & $7.0\times10^{-3}$\\
 & ES-BGK & 20, 20 & $5.8\times10^{-5}$ & $9.2\times10^{-7}$ & 23 & $1.2\times10^{-6}$ & 19 & $1.1\times10^{-5}$ & $7.2\times10^{-3}$\\
 & Boltzmann & 14, 14 & $6.8\times10^{-6}$ & $3.4\times10^{-6}$ & 13 & $2.9\times10^{-6}$ & 11 & $2.0\times10^{-4}$ & $6.9\times10^{-3}$\\
\midrule
Couette & BGK & 18, 18 & $9.4\times10^{-4}$ & $4.4\times10^{-6}$ & 18 & $9.5\times10^{-5}$ & 37 & $1.5\times10^{-4}$ & $1.5\times10^{-4}$\\
 & Shakhov & 14, 14 & $1.0\times10^{-2}$ & $3.0\times10^{-6}$ & 24 & $9.6\times10^{-5}$ & 21 & $2.0\times10^{-4}$ & $2.9\times10^{-2}$\\
 & ES-BGK & 15, 15 & $5.6\times10^{-3}$ & $3.0\times10^{-6}$ & 23 & $9.7\times10^{-5}$ & 25 & $2.1\times10^{-4}$ & $2.9\times10^{-2}$\\
 & Boltzmann & 12, 12 & $3.8\times10^{-4}$ & $1.1\times10^{-5}$ & 20 & $1.0\times10^{-4}$ & 18 & $9.3\times10^{-4}$ & $1.6\times10^{-2}$\\
\bottomrule
\end{tabular}
}
\caption{Leading part of the spectrum with the four operators, $Q=20$, $N=96$: in-plane, spanwise and thermodynamic families.}
\label{tab:spec_models}
\end{table}

\begin{figure}[!htbp]
\centering
\begin{tabular}{@{}c@{\hspace{0.02\textwidth}}c@{}}
\includegraphics[width=0.48\textwidth]{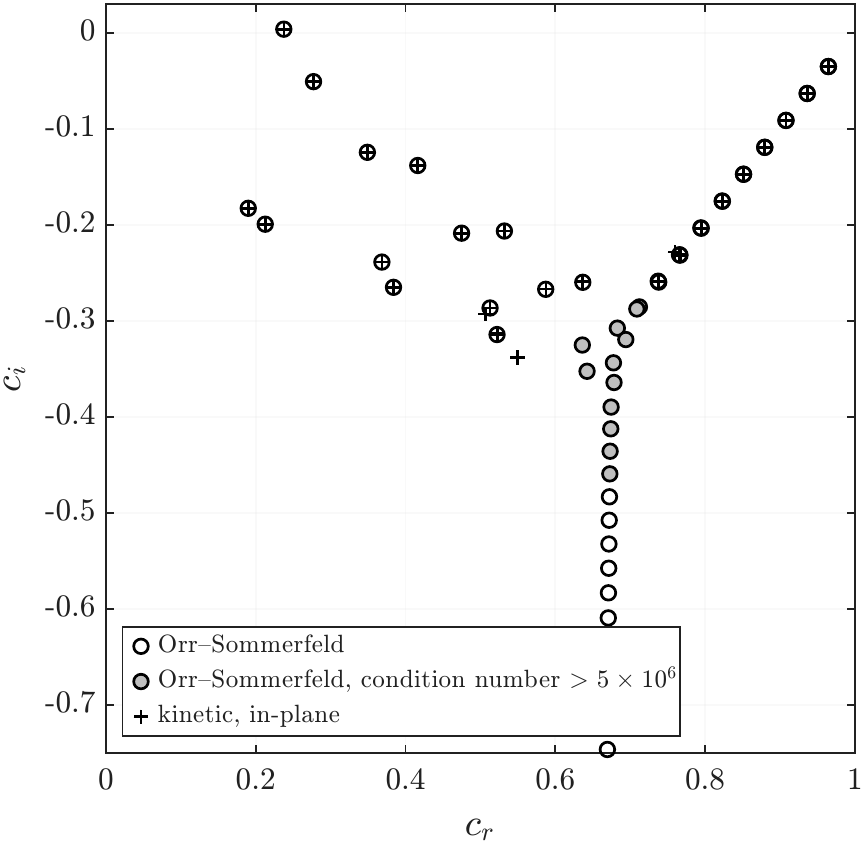} &
\includegraphics[width=0.48\textwidth]{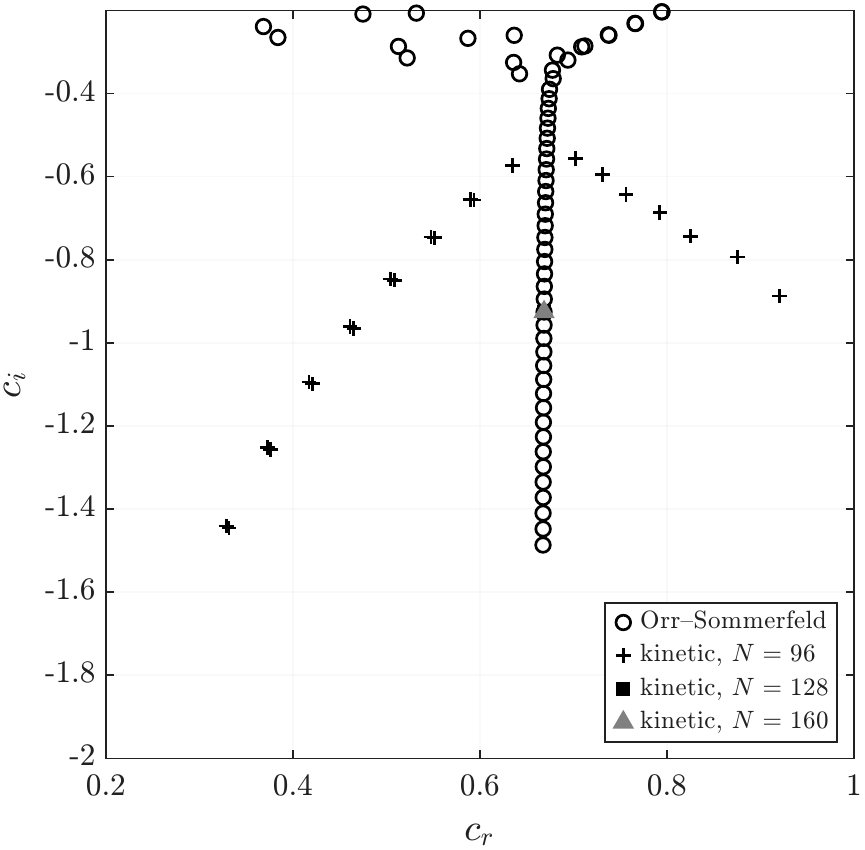}\\
(\textit{a}) & (\textit{b})
\end{tabular}
\caption{Plane Poiseuille flow, BGK, $(\Rey,\Ma,\alpha,\beta)=(10^4,0.1,1,0)$, $Q=20$. (\textit{a}) Deep spectrum, $N=96$; (\textit{b}) weighted iteration at $c=0.669-0.926\,\ii$: the fork at $N=96$ and the S-branch mode at $N=128$, $160$.}
\label{fig:spec_deep}
\end{figure}

\paragraph{The junction of the branches and the S branch}
The region where the A, P and S branches of the Poiseuille spectrum meet is beyond the reach of the full operator and was approached with BGK at $\Ma=0.1$, with one shift ($m=24$) next to each Orr--Sommerfeld value of modes 21 to 64. Figure~\ref{fig:spec_deep}(\textit{a}) shows the Orr--Sommerfeld spectrum down to $c_i=-0.75$, filled circles marking modes with condition numbers above $5\times10^{6}$, and the kinetic in-plane eigenvalues at $N=96$ that pass the acceptance test of \S\ref{sec:evpsol}. One accepted eigenvalue, $c=0.5495-0.3382\,\ii$ (residual $3\times10^{-7}$, no definite parity), has no Orr--Sommerfeld partner and is presumably a pseudo-eigenvalue that the test does not catch. The kinetic eigenvalue is obtained for the twelve modes whose Orr--Sommerfeld condition number is at most $1.8\times10^{6}$, namely the continuation of the A branch (modes 21, 22, 25, 29, 32, 35, within $2$ to $10\times10^{-5}$ of Orr--Sommerfeld, which at this Mach number includes the Mach-number offset), and the P-branch pairs and first junction modes 23, 24, 26, 27, 28 and 30 (within $2\times10^{-4}$ to $1.3\times10^{-3}$, residuals of $9\times10^{-10}$ to $8\times10^{-9}$). None of the modes with $\varkappa\ge7\times10^{6}$ is obtained, nor any of the upper S branch (modes 43 to 64), although their condition numbers fall again to $3\times10^4$. At the junction the preconditioner loses its accuracy, its self-check rising from $10^{-8}$ at the leading mode to $10^{-1}$ at $c=0.636-0.325\,\ii$, and on the upper S branch the outer iteration returns clusters of Ritz pairs around the shift, of mixed in-plane and thermodynamic content, for instance at $c=0.6730-0.4592\,\ii$ seven pairs within $5\times10^{-4}$ of the shift, with true residuals between $3\times10^{-9}$ and $2\times10^{-6}$. A pair $(\omega,\boldsymbol u)$ with residual $r$ implies $\|(\mathsf A+\ii\omega\mathsf I)^{-1}\|\ge1/r$, so that the resolvent norm of the kinetic operator there is at least $3\times10^{8}$, and the individual eigenvalues are not determined at the accuracy of the present iteration. The modes that can and cannot be resolved are separated by the condition number of the corresponding Orr--Sommerfeld eigenvalue.

The iteration was repeated in the inner product weighted by $1/M$, the inner product of the energy norm \eqref{eq:Edisc}, in which the linearized collision operator is self-adjoint, although the operator as a whole remains non-normal, since streaming and shear are untouched. The unstable eigenvalue is reproduced to ten digits. On the lower S branch the change is decisive, since with the shift at $c=0.669-0.926\,\ii$ and $N=96$, the Euclidean iteration with $m=40$ converges no Ritz pair at all, although the linear solves are accurate there, while the weighted one with $m=140$ converges 72, the in-plane ones to residuals of $6\times10^{-10}$ to $4\times10^{-9}$. Figure~\ref{fig:spec_deep}(\textit{b}) shows these eigenvalues, with residuals below $10^{-8}$. They are not on the S branch but form a fork, namely an arm of even and odd pairs running from $c=0.635-0.573\,\ii$ to $0.33-1.45\,\ii$, and an arm of single modes of alternating parity from $0.703-0.556\,\ii$ to $0.921-0.888\,\ii$, and the spanwise and thermodynamic families fork in the same way. At $N=64$ the same shift returns converged eigenvalues in entirely different places. At $N=128$ the fork is gone and the iteration returns $c=0.668665-0.925775\,\ii$ (even, residual $3\times10^{-10}$), within $5\times10^{-5}$ of Orr--Sommerfeld mode 61, $0.668680-0.925822\,\ii$, together with its spanwise and thermodynamic counterparts, and at $N=160$ it returns $0.668668-0.925776\,\ii$, a change of $3\times10^{-6}$. The fork is therefore an artifact of the cross-stream resolution, because the discretization error, $10^{-7}$ for the leading mode, is amplified by the condition number of the deep modes, and the kinetic spectrum below $c_i\approx-0.3$ is not converged at $N\le96$. It resembles the split S branch familiar from under-resolved Orr--Sommerfeld computations. At $N=128$ only these three eigenvalues converge, whereas the other 29 accepted Ritz pairs, with residuals of $2\times10^{-8}$ to $10^{-5}$, almost all of the thermodynamic family, scatter over $0.60\le c_r\le0.72$ and are pseudo-eigenvalues. With the S branch in place the shifted operator is nearly singular, the preconditioner self-check rises from $6\times10^{-9}$ at $N=96$ to $4\times10^{-3}$, and the inner iteration no longer reaches its tolerance. Resolving the S branch mode by mode needs both $N\ge128$ and linear solves more accurate than the present double-precision ones. Neither is needed for the pseudospectra, which are insensitive to this ill-conditioning.

\paragraph{Reynolds-number and wavenumber dependence of the open flows}
Table~\ref{tab:Re} follows the least-stable eigenvalue of the two open flows with the Reynolds number for the four operators, namely the mixing layer at $\alpha=0.4446$, $N=96$, from $\Rey=10^3$ to $10^5$, as limits $\Ma\to0$ of the runs at $\Ma=0.1$, $0.05$, $0.025$ (stationary mode, $c_i$ only), and the Tollmien--Schlichting wave of the Blasius boundary layer at $\alpha=0.179$, $N=128$, from $\Rey=1000$ to $10^4$, as limits through the same Mach numbers. The base cases $\Rey=100$ and $580$ are in table~\ref{tab:modal}, distances are from the Orr--Sommerfeld value in the last line of each block, and n.c.\ marks runs in which the inner iteration of the full operator did not converge. In the mixing layer the limits of the relaxation models agree with the Orr--Sommerfeld values to between $10^{-8}$ and $1.3\times10^{-6}$, and those of the full operator to between $2\times10^{-7}$ and $9\times10^{-6}$. At $\Rey=10^5$, where the collision rate at $\Ma=0.025$ is $\nu_\mu=3\times10^{6}$, more than eight orders of magnitude above the growth rate, the relaxation models still take two or three inner iterations, while the full operator takes $22$ at $\Rey=100$, $39$ at $10^4$ and $62$ at $10^5$, the cost of the three runs behind one limit rising from four minutes to twenty-five. In the Blasius boundary layer the wave passes from amplification into damping. For the relaxation models the agreement is between $3\times10^{-7}$ and $2\times10^{-6}$ throughout, and the full operator agrees to $2.2\times10^{-6}$ at $\Rey=580$ and $1000$ and to $1.1\times10^{-5}$ at $\Rey=2000$, with an inner iteration count that grows from $26$ to $58$, whereas at $\Rey=5000$ and $10^4$ its inner iteration does not converge within 150 steps at the smaller Mach numbers. In both flows the distance of the kinetic eigenvalue from the reference does not change with the Reynolds number on a fixed grid, consistently with table~\ref{tab:cross_Re}.

\begin{table}[!htbp]
\centering\scriptsize\setlength{\tabcolsep}{4pt}
\adjustbox{max width=\textwidth}{\begin{tabular}{@{}lllrrrrr@{}}
\toprule
mixing layer & operator & $\Rey$ & $10^{3}$ & $10^{4}$ & $10^{5}$\\
\midrule
 & BGK & $c_i$ & $0.4230427$ & $0.4263126$ & $0.4266422$\\
 &  & $|c-c_{\mathrm{OS}}|$ & $7.3\times10^{-9}$ & $1.4\times10^{-7}$ & $1.2\times10^{-6}$\\
 & ES-BGK & $c_i$ & $0.4230427$ & $0.4263126$ & $0.4266426$\\
 &  & $|c-c_{\mathrm{OS}}|$ & $3.2\times10^{-9}$ & $1.0\times10^{-7}$ & $8.4\times10^{-7}$\\
 & Shakhov & $c_i$ & $0.4230427$ & $0.4263126$ & $0.4266422$\\
 &  & $|c-c_{\mathrm{OS}}|$ & $7.4\times10^{-9}$ & $1.4\times10^{-7}$ & $1.3\times10^{-6}$\\
 & full Boltzmann (HS) & $c_i$ & $0.4230429$ & $0.4263120$ & $0.4266501$\\
 &  & $|c-c_{\mathrm{OS}}|$ & $1.8\times10^{-7}$ & $7.4\times10^{-7}$ & $6.8\times10^{-6}$\\
 & Orr--Sommerfeld & $c_i$ & $0.4230427$ & $0.4263127$ & $0.4266434$\\
\midrule
Blasius & operator & $\Rey$ & 1000 & 2000 & 5000 & 10000\\
\midrule
 & BGK & $c_r$ & $0.3382703$ & $0.3089212$ & $0.3282729$ & $0.3250401$\\
 &  & $c_i$ & $0.0048347$ & $-0.0165673$ & $-0.0293872$ & $-0.0324756$\\
 &  & $|c-c_{\mathrm{OS}}|$ & $5.0\times10^{-7}$ & $2.0\times10^{-6}$ & $6.1\times10^{-7}$ & $3.4\times10^{-7}$\\
 & ES-BGK & $c_r$ & $0.3382703$ & $0.3089212$ & $0.3282729$ & $0.3250401$\\
 &  & $c_i$ & $0.0048347$ & $-0.0165673$ & $-0.0293870$ & $-0.0324758$\\
 &  & $|c-c_{\mathrm{OS}}|$ & $4.9\times10^{-7}$ & $1.9\times10^{-6}$ & $4.0\times10^{-7}$ & $5.0\times10^{-7}$\\
 & Shakhov & $c_r$ & $0.3382703$ & $0.3089212$ & $0.3282729$ & $0.3250400$\\
 &  & $c_i$ & $0.0048347$ & $-0.0165673$ & $-0.0293871$ & $-0.0324758$\\
 &  & $|c-c_{\mathrm{OS}}|$ & $5.0\times10^{-7}$ & $2.0\times10^{-6}$ & $5.9\times10^{-7}$ & $4.2\times10^{-7}$\\
 & full Boltzmann (HS) & $c_r$ & $0.3382685$ & $0.3089171$ & n.c. & n.c.\\
 &  & $c_i$ & $0.0048343$ & $-0.0165759$ & n.c. & n.c.\\
 &  & $|c-c_{\mathrm{OS}}|$ & $2.2\times10^{-6}$ & $1.1\times10^{-5}$ &  & \\
 & Orr--Sommerfeld & $c_r$ & $0.3382706$ & $0.3089218$ & $0.3282728$ & $0.3250398$\\
 &  & $c_i$ & $0.0048351$ & $-0.0165654$ & $-0.0293866$ & $-0.0324754$\\
\bottomrule
\end{tabular}
}
\caption{Reynolds-number dependence of the least-stable eigenvalue, limits $\Ma\to0$: mixing layer, $\alpha=0.4446$; Blasius boundary layer, $\alpha=0.179$; n.c., not converged.}
\label{tab:Re}
\end{table}

Table~\ref{tab:mixing_alpha} gives the growth parameter of the mixing layer across the unstable band at $\Rey=100$ and $10^5$ for the four operators, each entry the limit $\Ma\to0$ from $\Ma=0.1$ and $0.05$, with the distance from the Orr--Sommerfeld value in the last line of each block. For $\alpha<0.3$ the domain is $y_\infty=160$ with $b_y=7.5$ and $N=128$, since long waves decay slowly in the free stream, as $\exp(-\alpha|y|)$, and with $y_\infty=40$ the Orr--Sommerfeld value at $\alpha=0.1$ is itself in error by $9\times10^{-5}$. The mode is stationary in every run, $|c_r|<10^{-8}$. These are the data of figure~\ref{fig:modal_open}(\textit{a}). At $\Rey=100$ the full operator takes $20$ or $21$ inner iterations at every wavenumber, whereas at $\Rey=10^5$ it takes between $39$ and $93$, without a clear dependence on the wavenumber, and $10$ to $18$ minutes for the two runs behind each limit.

\begin{table}[!htbp]
\centering\scriptsize\setlength{\tabcolsep}{2.5pt}
\adjustbox{max width=\textwidth}{\begin{tabular}{@{}lllrrrrrrrrr@{}}
\toprule
$\Rey$ & operator & & 0.1 & 0.2 & 0.3 & 0.4 & 0.5 & 0.6 & 0.7 & 0.8 & 0.9\\
\midrule
$10^2$ & BGK & $c_i$ & $0.778123$ & $0.656167$ & $0.541028$ & $0.436645$ & $0.341821$ & $0.254950$ & $0.174625$ & $0.099694$ & $0.029233$\\
 & & $|c-c_{\mathrm{OS}}|$ & $2.0\times10^{-5}$ & $1.2\times10^{-5}$ & $7.2\times10^{-6}$ & $4.0\times10^{-6}$ & $1.9\times10^{-6}$ & $4.4\times10^{-7}$ & $5.3\times10^{-7}$ & $1.2\times10^{-6}$ & $1.8\times10^{-6}$\\
 & ES-BGK & $c_i$ & $0.778123$ & $0.656167$ & $0.541028$ & $0.436645$ & $0.341821$ & $0.254950$ & $0.174625$ & $0.099694$ & $0.029233$\\
 & & $|c-c_{\mathrm{OS}}|$ & $2.0\times10^{-5}$ & $1.2\times10^{-5}$ & $7.3\times10^{-6}$ & $4.0\times10^{-6}$ & $1.9\times10^{-6}$ & $4.7\times10^{-7}$ & $5.1\times10^{-7}$ & $1.2\times10^{-6}$ & $1.8\times10^{-6}$\\
 & Shakhov & $c_i$ & $0.778123$ & $0.656167$ & $0.541028$ & $0.436645$ & $0.341821$ & $0.254950$ & $0.174625$ & $0.099694$ & $0.029233$\\
 & & $|c-c_{\mathrm{OS}}|$ & $2.0\times10^{-5}$ & $1.2\times10^{-5}$ & $7.3\times10^{-6}$ & $4.0\times10^{-6}$ & $1.9\times10^{-6}$ & $4.7\times10^{-7}$ & $5.1\times10^{-7}$ & $1.2\times10^{-6}$ & $1.8\times10^{-6}$\\
 & full Boltzmann (HS) & $c_i$ & $0.778126$ & $0.656172$ & $0.541034$ & $0.436653$ & $0.341830$ & $0.254961$ & $0.174638$ & $0.099708$ & $0.029249$\\
 & & $|c-c_{\mathrm{OS}}|$ & $1.7\times10^{-5}$ & $7.7\times10^{-6}$ & $9.0\times10^{-7}$ & $4.0\times10^{-6}$ & $7.7\times10^{-6}$ & $1.1\times10^{-5}$ & $1.3\times10^{-5}$ & $1.5\times10^{-5}$ & $1.7\times10^{-5}$\\
 & Orr--Sommerfeld & $c_i$ & $0.778144$ & $0.656179$ & $0.541035$ & $0.436649$ & $0.341823$ & $0.254951$ & $0.174625$ & $0.099693$ & $0.029232$\\
\midrule
$10^5$ & BGK & $c_i$ & $0.836344$ & $0.697393$ & $0.576846$ & $0.470408$ & $0.374983$ & $0.288287$ & $0.208606$ & $0.134632$ & $0.065350$\\
 & & $|c-c_{\mathrm{OS}}|$ & $2.3\times10^{-5}$ & $1.4\times10^{-5}$ & $8.4\times10^{-6}$ & $4.9\times10^{-6}$ & $2.6\times10^{-6}$ & $1.1\times10^{-6}$ & $8.2\times10^{-8}$ & $6.8\times10^{-7}$ & $1.3\times10^{-6}$\\
 & ES-BGK & $c_i$ & $0.836344$ & $0.697394$ & $0.576846$ & $0.470408$ & $0.374983$ & $0.288287$ & $0.208606$ & $0.134632$ & $0.065350$\\
 & & $|c-c_{\mathrm{OS}}|$ & $2.3\times10^{-5}$ & $1.4\times10^{-5}$ & $8.2\times10^{-6}$ & $4.8\times10^{-6}$ & $2.5\times10^{-6}$ & $1.1\times10^{-6}$ & $2.1\times10^{-8}$ & $7.3\times10^{-7}$ & $1.3\times10^{-6}$\\
 & Shakhov & $c_i$ & $0.836344$ & $0.697393$ & $0.576846$ & $0.470408$ & $0.374983$ & $0.288287$ & $0.208606$ & $0.134632$ & $0.065350$\\
 & & $|c-c_{\mathrm{OS}}|$ & $2.3\times10^{-5}$ & $1.4\times10^{-5}$ & $8.4\times10^{-6}$ & $4.9\times10^{-6}$ & $2.6\times10^{-6}$ & $1.1\times10^{-6}$ & $8.7\times10^{-8}$ & $6.8\times10^{-7}$ & $1.3\times10^{-6}$\\
 & full Boltzmann (HS) & $c_i$ & $0.836349$ & $0.697393$ & $0.576845$ & $0.470408$ & $0.374982$ & $0.288286$ & $0.208605$ & $0.134631$ & $0.065352$\\
 & & $|c-c_{\mathrm{OS}}|$ & $1.9\times10^{-5}$ & $1.4\times10^{-5}$ & $8.8\times10^{-6}$ & $5.4\times10^{-6}$ & $3.2\times10^{-6}$ & $1.7\times10^{-6}$ & $1.6\times10^{-6}$ & $5.9\times10^{-8}$ & $7.2\times10^{-6}$\\
 & Orr--Sommerfeld & $c_i$ & $0.836367$ & $0.697407$ & $0.576854$ & $0.470413$ & $0.374986$ & $0.288288$ & $0.208606$ & $0.134631$ & $0.065348$\\
\bottomrule
\end{tabular}
}
\caption{Mixing layer: growth parameter $c_i$ against $\alpha$ at $\Rey=100$ and $10^5$, limits $\Ma\to0$.}
\label{tab:mixing_alpha}
\end{table}

\FloatBarrier

\subsection{Non-modal analysis}\label{sec:res_nonmodal}
The non-modal results are for streamwise-invariant disturbances, $\alpha=0$, $\beta\neq0$, which are known to result in the largest transient energy growth.

\begin{table}[!htbp]
\centering\small\setlength{\tabcolsep}{5pt}
\adjustbox{max width=\textwidth}{\begin{tabular}{@{}lrrrrrr@{}}
\toprule
flow & $\Rey$ & $\beta$ & $t_{\max}$ & OS--Squire & kLST & rel.\ difference\\
\midrule
plane Couette & 1000 & 1.66 & 138 & 1165.95 & 1166.33 & $3.3\times10^{-4}$\\
plane Poiseuille & 1000 & 2.04 & 76 & 196.17 & 196.34 & $8.7\times10^{-4}$\\
mixing layer & 100 & 0.5 & 276 & 1350.35 & 1351.71 & $1.0\times10^{-3}$\\
Blasius & $1000^{\dagger}$ & $0.65^{\dagger}$ & $782^{\dagger}$ & 1515.49 & 1515.42 & $4.6\times10^{-5}$\\
\bottomrule
\end{tabular}
}
\caption{Maximum transient growth of streamwise-invariant disturbances: kLST (BGK, $\Ma=0.1$) against Orr--Sommerfeld--Squire. $^{\dagger}$In units of the displacement thickness $\delta^*$.}
\label{tab:tg}
\end{table}

\begin{table}[!htbp]
\centering\small\setlength{\tabcolsep}{4pt}
\adjustbox{max width=\textwidth}{\begin{tabular}{@{}llrrrrrr@{}}
\toprule
flow & $\Rey$ & $\alpha h$ & $\beta h$ & $tU/h$ & published & OS--Squire & kLST\\
\midrule
\multicolumn{8}{@{}l}{\textit{Maximum energy growth $G$} \citep{Butler1992}}\\
Couette & 1000 & 0.035 & 1.60 & 117 & 1185 & 1184.56 & 1184.83\\
 & & 0 & 1.66 & 138 & 1166 & 1165.95 & 1166.33\\
 & & 0.46 & 1.9 & 20 & 312 & 312.12 & 312.03\\
 & & 1.6 & 2.9 & 5 & 26.4 & 26.437 & 26.424\\
 & & 1.21 & 0 & 8.7 & 13.0 & 12.958 & 12.963\\
Poiseuille & 5000 & 0 & 2.044 & 379 & 4897 & 4896.95 & 4896.76\\
 & & 1.48 & 0 & 14.1 & 45.7 & 45.713 & 45.756\\
Blasius & $1000^{\dagger}$ & 0 & $0.65^{\dagger}$ & $778^{\dagger}$ & 1514 & 1515.49 & 1515.42\\
\midrule
\multicolumn{8}{@{}l}{\textit{Maximal resolvent norm, $\alpha=0$, $\omega=0$} \citep{TrefethenScience}}\\
Couette & 1000 & 0 & 1.18 & -- & $(\Rey/8.12)^2=15\,167$ & 15\,182 & 15\,193\\
Poiseuille & 1000 & 0 & 1.62 & -- & $(\Rey/17.4)^2=3303$ & 3322 & 3324\\
\bottomrule
\end{tabular}
}
\caption{Optimal energy growth and maximal resolvent norm: kLST (BGK, $\Ma=0.1$) against Orr--Sommerfeld--Squire and published values. $^{\dagger}$In units of the displacement thickness $\delta^*$.}
\label{tab:tglit}
\end{table}

\begin{figure}[!htbp]
\centering
\includegraphics[width=\textwidth]{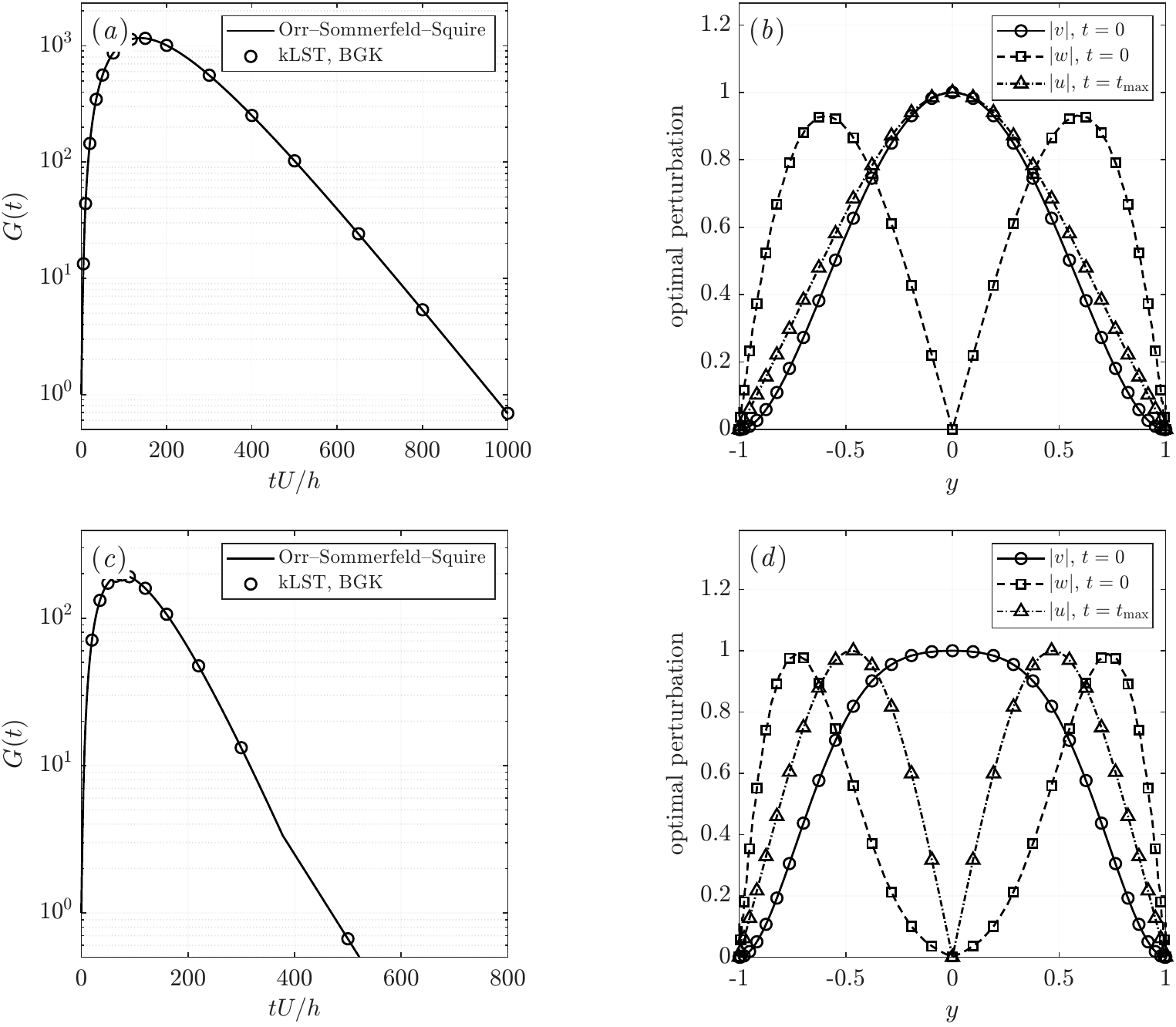}
\caption{Non-modal kLST of the channel flows, BGK, $(\Rey,\Ma,\alpha)=(1000,0.1,0)$. (\textit{a}) $G(t)$ and (\textit{b}) optimal disturbance and response of plane Couette flow, $\beta=1.66$; (\textit{c}, \textit{d}) the same for plane Poiseuille flow, $\beta=2.04$. Lines, Orr--Sommerfeld--Squire; symbols, kLST.}
\label{fig:nm_ch}
\end{figure}

\begin{figure}[!htbp]
\centering
\includegraphics[width=\textwidth]{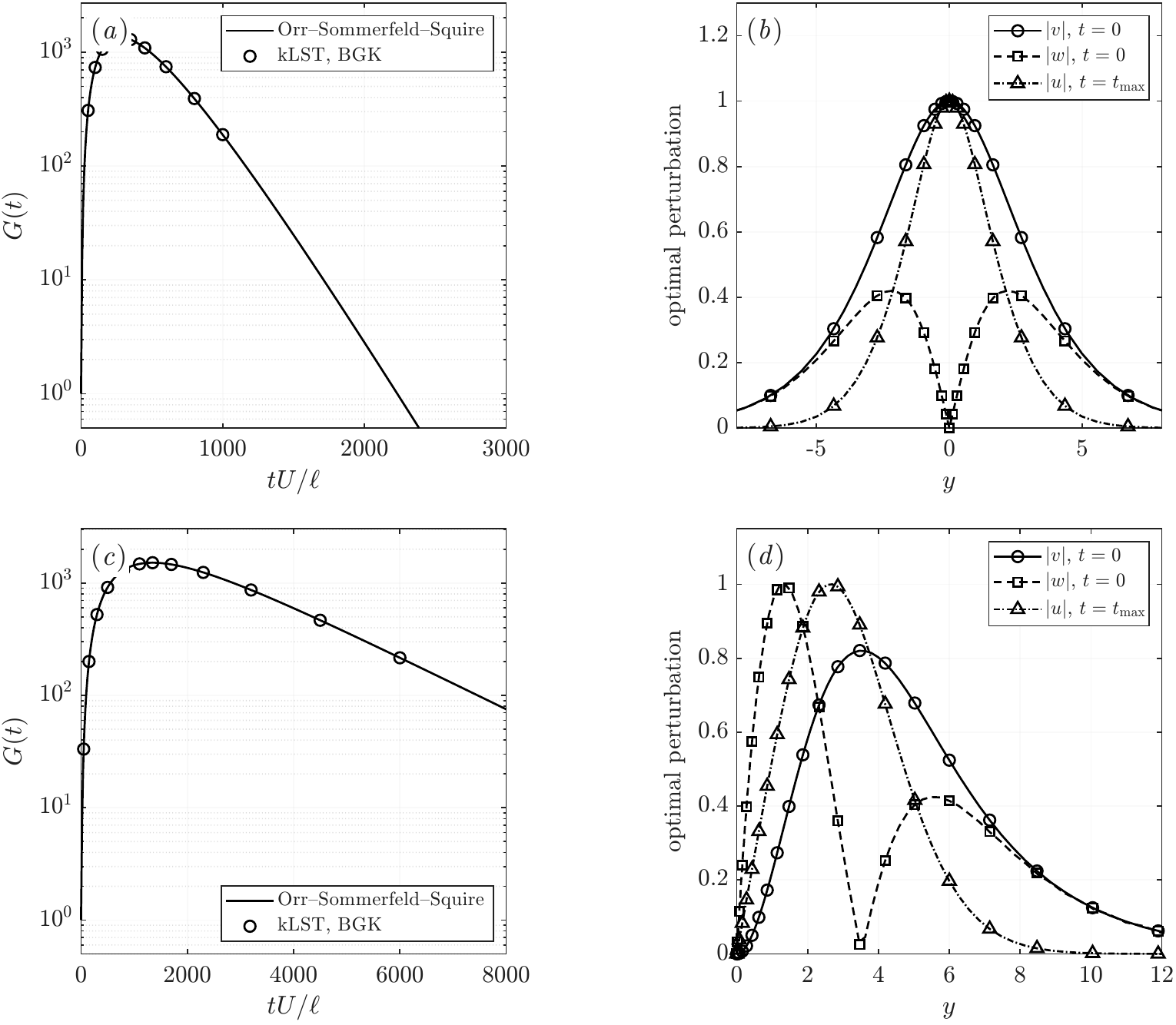}
\caption{Non-modal kLST of the open flows, BGK, $\Ma=0.1$, $\alpha=0$. (\textit{a}) $G(t)$ and (\textit{b}) optimal disturbance and response of the mixing layer, $(\Rey,\beta)=(100,0.5)$; (\textit{c}, \textit{d}) the same for the Blasius boundary layer, $(\Rey_{\delta^*},\beta\delta^*)=(1000,0.65)$. Lines, Orr--Sommerfeld--Squire; symbols, kLST.}
\label{fig:nm_open}
\end{figure}

In figure~\ref{fig:nm_ch} the non-modal results with the BGK operator for the channel flows are shown at $\Rey=1000$, $\Ma=0.1$, $Q=16$, $N=32$. Figure~\ref{fig:nm_ch}(\textit{a}) shows $G(t)$ of plane Couette flow at $(\alpha,\beta)=(0,1.66)$, Orr--Sommerfeld--Squire (line) against kLST (symbols), and figure~\ref{fig:nm_ch}(\textit{b}) the optimal initial disturbance ($|v|$, $|w|$) and the optimal response ($|u|$) at $t=t_{\max}$, each normalized by its maximum, Orr--Sommerfeld--Squire (lines) against the velocity moments of the kinetic optimal distribution (symbols). Figures~\ref{fig:nm_ch}(\textit{c}) and (\textit{d}) show the same for plane Poiseuille flow at $(\alpha,\beta)=(0,2.04)$. Figure~\ref{fig:nm_open} shows the open flows in the same way, kLST with the BGK operator at $\Ma=0.1$, $Q=16$, $N=64$, with the mixing layer at $(\Rey,\alpha,\beta)=(100,0,0.5)$ in (\textit{a}) and (\textit{b}), and the Blasius boundary layer at $\Rey_{\delta^*}=1000$, $\beta\delta^*=0.65$, that is $\Rey=581.1$, $\beta=0.3777$ in units of $\ell$, in (\textit{c}) and (\textit{d}).

\begin{figure}[!htbp]
\centering
\begin{tabular}{@{}cc@{}}
\includegraphics[width=0.47\textwidth]{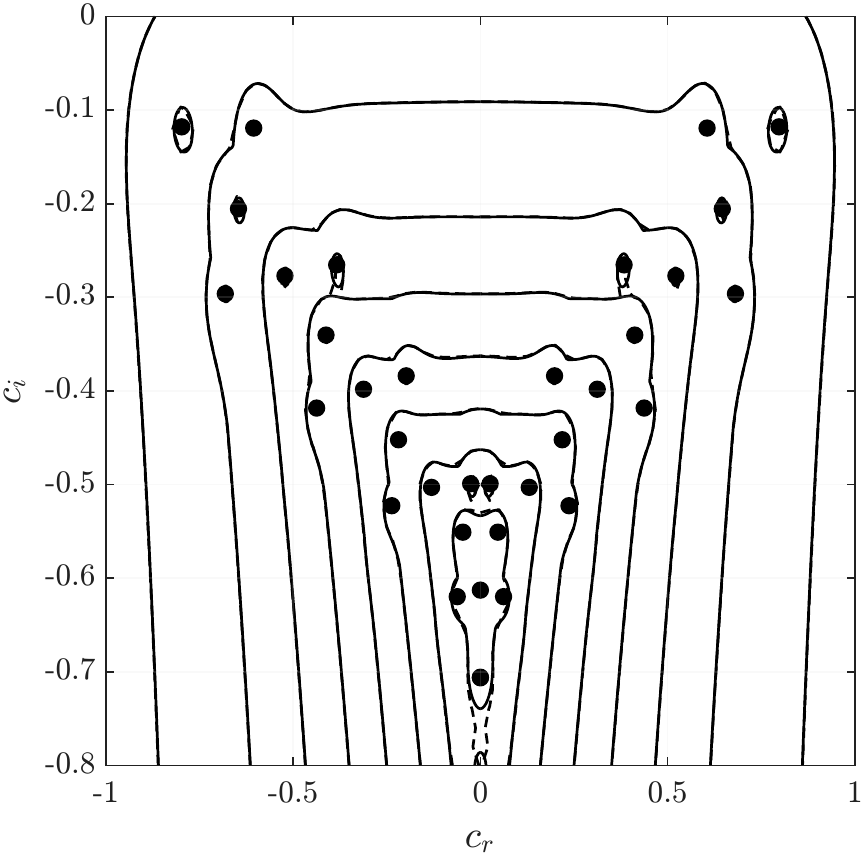} &
\includegraphics[width=0.47\textwidth]{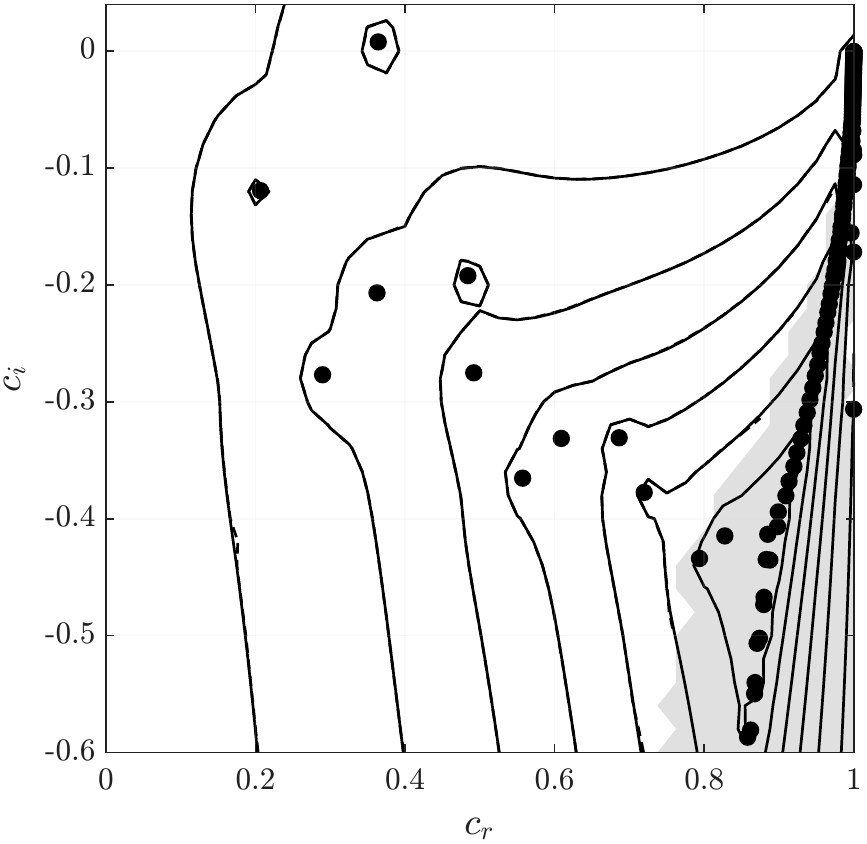}\\
(\textit{a}) & (\textit{b})
\end{tabular}
\caption{Pseudospectra in the energy norm, contours $\log_{10}\varepsilon=-1,\dots,-8$: (\textit{a}) plane Couette flow at $(\Rey,\alpha,\beta)=(1000,1,0)$; (\textit{b}) Blasius boundary layer at $(580,0.179,0)$. Solid, Orr--Sommerfeld--Squire; dashed, kLST; dots, eigenvalues.}
\label{fig:pseudo}
\end{figure}

Figure~\ref{fig:pseudo} shows the pseudospectra in the energy norm, contours $\log_{10}\varepsilon=-1,\dots,-8$, Orr--Sommerfeld--Squire (solid) against kLST with the BGK operator at $\Ma=0.1$, $Q=16$ (dashed), with the reference eigenvalues as dots. In figure~\ref{fig:pseudo}(\textit{a}) plane Couette flow is shown at $(\Rey,\alpha,\beta)=(1000,1,0)$. The reference uses 64 polynomials and $401\times321$ points, while the kLST computation employs $N=48$ and $41\times41$ points in $c_r\ge0$, extended by the symmetry $c_r\to-c_r$ of the Couette problem. In figure~\ref{fig:pseudo}(\textit{b}) the Blasius boundary layer is shown at $(\Rey,\alpha,\beta)=(580,0.179,0)$, on the domain $0\le y\le60$ of the modal results and on $41\times33$ points in $0\le c_r\le1$, $-0.6\le c_i\le0.04$. kLST uses $N=64$, and $N=96$ near the continuous spectrum, in $c_r\ge0.5$, $c_i\le-0.2$ and in $c_r\ge0.875$. The reference, computed directly from the resolvent with $N=160$ on the same mapped domain, is converged in $N$ to $10^{-3}$ in $\log_{10}\varepsilon$. Outside the shaded region the two agree to better than $0.02$ in $\log_{10}\varepsilon$, with a median difference of $1.4\times10^{-3}$. In the shaded region, which borders the continuous spectrum at $c_r=1$, they differ by more than $0.1$ in $\log_{10}\varepsilon$, because there the resolvent is dominated by the discretized continuous spectrum, which depends on how the free stream is truncated at $y=60$, through the incoming molecules in kLST and through $v=v'=\eta=0$ in the continuum problem.

The kinetic values of $G(t)$ lie on the Orr--Sommerfeld--Squire curves throughout, including the maxima of table~\ref{tab:tg}, which kLST reproduces to $3.3\times10^{-4}$ (Couette) and $8.7\times10^{-4}$ (Poiseuille). The relative differences, $2\times10^{-4}$ to $2\times10^{-3}$ at the twenty times computed in the channel flows, are of the order $\Ma^2$ by which the kinetic problem at $\Ma=0.1$ differs from the incompressible one, and grow slowly with $t$, as an $\mathcal{O}(\Ma^2)$ difference of the decay rates makes them. For the Blasius boundary layer the Orr--Sommerfeld--Squire maximum is $G_{\max}=1515.5$ at $tU_\infty/\delta^*=782$, against $1514$ at $778$ given by \citet{Butler1992}, and the kinetic value at that time is $1515.4$. At the twelve times computed the relative differences are $3\times10^{-6}$ to $6\times10^{-4}$. For the mixing layer the maximum is $G_{\max}=1350.4$ at $tU/\ell=276$ and the kinetic value $1351.7$, a relative difference of $1.0\times10^{-3}$. At the ten times computed the differences are $1.4\times10^{-4}$ to $4.1\times10^{-3}$, growing with $t$ as in the channel flows. The Orr--Sommerfeld--Squire values of the reference solver reproduce the optimal growth of \citet{Butler1992} in table~\ref{tab:tglit} to the digits given, and the resolvent-norm fits of \citet{TrefethenScience} to $0.1\%$ (Couette) and $0.6\%$ (Poiseuille). The kinetic values agree with the reference solver to $1\times10^{-3}$ or better in every case, from the oblique and two-dimensional optima at $tU/h<10$ to the streamwise-invariant optimum of Poiseuille flow at $\Rey=5000$, $tU/h=379$, where $G\approx4.9\times10^{3}$, and for the maximal resolvent norms.

In every case the optimal initial disturbance is a streamwise vortex, carried by the transverse velocity components $v$ and $w$, and the optimal response is a streak in the streamwise velocity component $u$. This is the lift-up mechanism, which the kinetic operator contains through the shear term $U'c_y\partial_{c_x}$. The normalized kinetic profiles agree with the reference to $4\times10^{-4}$ to $1.5\times10^{-3}$ (Couette), $9\times10^{-4}$ to $2.7\times10^{-3}$ (Poiseuille) and $9\times10^{-4}$ to $1.2\times10^{-3}$ (mixing layer), and for the Blasius boundary layer to $1.3\times10^{-3}$ in $|v|$ and $|u|$ and $8\times10^{-3}$ in $|w|$, at its zero. The optimal initial distribution carries $99.9\%$ or more of its energy in $v$ and $w$, the response $99.9\%$ or more in $u$. In the channel flows the thermodynamic perturbations carry $4\times10^{-6}$ of the initial and $2\times10^{-4}$ of the final energy, and the non-equilibrium part $(\mathsf I-\Pi_h)\hat f$ of \eqref{eq:Pih} a fraction $1.6\times10^{-7}$ to $2.4\times10^{-7}$ of either. This is the square of the Knudsen number, $\Kn=8\times10^{-5}$, times the square of the inverse length scale of the disturbance in units of the gap, as the first-order Chapman--Enskog non-equilibrium part implies. The fraction is $3\times10^{-8}$ and $2\times10^{-8}$ for the Blasius boundary layer, and $6\times10^{-7}$ and $7\times10^{-7}$ for the mixing layer, whose Knudsen number is larger by an order of magnitude. The kinetic optimal disturbances are thus local Maxwellians of the continuum optimal disturbances, to the accuracy that the Chapman--Enskog expansion predicts.

In plane Couette flow the kinetic and Orr--Sommerfeld--Squire contours coincide. At the 1681 points of the kinetic grid $\log_{10}\varepsilon$ agrees to $1.6\times10^{-3}$ (median) and to $1.1\times10^{-2}$ at 95\% of the points. The largest differences, up to $0.7$ in $\log_{10}\varepsilon$, occur on the stem where $\varepsilon<10^{-8}$, so close to an eigenvalue that the $\mathcal{O}(\Ma^2)$ displacement of the kinetic eigenvalues is visible. The agreement extends to the stem of the Y-shaped spectrum, where the resolvent norm exceeds $10^{8}$ and where two of the eigenvalues could not be determined (\S\ref{sec:res_modal}), because the eigenvalues there are ill-conditioned while the resolvent norm is not. In the Blasius boundary layer the contours coincide around the Tollmien--Schlichting wave and the damped discrete modes, down to $\varepsilon=10^{-8}$, and differ only next to the continuous spectrum, where the resolvent depends on the truncation of the free stream in both problems. The cross-stream resolution required is that of the Orr--Sommerfeld--Squire problem itself, $N\ge48$ for Couette flow and $N=96$ near the continuous spectrum of the Blasius boundary layer.

Taken together, these results constitute the first modal and non-modal stability analysis of shear flows with the linearized Boltzmann equation and its full collision operator. For all four flows and all four operators kLST recovers the Orr--Sommerfeld eigenvalues in the incompressible limit to a few parts in $10^7$ for the relaxation models and in $10^6$ for the full operator, and, with the BGK operator, the transient energy growth and the resolvent norms of Orr--Sommerfeld--Squire theory to $0.1\%$ and its pseudospectra to a few tenths of a percent in the median, so that kinetic linear stability theory attains the accuracy of continuum linear stability theory.

\FloatBarrier

\section{Discussion}\label{sec:conclusions}
The first shear-flow stability analysis using the full Boltzmann equation has been presented. The full Boltzmann operator for hard spheres, made affordable by the fast spectral method of \citet{GHHH2017}, has been placed in the kLST eigenvalue and initial-value problems of plane Couette and Poiseuille flow, the mixing layer and the Blasius boundary layer, at Reynolds numbers of practical interest, and solved in one discretization with the BGK, ES-BGK and Shakhov models, so that the collision operators of kinetic theory have been compared with one another, and with the continuum, on the same stability problem. The disturbance is three-dimensional and fully kinetic, and the spanwise and thermal families appear beside the in-plane modes. The base profile may be given as nodal values, and no-slip and the decay of the disturbance in the free stream follow from conditions on the incoming molecules alone. The discretization inherits the energy inequality of the continuous operator, and the eigenvalues are computed by shift-invert Arnoldi iteration with an exact relaxation-model preconditioner, without the stability matrix being formed. A manufactured solution verifies every term of the discrete operator, and the solver, independently of the comparison with continuum theory, and shows both to be spectrally accurate. A non-modal kLST has also been formulated, comprising an energy norm whose hydrodynamic part is the disturbance energy of \citet{Chu1965}, a kinetic Reynolds--Orr equation in which the mean shear is the only source of disturbance energy, the adjoint operator, and transient growth and pseudospectra computed from the same preconditioned shifted solves as the eigenvalues.

In the limit of small Mach number kLST with the relaxation models reproduces the Orr--Sommerfeld eigenvalues of all four flows to a few parts in $10^{7}$, and the full operator to its velocity-grid error, a few parts in $10^{6}$, which falls to the level of the models on a velocity grid finer by about four nodes per direction. Earlier kinetic analyses of the Blasius boundary layer reproduced its discrete modes to about $1\%$ \cite{Ma2024}, whereas kLST recovers them to $3\times10^{-6}$. The mixing layer approaches its inviscid Rayleigh limit at high Reynolds number, although nothing in the kinetic equation is inviscid. The transient growth of the classical optimal disturbances of \citet{Butler1992}, streamwise vortices that grow into streaks by lift-up, is reproduced by kLST to $0.1\%$, and the resolvent norms of \citet{TrefethenScience} agree with the Orr--Sommerfeld--Squire computation to $0.1\%$. The kLST optimal disturbances are local Maxwellians whose non-equilibrium part carries only a few parts in $10^{7}$ of their energy, as the Chapman--Enskog expansion predicts. At finite Mach number the kLST eigenvalues depart from the continuum ones quadratically through compressibility and, wherever there is a wall, linearly through rarefaction. The collision operators differ only in these departures. The three relaxation models share the linear coefficient, and the full Boltzmann operator for hard spheres departs from it by about $2\%$, while in the quadratic coefficient and in the thermal modes the operators group by their Prandtl number. Near the junction of the branches of the channel spectra the individual eigenvalues are too ill-conditioned to be computed reliably with any operator, but the kLST and Orr--Sommerfeld--Squire pseudospectra agree there to a median of $0.4\%$ in $\varepsilon$, which resolves the non-convergence reported by \citet{Zou2023}.

Questions that the present work leaves open begin with the full operator. The relaxation-model preconditioner loses its effectiveness deep in the spectrum, for strongly damped modes and at large Reynolds numbers, and a preconditioner that keeps the full collision operator within reach there has yet to be found. The eigenvalues of the most non-normal parts of the spectrum are not determined by double-precision linear solves. The difference between the slip coefficients of the full operator and of the models, and a small residual difference between the thermal modes of the full operator and the spectrum of the energy equation at its Prandtl number, are measured but not explained. Near the continuous spectrum of the open flows the pseudospectra depend on the truncation of the free stream, in the kinetic as in the continuum problem. And the results are confined to near-continuum, nearly incompressible base flows of uniform density and temperature, although the perturbation itself is fully compressible and kinetic.

\section*{CRediT authorship contribution statement}
\textbf{V. Theofilis:} Conceptualization, Methodology, Software, Validation, Formal analysis, Investigation, Writing -- original draft, Writing -- review \& editing, Funding acquisition.

\section*{Declaration of competing interest}
The author declares that he has no known competing financial interests or personal relationships that could have appeared to influence the work reported in this paper.

\section*{Acknowledgments}
Work supported by Office of Naval Research Grant No.~N00014-23-1-2839 ``Kinetic Treatment of Sources and Mechanisms that Drive Unsteady Shock-dominated Flow Instability'' with Dr.~Eric Marineau as Program Officer.

\appendix
\section{Linearized collision operators}\label{app:linops}
\setcounter{equation}{0}\renewcommand{\theequation}{A.\arabic{equation}}
Here details on the operators $\mathcal{L}_M$ and $\mathcal{L}_{\bar g}$ of \eqref{eq:Ldef} are presented for the full collision operator and its three relaxation models.

\paragraph{Full operator} $\mathcal{Q}$ is bilinear, so the expansion terminates and the split is exact for any $\bar g$,
\begin{equation}
 \begin{gathered}
 \mathcal{J}(\bar f+\hat f)-\mathcal{J}(\bar f)=\kappa\bigl[\mathcal{Q}(\bar f,\hat f)+\mathcal{Q}(\hat f,\bar f)\bigr]+\kappa\,\mathcal{Q}(\hat f,\hat f),\\
 \mathcal{L}_M\hat f=\kappa\bigl[\mathcal{Q}(M,\hat f)+\mathcal{Q}(\hat f,M)\bigr],\qquad
 \mathcal{L}_{\bar g}\hat f=\kappa\bigl[\mathcal{Q}(\bar g,\hat f)+\mathcal{Q}(\hat f,\bar g)\bigr].
 \end{gathered}
\end{equation}

\paragraph{Relaxation models} With the target $\mathcal{G}[f]$ ($\mathcal{M}$, $\mathcal{E}$ or $\mathcal{S}$ of Table~\ref{tab:collops}) and $\mathcal{J}(f)=\nu[f]\,(\mathcal{G}[f]-f)$, the derivative has two parts,
\begin{equation}
 \mathcal{J}'(\bar f)\,\hat f=\delta\nu\,\bigl(\mathcal{G}[\bar f]-\bar f\bigr)+\nu[\bar f]\,\bigl(\mathcal{G}'[\bar f]\,\hat f-\hat f\bigr),
 \qquad \delta\nu=\nu_\mu\bigl[\hat\rho+(1-\eta)\hat T\bigr],
\end{equation}
$\delta\nu$ being the variation of $\nu=\nu_\mu nT^{1-\eta}$ (times $\mathit{Pr}$ for ES-BGK). At the Maxwellian, $\mathcal{G}[M]=M$, and the variation of the target is a projection in the inner product $(\phi,\hat f)_c=\int\phi^*\hat f/M\,\mathrm{d}\boldsymbol c$, onto the perturbed Maxwellian,
\begin{equation}
 \Pi\hat f=M\bigl[\hat\rho+\boldsymbol c\cdot\hat{\boldsymbol u}+\hat T\bigl(\tfrac12|\boldsymbol c|^2-\tfrac32\bigr)\bigr],
\end{equation}
and, for ES-BGK and Shakhov, also onto the stress and the heat-flux functions,
\begin{equation}
 \Pi_\sigma\hat f=\tfrac12\,M\,\hat\sigma_{ij}c_ic_j,\qquad
 \Pi_\chi\hat f=\tfrac25\,M\,\hat{\boldsymbol q}\cdot\boldsymbol c\,\bigl(\tfrac12|\boldsymbol c|^2-\tfrac52\bigr),
\end{equation}
where $\hat\rho$, $\hat{\boldsymbol u}$, $\hat T$ and $\hat{\boldsymbol q}$ are the moments \eqref{eq:moments} of $\hat f$ and $\hat\sigma_{ij}=\hat P_{ij}-\tfrac13\delta_{ij}\hat P_{kk}$ is its deviatoric stress. Hence
\begin{equation}
 \begin{aligned}
 &\text{BGK:} &\mathcal{L}_M\hat f&=\nu_\mu\bigl(\Pi\hat f-\hat f\bigr),\\
 &\text{ES-BGK:} &\mathcal{L}_M\hat f&=\mathit{Pr}\,\nu_\mu\bigl(\Pi\hat f+b\,\Pi_\sigma\hat f-\hat f\bigr),\qquad b=1-1/\mathit{Pr},\\
 &\text{Shakhov:} &\mathcal{L}_M\hat f&=\nu_\mu\bigl(\Pi\hat f+(1-\mathit{Pr})\,\Pi_\chi\hat f-\hat f\bigr).
 \end{aligned}
\end{equation}
In the base state, $\bar g$ carries no mass, momentum or energy, so $\nu[\bar f]=\nu[M]$, and with the base-state $\bar g=-c_xc_yM/\nu_\mu$ of the relaxation models the first part gives $U'[\hat\rho+(1-\eta)\hat T]\,c_xc_yM$ for all three models, because for BGK and Shakhov $\mathcal{G}[\bar f]-\bar f=-U'\bar g$, and for ES-BGK the factor $\mathit{Pr}$ in $\nu$ is cancelled by the anisotropy of $\mathcal{E}[\bar f]$ given below. For BGK the Maxwellian target depends only on $n$, $\boldsymbol u$, $T$, which $\bar g$ leaves unchanged, and this is the whole of $\mathcal{L}_{\bar g}$. For ES-BGK the Gaussian of the stressed base state is anisotropic, $\mathcal{E}[\bar f]=M\,[1+b\,\sigma_{b,xy}c_xc_y]$ with $\sigma_{b,xy}=-\mu_0U'$, and its variation with $\hat{\boldsymbol u}$ and $\hat T$ adds a second group. For Shakhov the heat flux is defined about the perturbed mean velocity, which in the stressed base state contributes $\delta\boldsymbol q=-\boldsymbol\sigma_b\cdot\hat{\boldsymbol u}$. The three operators are therefore
\begin{equation}
 \begin{aligned}
 &\text{BGK:} &\mathcal{L}_{\bar g}\hat f&=\bigl[\hat\rho+(1-\eta)\hat T\bigr]c_xc_yM,\\
 &\text{ES-BGK:} &\mathcal{L}_{\bar g}\hat f&=\bigl[\hat\rho+(1-\eta)\hat T\bigr]c_xc_yM\\
 & & &\quad+(1-\mathit{Pr})\,M\Bigl\{c_xc_y\,(\boldsymbol c\cdot\hat{\boldsymbol u})-\hat u_xc_y-\hat u_yc_x+\hat Tc_xc_y\bigl(\tfrac12|\boldsymbol c|^2-\tfrac72\bigr)\Bigr\},\\
 &\text{Shakhov:} &\mathcal{L}_{\bar g}\hat f&=\bigl[\hat\rho+(1-\eta)\hat T\bigr]c_xc_yM
 +\frac{1-\mathit{Pr}}{5}\,M\,\bigl(\hat u_yc_x+\hat u_xc_y\bigr)\bigl(|\boldsymbol c|^2-5\bigr).
 \end{aligned}
\end{equation}

\section{The relaxation-model preconditioner}\label{app:precond}
\setcounter{equation}{0}\renewcommand{\theequation}{B.\arabic{equation}}
The preconditioner of \S\ref{sec:shsolve} is the exact inverse of a preconditioning operator $\mathsf A_{\mathrm p}$, which is $\mathsf A$ with two changes. First, the relaxation model replaces the collision operator, the model itself for BGK, ES-BGK and Shakhov and the Shakhov model at the same viscosity for the full operator. Second, the shear, body-force and $\mathcal{L}_{\bar g}$ terms act on the hydrodynamic projection $\Pi_h\hat f$ only. Then
\begin{equation}
 \mathsf A_{\mathrm p}-z_0\mathsf I=\mathsf B+\mathsf U\,\mathsf V^{T},
 \label{eq:splitting}
\end{equation}
where $\mathsf B$ does not couple different velocity nodes and $\mathsf U\mathsf V^{T}$ is of low rank. For velocity node $j$, $\mathsf B$ is the $(N+1)\times(N+1)$ block
\begin{equation}
 \mathsf B^{(j)}=-c_{y,j}\,\mathsf D-\mathrm{diag}\bigl[\nu_{\mathrm{rel}}+\ii\alpha\,(c_{x,j}+U)+\ii\beta\,c_{z,j}+z_0\bigr],
 \label{eq:Bblock}
\end{equation}
which contains streaming, the Doppler-shifted phase, the loss term of the relaxation model, at the rate $\nu_{\mathrm{rel}}$ ($\nu_\mu$, or $\mathit{Pr}\,\nu_\mu$ for ES-BGK), and the shift. The penalty $-|c_{y,j}|/w$ of \eqref{eq:sat}, $w$ being the quadrature weight of the boundary point, is added on the diagonal at the boundary point at which molecule $j$ enters the domain. For $\beta=0$ the block depends on the node only through $(c_x,c_y)$, so that $Q^2$ blocks are inverted instead of $n_b$.

The rows of $\mathsf V^{T}$ are the $n_m$ moments $\boldsymbol b_m^{T}\hat f$ of the relaxation model at each cross-stream point, $\boldsymbol b_m$ being the weights of the collision invariants and, for ES-BGK and Shakhov, of the stress and the heat flux, and the two re-emitted densities $\rho_w$ of \eqref{eq:sat}, weighted sums of the incident values at the two boundary points. The corresponding columns of $\mathsf U$ at the point $y_i$ are
\begin{equation}
 \nu_{\mathrm{rel}}\,\boldsymbol p_m+U'(y_i)\,\boldsymbol q^{(1)}_m+a(y_i)\,\boldsymbol q^{(2)}_m,\qquad
 \boldsymbol q^{(1)}_m=c_y\,\mathsf D_{c_x}\boldsymbol p_m+\mathcal{L}_{\bar g}\boldsymbol p_m,\qquad
 \boldsymbol q^{(2)}_m=-\mathsf D_{c_x}\boldsymbol p_m ,
 \label{eq:Ucols}
\end{equation}
where $\boldsymbol p_m$ are the functions of $\boldsymbol c$ onto which the model projects, so that the gain of the model is $\nu_{\mathrm{rel}}\sum_m\boldsymbol p_m\boldsymbol b_m^{T}\hat f$. For the two re-emitted densities they are the re-emission columns of the penalty, $|c_y|\,M/w$ on the entries of the incoming molecules at the boundary point. The rank is $r=n_m(N+1)+2$, for example $r=487$ for BGK and $r=778$ for Shakhov at $N=96$. The Woodbury identity \citep{Woodbury1950,Hager1989} then gives
\begin{equation}
 (\mathsf A_{\mathrm p}-z_0\mathsf I)^{-1}=\mathsf B^{-1}-\mathsf B^{-1}\mathsf U\,\mathsf S^{-1}\,\mathsf V^{T}\mathsf B^{-1},\qquad
 \mathsf S=\mathsf I_r+\mathsf V^{T}\mathsf B^{-1}\mathsf U ,
 \label{eq:woodbury}
\end{equation}
with the dense capacitance matrix $\mathsf S$ of order $r$, assembled from the block inverses by one matrix product. For each shift the setup inverts the $Q^2$ blocks and assembles and factors $\mathsf S$. One application of \eqref{eq:woodbury} costs two sweeps of block products and one back-substitution with $\mathsf S$. The adjoint preconditioner, $\mathsf B^{-H}-\mathsf B^{-H}\mathsf V\,\mathsf S^{-H}\mathsf U^{H}\mathsf B^{-H}$, is applied with the same factors.

For a relaxation model at rest $\mathsf A_{\mathrm p}=\mathsf A$, so that the preconditioner is the exact inverse, to a self-check residual of $10^{-14}$, and shift-invert Arnoldi iteration is a direct method at any Knudsen number. With shear, $\mathsf A_{\mathrm p}$ omits only the shear term acting on the non-equilibrium part of $\hat f$, of relative size $\mathit{Kn}$, and GMRES converges in three iterations. For the full operator GMRES resolves the difference between the full and the Shakhov operators on the non-hydrodynamic subspace.

\section{Cross-check against rarefied test cases}\label{app:wu}
\setcounter{equation}{0}\renewcommand{\theequation}{C.\arabic{equation}}
\citet{Wu2025} characterizes the rarefaction by the parameter $\delta=\sqrt\pi/(2\mathit{Kn})$, with $\mathit{Kn}=(\mu_0/p_0H)\sqrt{\pi RT_0/2}$, $H$ being the distance between the walls, so that $\nu_\mu=\sqrt2\,\delta$ in the present units. Velocities there are scaled by the most probable speed $\sqrt{2RT_0}$, the perturbation is written as $\exp(\ii Kx-\ii\varpi t)$, and the eigenvalue reported is $s=-\ii\varpi$, so that $s=-\ii\omega/\sqrt2$ in terms of the eigenvalue $\omega$ of \eqref{eq:harmonic}.

For the plane Couette flow of \citet[\S5.1]{Wu2025} at $\delta=1$ and $K=\alpha=1$, with the BGK model, the present eigenvalue is $s=-0.6466$, against $-0.6564+0.0002\,\ii$ reported there. The difference, $1.5\%$, is converged in $Q$ and $N$ on the present side and is larger than the accuracy of $10^{-4}$ estimated by \citet{Wu2025}. In \citet{Wu2025} the walls move at $0.01$ times the most probable speed, whereas they are at rest here, and whether this accounts for the difference has not been checked. For the plane sound wave at $K=\delta=1$ the hard-sphere traveling wave is $s=-0.457\pm0.965\,\ii$ at $Q=20$, against $-0.4662\pm0.9718\,\ii$ \citep[Table~2]{Wu2025}, a difference of $2\%$ in the damping rate and below $1\%$ in the frequency. The entropy wave, the non-propagating mode with a real eigenvalue, is not reproduced ($-0.423$ against $-0.5543$). These modes lie in the kinetic regime, with a mean free path comparable to the wavelength, for which the velocity grid of the present computations, $\Delta v=0.66$, chosen for near-continuum flows, is coarse.

\section*{Data availability}
Data will be made available on request.

\bibliographystyle{elsarticle-num-names}
\bibliography{refs}

\begin{thebibliography}{77}
\expandafter\ifx\csname natexlab\endcsname\relax\def\natexlab#1{#1}\fi
\providecommand{\url}[1]{\texttt{#1}}
\providecommand{\href}[2]{#2}
\providecommand{\path}[1]{#1}
\providecommand{\DOIprefix}{doi:}
\providecommand{\ArXivprefix}{arXiv:}
\providecommand{\URLprefix}{URL: }
\providecommand{\Pubmedprefix}{pmid:}
\providecommand{\doi}[1]{\href{http://dx.doi.org/#1}{\path{#1}}}
\providecommand{\Pubmed}[1]{\href{pmid:#1}{\path{#1}}}
\providecommand{\bibinfo}[2]{#2}
\ifx\xfnm\relax \def\xfnm[#1]{\unskip,\space#1}\fi
\bibitem[{Reed et~al.(1996)Reed, Saric, and Arnal}]{Reed1996}
\bibinfo{author}{H.~L. Reed}, \bibinfo{author}{W.~S. Saric},
  \bibinfo{author}{D.~Arnal},
\newblock \bibinfo{title}{Linear stability theory applied to boundary layers},
\newblock \bibinfo{journal}{Annu. Rev. Fluid Mech.} \bibinfo{volume}{28}
  (\bibinfo{year}{1996}) \bibinfo{pages}{389--428}.
\bibitem[{Smith and Gamberoni(1956)}]{SmithGamberoni1956}
\bibinfo{author}{A.~M.~O. Smith}, \bibinfo{author}{N.~Gamberoni},
  \bibinfo{title}{Transition, pressure gradient and stability theory},
  \bibinfo{type}{Technical Report} \bibinfo{number}{ES 26388}, Douglas Aircraft
  Company, \bibinfo{year}{1956}.
\bibitem[{van Ingen(1956)}]{vanIngen1956}
\bibinfo{author}{J.~L. van Ingen}, \bibinfo{title}{A suggested semi-empirical
  method for the calculation of the boundary layer transition region},
  \bibinfo{type}{Technical Report} \bibinfo{number}{VTH-74}, Department of
  Aeronautical Engineering, Technische Hogeschool Delft, \bibinfo{year}{1956}.
\bibitem[{Schlichting(1933)}]{Schlichting1933}
\bibinfo{author}{H.~Schlichting},
\newblock \bibinfo{title}{Zur {E}ntstehung der {T}urbulenz bei der
  {P}lattenstr{\"o}mung},
\newblock \bibinfo{journal}{Nachr. Ges. Wiss. G{\"o}ttingen, Math.-Phys. Kl.}
  (\bibinfo{year}{1933}) \bibinfo{pages}{181--208}.
\bibitem[{Orszag(1971)}]{Orszag1971}
\bibinfo{author}{S.~A. Orszag},
\newblock \bibinfo{title}{Accurate solution of the {Orr--Sommerfeld} stability
  equation},
\newblock \bibinfo{journal}{J. Fluid Mech.} \bibinfo{volume}{50}
  (\bibinfo{year}{1971}) \bibinfo{pages}{689--703}.
\bibitem[{Cercignani and Stefanov(1992)}]{CercignaniStefanov1992}
\bibinfo{author}{C.~Cercignani}, \bibinfo{author}{S.~Stefanov},
\newblock \bibinfo{title}{{B\'enard}'s instability in kinetic theory},
\newblock \bibinfo{journal}{Transp. Theory Stat. Phys.} \bibinfo{volume}{21}
  (\bibinfo{year}{1992}) \bibinfo{pages}{371--381}.
\bibitem[{Stefanov and Cercignani(1993)}]{StefanovCercignani1993}
\bibinfo{author}{S.~Stefanov}, \bibinfo{author}{C.~Cercignani},
\newblock \bibinfo{title}{{Monte Carlo} simulation of the {Taylor}--{Couette}
  flow of a rarefied gas},
\newblock \bibinfo{journal}{J. Fluid Mech.} \bibinfo{volume}{256}
  (\bibinfo{year}{1993}) \bibinfo{pages}{199--213}.
\bibitem[{Stefanov et~al.(2002)Stefanov, Roussinov, and
  Cercignani}]{StefanovRoussinovCercignani2002}
\bibinfo{author}{S.~Stefanov}, \bibinfo{author}{V.~Roussinov},
  \bibinfo{author}{C.~Cercignani},
\newblock \bibinfo{title}{{Rayleigh--B\'enard} flow of a rarefied gas and its
  attractors. i. convection regime},
\newblock \bibinfo{journal}{Phys. Fluids} \bibinfo{volume}{14}
  (\bibinfo{year}{2002}) \bibinfo{pages}{2255}.
\bibitem[{Sone et~al.(1997)Sone, Aoki, and Sugimoto}]{SoneAokiSugimoto1997}
\bibinfo{author}{Y.~Sone}, \bibinfo{author}{K.~Aoki},
  \bibinfo{author}{H.~Sugimoto},
\newblock \bibinfo{title}{The {B\'enard} problem for a rarefied gas: formation
  of steady flow patterns and stability of array of rolls},
\newblock \bibinfo{journal}{Phys. Fluids} \bibinfo{volume}{9}
  (\bibinfo{year}{1997}) \bibinfo{pages}{3898}.
\bibitem[{Manela and Frankel(2005)}]{ManelaFrankel2005}
\bibinfo{author}{A.~Manela}, \bibinfo{author}{I.~Frankel},
\newblock \bibinfo{title}{On the {Rayleigh--B\'enard} problem in the continuum
  limit},
\newblock \bibinfo{journal}{Phys. Fluids} \bibinfo{volume}{17}
  (\bibinfo{year}{2005}) \bibinfo{pages}{036101}.
\bibitem[{Manela and Frankel(2007)}]{ManelaFrankel2007}
\bibinfo{author}{A.~Manela}, \bibinfo{author}{I.~Frankel},
\newblock \bibinfo{title}{On the compressible {Taylor}--{Couette} problem},
\newblock \bibinfo{journal}{J. Fluid Mech.} \bibinfo{volume}{588}
  (\bibinfo{year}{2007}) \bibinfo{pages}{59--74}.
\bibitem[{Ben-Ami and Manela(2019)}]{BenAmiManela2019}
\bibinfo{author}{Y.~Ben-Ami}, \bibinfo{author}{A.~Manela},
\newblock \bibinfo{title}{Effect of heat-flux boundary conditions on the
  {Rayleigh--B\'enard} instability in a rarefied gas},
\newblock \bibinfo{journal}{Phys. Rev. Fluids} \bibinfo{volume}{4}
  (\bibinfo{year}{2019}) \bibinfo{pages}{033402}.
\bibitem[{Garc\'ia and Penland(1991)}]{GarciaPenland1991}
\bibinfo{author}{A.~L. Garc\'ia}, \bibinfo{author}{C.~Penland},
\newblock \bibinfo{title}{Fluctuating hydrodynamics and principal oscillation
  pattern analysis},
\newblock \bibinfo{journal}{J. Stat. Phys.} \bibinfo{volume}{64}
  (\bibinfo{year}{1991}) \bibinfo{pages}{1121--1132}.
\bibitem[{Gallis et~al.(2016)Gallis, Koehler, Torczynski, and
  Plimpton}]{Gallis2016}
\bibinfo{author}{M.~A. Gallis}, \bibinfo{author}{T.~P. Koehler},
  \bibinfo{author}{J.~R. Torczynski}, \bibinfo{author}{S.~J. Plimpton},
\newblock \bibinfo{title}{Direct simulation {Monte Carlo} investigation of the
  {Rayleigh--Taylor} instability},
\newblock \bibinfo{journal}{Phys. Rev. Fluids} \bibinfo{volume}{1}
  (\bibinfo{year}{2016}) \bibinfo{pages}{043403}.
\bibitem[{Gallis et~al.(2018)Gallis, Torczynski, Bitter, Koehler, Plimpton, and
  Papadakis}]{Gallis2018}
\bibinfo{author}{M.~A. Gallis}, \bibinfo{author}{J.~R. Torczynski},
  \bibinfo{author}{N.~P. Bitter}, \bibinfo{author}{T.~P. Koehler},
  \bibinfo{author}{S.~J. Plimpton}, \bibinfo{author}{G.~Papadakis},
\newblock \bibinfo{title}{Gas-kinetic simulation of sustained turbulence in
  minimal {Couette} flow},
\newblock \bibinfo{journal}{Phys. Rev. Fluids} \bibinfo{volume}{3}
  (\bibinfo{year}{2018}) \bibinfo{pages}{071402(R)}.
\bibitem[{Mohan et~al.(2022)Mohan, Sameen, Srinivasan, and
  Girimaji}]{Mohan2022}
\bibinfo{author}{V.~Mohan}, \bibinfo{author}{A.~Sameen},
  \bibinfo{author}{B.~Srinivasan}, \bibinfo{author}{S.~S. Girimaji},
\newblock \bibinfo{title}{Continuum breakdown in compressible mixing layers},
\newblock \bibinfo{journal}{Phys. Rev. E} \bibinfo{volume}{105}
  (\bibinfo{year}{2022}) \bibinfo{pages}{065102}.
\bibitem[{Mohan et~al.(2023)Mohan, Sameen, Srinivasan, and
  Girimaji}]{Mohan2023}
\bibinfo{author}{V.~Mohan}, \bibinfo{author}{A.~Sameen},
  \bibinfo{author}{B.~Srinivasan}, \bibinfo{author}{S.~S. Girimaji},
\newblock \bibinfo{title}{Instability of mixing layers: momentum and thermal
  transport in the continuum breakdown regime},
\newblock \bibinfo{journal}{Phys. Rev. E} \bibinfo{volume}{108}
  (\bibinfo{year}{2023}) \bibinfo{pages}{L053101}.
\bibitem[{Kudryavtsev et~al.(2019)Kudryavtsev, Poleshkin, and
  Shershnev}]{Kudryavtsev2019}
\bibinfo{author}{A.~N. Kudryavtsev}, \bibinfo{author}{S.~O. Poleshkin},
  \bibinfo{author}{A.~A. Shershnev},
\newblock \bibinfo{title}{Numerical simulation of the instability development
  in a compressible mixing layer using kinetic and continuum approaches},
\newblock \bibinfo{journal}{AIP Conf. Proc.} \bibinfo{volume}{2125}
  (\bibinfo{year}{2019}) \bibinfo{pages}{030033}.
\bibitem[{Shershnev et~al.(2024)Shershnev, Kudryavtsev, Kashkovsky, Borisov,
  and Poleshkin}]{Shershnev2024}
\bibinfo{author}{A.~A. Shershnev}, \bibinfo{author}{A.~N. Kudryavtsev},
  \bibinfo{author}{A.~V. Kashkovsky}, \bibinfo{author}{S.~P. Borisov},
  \bibinfo{author}{S.~O. Poleshkin},
\newblock \bibinfo{title}{Numerical simulation of rarefied flow instabilities
  using kinetic approach},
\newblock \bibinfo{journal}{Supercomput. Front. Innov.} \bibinfo{volume}{11}
  (\bibinfo{year}{2024}) \bibinfo{pages}{65--77}.
\bibitem[{Tumuklu et~al.(2018{\natexlab{a}})Tumuklu, Levin, and
  Theofilis}]{Tumuklu2018a}
\bibinfo{author}{O.~Tumuklu}, \bibinfo{author}{D.~A. Levin},
  \bibinfo{author}{V.~Theofilis},
\newblock \bibinfo{title}{Investigation of unsteady, hypersonic, laminar
  separated flows over a double cone geometry using a kinetic approach},
\newblock \bibinfo{journal}{Phys. Fluids} \bibinfo{volume}{30}
  (\bibinfo{year}{2018}{\natexlab{a}}) \bibinfo{pages}{046103}.
\bibitem[{Tumuklu et~al.(2018{\natexlab{b}})Tumuklu, Theofilis, and
  Levin}]{Tumuklu2018b}
\bibinfo{author}{O.~Tumuklu}, \bibinfo{author}{V.~Theofilis},
  \bibinfo{author}{D.~A. Levin},
\newblock \bibinfo{title}{On the unsteadiness of shock--laminar boundary layer
  interactions of hypersonic flows over a double cone},
\newblock \bibinfo{journal}{Phys. Fluids} \bibinfo{volume}{30}
  (\bibinfo{year}{2018}{\natexlab{b}}) \bibinfo{pages}{106111}.
\bibitem[{Tumuklu et~al.(2019)Tumuklu, Levin, and Theofilis}]{Tumuklu2019}
\bibinfo{author}{O.~Tumuklu}, \bibinfo{author}{D.~A. Levin},
  \bibinfo{author}{V.~Theofilis},
\newblock \bibinfo{title}{Modal analysis with proper orthogonal decomposition
  of hypersonic separated flows over a double wedge},
\newblock \bibinfo{journal}{Phys. Rev. Fluids} \bibinfo{volume}{4}
  (\bibinfo{year}{2019}) \bibinfo{pages}{033403}.
\bibitem[{Sawant et~al.(2021)Sawant, Levin, and Theofilis}]{Sawant2021}
\bibinfo{author}{S.~S. Sawant}, \bibinfo{author}{D.~A. Levin},
  \bibinfo{author}{V.~Theofilis},
\newblock \bibinfo{title}{A kinetic approach to studying low-frequency
  molecular fluctuations in a one-dimensional shock},
\newblock \bibinfo{journal}{Phys. Fluids} \bibinfo{volume}{33}
  (\bibinfo{year}{2021}) \bibinfo{pages}{104106}.
\bibitem[{Sawant et~al.(2022)Sawant, Theofilis, and Levin}]{Sawant2022}
\bibinfo{author}{S.~S. Sawant}, \bibinfo{author}{V.~Theofilis},
  \bibinfo{author}{D.~A. Levin},
\newblock \bibinfo{title}{On the synchronisation of three-dimensional shock
  layer and laminar separation bubble instabilities in hypersonic flow over a
  double wedge},
\newblock \bibinfo{journal}{J. Fluid Mech.} \bibinfo{volume}{941}
  (\bibinfo{year}{2022}) \bibinfo{pages}{A7}.
\bibitem[{Theofilis(2011)}]{Theofilis2011}
\bibinfo{author}{V.~Theofilis},
\newblock \bibinfo{title}{Global linear instability},
\newblock \bibinfo{journal}{Annu. Rev. Fluid Mech.} \bibinfo{volume}{43}
  (\bibinfo{year}{2011}) \bibinfo{pages}{319--352}.
  \DOIprefix\doi{10.1146/annurev-fluid-122109-160705}.
\bibitem[{Klothakis et~al.(2022)Klothakis, {Quintanilha Jr.}, Sawant,
  Protopapadakis, Theofilis, and Levin}]{Klothakis2022}
\bibinfo{author}{A.~Klothakis}, \bibinfo{author}{H.~{Quintanilha Jr.}},
  \bibinfo{author}{S.~S. Sawant}, \bibinfo{author}{E.~Protopapadakis},
  \bibinfo{author}{V.~Theofilis}, \bibinfo{author}{D.~A. Levin},
\newblock \bibinfo{title}{Linear stability analysis of hypersonic boundary
  layers computed by a kinetic approach: a semi-infinite flat plate at $4.5\le
  m_\infty\le9$},
\newblock \bibinfo{journal}{Theor. Comput. Fluid Dyn.} \bibinfo{volume}{36}
  (\bibinfo{year}{2022}) \bibinfo{pages}{117--139}.
\bibitem[{Karpuzcu et~al.(2026)Karpuzcu, Pezlar, Levin, and
  Theofilis}]{Karpuzcu2026}
\bibinfo{author}{I.~T. Karpuzcu}, \bibinfo{author}{V.~Pezlar},
  \bibinfo{author}{D.~A. Levin}, \bibinfo{author}{V.~Theofilis},
\newblock \bibinfo{title}{On linear global instability at the leading edge of a
  short compression corner with large ramp angle},
\newblock \bibinfo{journal}{Proc. R. Soc. A} \bibinfo{volume}{482}
  (\bibinfo{year}{2026}) \bibinfo{pages}{20250702}.
  \DOIprefix\doi{10.1098/rspa.2025.0702}.
\bibitem[{Shan(1997)}]{Shan1997}
\bibinfo{author}{X.~Shan},
\newblock \bibinfo{title}{Simulation of {Rayleigh--B\'enard} convection using a
  lattice {Boltzmann} method},
\newblock \bibinfo{journal}{Phys. Rev. E} \bibinfo{volume}{55}
  (\bibinfo{year}{1997}) \bibinfo{pages}{2780--2788}.
\bibitem[{Mendoza et~al.(2014)Mendoza, Debus, Succi, and
  Herrmann}]{Mendoza2014}
\bibinfo{author}{M.~Mendoza}, \bibinfo{author}{J.-D. Debus},
  \bibinfo{author}{S.~Succi}, \bibinfo{author}{H.~J. Herrmann},
\newblock \bibinfo{title}{Lattice kinetic scheme for generalized coordinates
  and curved spaces},
\newblock \bibinfo{journal}{Int. J. Mod. Phys. C} \bibinfo{volume}{25}
  (\bibinfo{year}{2014}) \bibinfo{pages}{1441001}.
\bibitem[{Fedorov(2011)}]{Fedorov2011}
\bibinfo{author}{A.~Fedorov},
\newblock \bibinfo{title}{Transition and stability of high-speed boundary
  layers},
\newblock \bibinfo{journal}{Annu. Rev. Fluid Mech.} \bibinfo{volume}{43}
  (\bibinfo{year}{2011}) \bibinfo{pages}{79--95}.
\bibitem[{Trefethen et~al.(1993)Trefethen, Trefethen, Reddy, and
  Driscoll}]{TrefethenScience}
\bibinfo{author}{L.~N. Trefethen}, \bibinfo{author}{A.~E. Trefethen},
  \bibinfo{author}{S.~C. Reddy}, \bibinfo{author}{T.~A. Driscoll},
\newblock \bibinfo{title}{Hydrodynamic stability without eigenvalues},
\newblock \bibinfo{journal}{Science} \bibinfo{volume}{261}
  (\bibinfo{year}{1993}) \bibinfo{pages}{578--584}.
\bibitem[{Grad(1963{\natexlab{a}})}]{Grad1963}
\bibinfo{author}{H.~Grad},
\newblock \bibinfo{title}{Asymptotic theory of the {Boltzmann} equation, {II}},
\newblock in: \bibinfo{editor}{J.~A. Laurmann} (Ed.),
  \bibinfo{booktitle}{Rarefied Gas Dynamics}, volume~\bibinfo{volume}{1},
  \bibinfo{publisher}{Academic Press}, \bibinfo{address}{New York},
  \bibinfo{year}{1963}{\natexlab{a}}, pp. \bibinfo{pages}{26--59}.
\bibitem[{Grad(1963{\natexlab{b}})}]{Grad1963PoF}
\bibinfo{author}{H.~Grad},
\newblock \bibinfo{title}{Asymptotic theory of the {Boltzmann} equation},
\newblock \bibinfo{journal}{Phys. Fluids} \bibinfo{volume}{6}
  (\bibinfo{year}{1963}{\natexlab{b}}) \bibinfo{pages}{147--181}.
\bibitem[{Cercignani(1988)}]{Cercignani1988}
\bibinfo{author}{C.~Cercignani}, \bibinfo{title}{The {Boltzmann} Equation and
  Its Applications}, \bibinfo{publisher}{Springer}, \bibinfo{address}{New
  York}, \bibinfo{year}{1988}.
\bibitem[{Kogelbauer and Karlin(2024)}]{KogelbauerKarlin2024}
\bibinfo{author}{F.~Kogelbauer}, \bibinfo{author}{I.~Karlin},
\newblock \bibinfo{title}{Exact hydrodynamic manifolds for the linear
  {Boltzmann} {BGK} equation {I}: spectral theory},
\newblock \bibinfo{journal}{Continuum Mech. Thermodyn.} \bibinfo{volume}{36}
  (\bibinfo{year}{2024}) \bibinfo{pages}{1685--1709}.
  \DOIprefix\doi{10.1007/s00161-024-01323-2}.
\bibitem[{Kogelbauer and Karlin(2025)}]{KogelbauerKarlin2025}
\bibinfo{author}{F.~Kogelbauer}, \bibinfo{author}{I.~Karlin},
\newblock \bibinfo{title}{Exact hydrodynamic manifolds for the linear
  {Boltzmann} {BGK} equation {II}: spectral closure},
\newblock \bibinfo{journal}{Continuum Mech. Thermodyn.} \bibinfo{volume}{37}
  (\bibinfo{year}{2025}) \bibinfo{pages}{45}.
  \DOIprefix\doi{10.1007/s00161-025-01379-8}.
\bibitem[{Ellis and Pinsky(1975)}]{EllisPinsky1975}
\bibinfo{author}{R.~S. Ellis}, \bibinfo{author}{M.~A. Pinsky},
\newblock \bibinfo{title}{The first and second fluid approximations to the
  linearized {Boltzmann} equation},
\newblock \bibinfo{journal}{J. Math. Pures Appl.} \bibinfo{volume}{54}
  (\bibinfo{year}{1975}) \bibinfo{pages}{125--156}.
\bibitem[{{Wang Chang} and Uhlenbeck(1952)}]{WangChangUhlenbeck1952}
\bibinfo{author}{C.~S. {Wang Chang}}, \bibinfo{author}{G.~E. Uhlenbeck},
  \bibinfo{title}{On the propagation of sound in monatomic gases},
  \bibinfo{type}{Report} \bibinfo{number}{ONR Contract N6onr-23222},
  Engineering Research Institute, University of Michigan, \bibinfo{year}{1952}.
\bibitem[{Sirovich and Thurber(1965)}]{SirovichThurber1965}
\bibinfo{author}{L.~Sirovich}, \bibinfo{author}{J.~K. Thurber},
\newblock \bibinfo{title}{Propagation of forced sound waves in rarefied
  gasdynamics},
\newblock \bibinfo{journal}{J. Acoust. Soc. Am.} \bibinfo{volume}{37}
  (\bibinfo{year}{1965}) \bibinfo{pages}{329--339}.
\bibitem[{Lees(1959)}]{Lees1959}
\bibinfo{author}{L.~Lees}, \bibinfo{title}{A kinetic theory description of
  rarefied gas flows}, \bibinfo{type}{Hypersonic Research Project Memorandum}
  \bibinfo{number}{51}, Guggenheim Aeronautical Laboratory, California
  Institute of Technology, \bibinfo{year}{1959}.
\bibitem[{Lee and Dufty(1997)}]{LeeDufty1997}
\bibinfo{author}{M.~Lee}, \bibinfo{author}{J.~W. Dufty},
\newblock \bibinfo{title}{Transport far from equilibrium: uniform shear flow},
\newblock \bibinfo{journal}{Phys. Rev. E} \bibinfo{volume}{56}
  (\bibinfo{year}{1997}) \bibinfo{pages}{1733--1745}.
\bibitem[{Garz\'o(2006)}]{Garzo2006}
\bibinfo{author}{V.~Garz\'o},
\newblock \bibinfo{title}{Transport coefficients for an inelastic gas around
  uniform shear flow: linear stability analysis},
\newblock \bibinfo{journal}{Phys. Rev. E} \bibinfo{volume}{73}
  (\bibinfo{year}{2006}) \bibinfo{pages}{021304}.
\bibitem[{Arkeryd et~al.(2010)Arkeryd, Esposito, Marra, and
  Nouri}]{ArkerydEspositoMarraNouri2010}
\bibinfo{author}{L.~Arkeryd}, \bibinfo{author}{R.~Esposito},
  \bibinfo{author}{R.~Marra}, \bibinfo{author}{A.~Nouri},
\newblock \bibinfo{title}{Stability for {Rayleigh--Benard} convective solutions
  of the {Boltzmann} equation},
\newblock \bibinfo{journal}{Arch. Ration. Mech. Anal.} \bibinfo{volume}{198}
  (\bibinfo{year}{2010}) \bibinfo{pages}{125--187}.
\bibitem[{Sone and Doi(2004)}]{SoneDoi2004}
\bibinfo{author}{Y.~Sone}, \bibinfo{author}{T.~Doi},
\newblock \bibinfo{title}{Ghost effect of infinitesimal curvature in the plane
  {Couette} flow of a gas in the continuum limit},
\newblock \bibinfo{journal}{Phys. Fluids} \bibinfo{volume}{16}
  (\bibinfo{year}{2004}) \bibinfo{pages}{952}.
\bibitem[{P\'erez et~al.(2017)P\'erez, Aguilar, and Theofilis}]{Perez2017}
\bibinfo{author}{J.~M. P\'erez}, \bibinfo{author}{A.~Aguilar},
  \bibinfo{author}{V.~Theofilis},
\newblock \bibinfo{title}{Lattice {Boltzmann} methods for global linear
  instability analysis},
\newblock \bibinfo{journal}{Theor. Comput. Fluid Dyn.} \bibinfo{volume}{31}
  (\bibinfo{year}{2017}) \bibinfo{pages}{643--664}.
\bibitem[{Jiang et~al.(2022)Jiang, Luo, Zhang, Wu, and Yi}]{Jiang2022}
\bibinfo{author}{H.-K. Jiang}, \bibinfo{author}{K.~Luo}, \bibinfo{author}{Z.-Y.
  Zhang}, \bibinfo{author}{J.~Wu}, \bibinfo{author}{H.-L. Yi},
\newblock \bibinfo{title}{Global linear instability analysis of thermal
  convective flow using the linearized lattice {Boltzmann} method},
\newblock \bibinfo{journal}{J. Fluid Mech.} \bibinfo{volume}{944}
  (\bibinfo{year}{2022}) \bibinfo{pages}{A31}.
\bibitem[{Luo et~al.(2023)Luo, Jiang, Wu, Zhang, and Yi}]{Luo2023}
\bibinfo{author}{K.~Luo}, \bibinfo{author}{H.-K. Jiang},
  \bibinfo{author}{J.~Wu}, \bibinfo{author}{M.~Zhang}, \bibinfo{author}{H.-L.
  Yi},
\newblock \bibinfo{title}{Stability and bifurcation of annular
  electro-thermo-convection},
\newblock \bibinfo{journal}{J. Fluid Mech.} \bibinfo{volume}{966}
  (\bibinfo{year}{2023}) \bibinfo{pages}{A13}.
\bibitem[{Zou et~al.(2022)Zou, Zhong, Bi, Yuan, and Tang}]{Zou2022}
\bibinfo{author}{S.~Zou}, \bibinfo{author}{C.~Zhong}, \bibinfo{author}{L.~Bi},
  \bibinfo{author}{X.~Yuan}, \bibinfo{author}{Z.~Tang},
\newblock \bibinfo{title}{A new linear stability analysis approach for
  microchannel flow based on the {Boltzmann} bhatnagar--gross--{Krook}
  equation},
\newblock \bibinfo{journal}{Phys. Fluids} \bibinfo{volume}{34}
  (\bibinfo{year}{2022}) \bibinfo{pages}{124114}.
\bibitem[{Zou et~al.(2023)Zou, Bi, Zhong, Yuan, and Tang}]{Zou2023}
\bibinfo{author}{S.~Zou}, \bibinfo{author}{L.~Bi}, \bibinfo{author}{C.~Zhong},
  \bibinfo{author}{X.~Yuan}, \bibinfo{author}{Z.~Tang},
\newblock \bibinfo{title}{A novel linear stability analysis method for plane
  {Couette} flow considering rarefaction effects},
\newblock \bibinfo{journal}{J. Fluid Mech.} \bibinfo{volume}{963}
  (\bibinfo{year}{2023}) \bibinfo{pages}{A33}.
\bibitem[{Zou et~al.(2024)Zou, Bi, Zhong, Yuan, and Tang}]{Zou2024}
\bibinfo{author}{S.~Zou}, \bibinfo{author}{L.~Bi}, \bibinfo{author}{C.~Zhong},
  \bibinfo{author}{X.~Yuan}, \bibinfo{author}{Z.~Tang},
\newblock \bibinfo{title}{A normal-mode approach for high-speed rarefied plane
  {Couette} flow},
\newblock \bibinfo{journal}{Phys. Fluids} \bibinfo{volume}{36}
  (\bibinfo{year}{2024}) \bibinfo{pages}{014121}.
\bibitem[{Wu et~al.(2024)Wu, Bi, Zhao, Tang, Yuan, and Wen}]{WuC2024}
\bibinfo{author}{C.~Wu}, \bibinfo{author}{L.~Bi}, \bibinfo{author}{J.~Zhao},
  \bibinfo{author}{Z.~Tang}, \bibinfo{author}{X.~Yuan},
  \bibinfo{author}{D.~Wen},
\newblock \bibinfo{title}{A novel linear stability method based on {Shakhov}
  bhatnagar--gross--{Krook} equations for {Rayleigh--B\'enard} convection in
  rarefied gases},
\newblock \bibinfo{journal}{Phys. Fluids} \bibinfo{volume}{36}
  (\bibinfo{year}{2024}) \bibinfo{pages}{124117}.
\bibitem[{Bi et~al.(2025)Bi, Ma, Gao, Lv, Wang, and Yuan}]{Bi2025}
\bibinfo{author}{L.~Bi}, \bibinfo{author}{Q.~Ma}, \bibinfo{author}{H.~Gao},
  \bibinfo{author}{J.~X. Lv}, \bibinfo{author}{H.~Wang}, \bibinfo{author}{X.~X.
  Yuan},
\newblock \bibinfo{title}{The relationship between {SBGK-LSE} and {NS-LSEs}
  under continuum assumption},
\newblock \bibinfo{journal}{Commun. Nonlinear Sci. Numer. Simul.}
  \bibinfo{volume}{143} (\bibinfo{year}{2025}) \bibinfo{pages}{108598}.
\bibitem[{Mack(1976)}]{Mack1976}
\bibinfo{author}{L.~M. Mack},
\newblock \bibinfo{title}{A numerical study of the temporal eigenvalue spectrum
  of the {Blasius} boundary layer},
\newblock \bibinfo{journal}{J. Fluid Mech.} \bibinfo{volume}{73}
  (\bibinfo{year}{1976}) \bibinfo{pages}{497--520}.
\bibitem[{Ma et~al.(2024)Ma, Lv, and Bi}]{Ma2024}
\bibinfo{author}{Q.~Ma}, \bibinfo{author}{J.~Lv}, \bibinfo{author}{L.~Bi},
\newblock \bibinfo{title}{The equivalence between {BE-LSE} and {NS-LSEs} under
  continuum assumption},
\newblock \bibinfo{journal}{Commun. Nonlinear Sci. Numer. Simul.}
  \bibinfo{volume}{128} (\bibinfo{year}{2024}) \bibinfo{pages}{107637}.
\bibitem[{Karpuzcu and Levin(2025)}]{KarpuzcuLevin2025}
\bibinfo{author}{I.~T. Karpuzcu}, \bibinfo{author}{D.~A. Levin},
  \bibinfo{title}{Linear stability analysis of canonical flows using
  {Boltzmann} {BGK} equation}, \bibinfo{howpublished}{AIAA Paper 2025-1313},
  \bibinfo{year}{2025}.
\bibitem[{Karpuzcu et~al.(2026)Karpuzcu, Levin, and
  Theofilis}]{KarpuzcuLevinTheofilis2026}
\bibinfo{author}{I.~T. Karpuzcu}, \bibinfo{author}{D.~A. Levin},
  \bibinfo{author}{V.~Theofilis}, \bibinfo{title}{Kinetic linear stability
  theory analysis of canonical high speed flows}, \bibinfo{howpublished}{AIAA
  Paper 2026-2143}, \bibinfo{year}{2026}.
\bibitem[{Aharoni et~al.(2026)Aharoni, Rovinsky, Karpuzcu, Burtsev, Theofilis,
  and Levin}]{Aharoni2026}
\bibinfo{author}{M.~Aharoni}, \bibinfo{author}{Y.~Rovinsky},
  \bibinfo{author}{I.~T. Karpuzcu}, \bibinfo{author}{A.~Burtsev},
  \bibinfo{author}{V.~Theofilis}, \bibinfo{author}{D.~A. Levin},
  \bibinfo{title}{Kinetic linear stability theory revisited. {Part II}: the
  {BGK} approximation of the {Boltzmann} equation}, \bibinfo{howpublished}{AIAA
  Paper 2026-4028}, \bibinfo{year}{2026}.
\bibitem[{Wu(2026)}]{Wu2025}
\bibinfo{author}{L.~Wu},
\newblock \bibinfo{title}{Efficient solutions of eigenvalue problems in
  rarefied gas flows},
\newblock \bibinfo{journal}{J. Comput. Phys.} \bibinfo{volume}{549}
  (\bibinfo{year}{2026}) \bibinfo{pages}{114607}.
\bibitem[{Reddy et~al.(1993)Reddy, Schmid, and Henningson}]{Reddy1993}
\bibinfo{author}{S.~C. Reddy}, \bibinfo{author}{P.~J. Schmid},
  \bibinfo{author}{D.~S. Henningson},
\newblock \bibinfo{title}{Pseudospectra of the {Orr--Sommerfeld} operator},
\newblock \bibinfo{journal}{SIAM J. Appl. Math.} \bibinfo{volume}{53}
  (\bibinfo{year}{1993}) \bibinfo{pages}{15--47}.
\bibitem[{Butler and Farrell(1992)}]{Butler1992}
\bibinfo{author}{K.~M. Butler}, \bibinfo{author}{B.~F. Farrell},
\newblock \bibinfo{title}{Three-dimensional optimal perturbations in viscous
  shear flow},
\newblock \bibinfo{journal}{Phys. Fluids A} \bibinfo{volume}{4}
  (\bibinfo{year}{1992}) \bibinfo{pages}{1637--1650}.
\bibitem[{Michalke(1964)}]{Michalke1964}
\bibinfo{author}{A.~Michalke},
\newblock \bibinfo{title}{On the inviscid instability of the hyperbolic-tangent
  velocity profile},
\newblock \bibinfo{journal}{J. Fluid Mech.} \bibinfo{volume}{19}
  (\bibinfo{year}{1964}) \bibinfo{pages}{543--556}.
\bibitem[{Woodbury(1950)}]{Woodbury1950}
\bibinfo{author}{M.~A. Woodbury}, \bibinfo{title}{Inverting modified matrices},
  \bibinfo{type}{Memorandum Report} \bibinfo{number}{42}, Statistical Research
  Group, Princeton University, \bibinfo{year}{1950}.
\bibitem[{Hager(1989)}]{Hager1989}
\bibinfo{author}{W.~W. Hager},
\newblock \bibinfo{title}{Updating the inverse of a matrix},
\newblock \bibinfo{journal}{SIAM Rev.} \bibinfo{volume}{31}
  (\bibinfo{year}{1989}) \bibinfo{pages}{221--239}.
\bibitem[{Gamba et~al.(2017)Gamba, Haack, Hauck, and Hu}]{GHHH2017}
\bibinfo{author}{I.~M. Gamba}, \bibinfo{author}{J.~R. Haack},
  \bibinfo{author}{C.~D. Hauck}, \bibinfo{author}{J.~Hu},
\newblock \bibinfo{title}{A fast spectral method for the {Boltzmann} collision
  operator with general collision kernels},
\newblock \bibinfo{journal}{SIAM J. Sci. Comput.} \bibinfo{volume}{39}
  (\bibinfo{year}{2017}) \bibinfo{pages}{B658--B674}.
\bibitem[{Chu(1965)}]{Chu1965}
\bibinfo{author}{B.-T. Chu},
\newblock \bibinfo{title}{On the energy transfer to small disturbances in fluid
  flow ({Part I})},
\newblock \bibinfo{journal}{Acta Mech.} \bibinfo{volume}{1}
  (\bibinfo{year}{1965}) \bibinfo{pages}{215--234}.
\bibitem[{Joseph(1976)}]{Joseph1976}
\bibinfo{author}{D.~D. Joseph}, \bibinfo{title}{Stability of Fluid Motions
  {I}}, volume~\bibinfo{volume}{27} of \textit{\bibinfo{series}{Springer Tracts
  in Natural Philosophy}}, \bibinfo{publisher}{Springer},
  \bibinfo{address}{Berlin}, \bibinfo{year}{1976}.
\bibitem[{Bhatnagar et~al.(1954)Bhatnagar, Gross, and Krook}]{BGK1954}
\bibinfo{author}{P.~L. Bhatnagar}, \bibinfo{author}{E.~P. Gross},
  \bibinfo{author}{M.~Krook},
\newblock \bibinfo{title}{A model for collision processes in gases. i. small
  amplitude processes in charged and neutral one-component systems},
\newblock \bibinfo{journal}{Phys. Rev.} \bibinfo{volume}{94}
  (\bibinfo{year}{1954}) \bibinfo{pages}{511--525}.
\bibitem[{Holway(1966)}]{Holway1966}
\bibinfo{author}{L.~H. Holway},
\newblock \bibinfo{title}{New statistical models for kinetic theory: methods of
  construction},
\newblock \bibinfo{journal}{Phys. Fluids} \bibinfo{volume}{9}
  (\bibinfo{year}{1966}) \bibinfo{pages}{1658--1673}.
\bibitem[{Andries et~al.(2000)Andries, {Le Tallec}, Perlat, and
  Perthame}]{Andries2000}
\bibinfo{author}{P.~Andries}, \bibinfo{author}{P.~{Le Tallec}},
  \bibinfo{author}{J.-P. Perlat}, \bibinfo{author}{B.~Perthame},
\newblock \bibinfo{title}{The {Gaussian-BGK} model of {Boltzmann} equation with
  small {Prandtl} number},
\newblock \bibinfo{journal}{Eur. J. Mech. B/Fluids} \bibinfo{volume}{19}
  (\bibinfo{year}{2000}) \bibinfo{pages}{813--830}.
\bibitem[{Shakhov(1968)}]{Shakhov1968}
\bibinfo{author}{E.~M. Shakhov},
\newblock \bibinfo{title}{Generalization of the {Krook} kinetic relaxation
  equation},
\newblock \bibinfo{journal}{Fluid Dyn.} \bibinfo{volume}{3}
  (\bibinfo{year}{1968}) \bibinfo{pages}{95--96}.
\bibitem[{Malik(1990)}]{Malik1990}
\bibinfo{author}{M.~R. Malik},
\newblock \bibinfo{title}{Numerical methods for hypersonic boundary layer
  stability},
\newblock \bibinfo{journal}{J. Comput. Phys.} \bibinfo{volume}{86}
  (\bibinfo{year}{1990}) \bibinfo{pages}{376--413}.
\bibitem[{Theofilis(1995)}]{Theofilis1995}
\bibinfo{author}{V.~Theofilis},
\newblock \bibinfo{title}{Spatial stability of incompressible attachment-line
  flow},
\newblock \bibinfo{journal}{Theor. Comput. Fluid Dyn.} \bibinfo{volume}{7}
  (\bibinfo{year}{1995}) \bibinfo{pages}{159--171}.
  \DOIprefix\doi{10.1007/BF00312360}.
\bibitem[{Carpenter et~al.(1994)Carpenter, Gottlieb, and
  Abarbanel}]{CarpenterGottliebAbarbanel1994}
\bibinfo{author}{M.~H. Carpenter}, \bibinfo{author}{D.~Gottlieb},
  \bibinfo{author}{S.~Abarbanel},
\newblock \bibinfo{title}{Time-stable boundary conditions for finite-difference
  schemes solving hyperbolic systems: methodology and application to high-order
  compact schemes},
\newblock \bibinfo{journal}{J. Comput. Phys.} \bibinfo{volume}{111}
  (\bibinfo{year}{1994}) \bibinfo{pages}{220--236}.
\bibitem[{Darroz\`es and Guiraud(1966)}]{DarrozesGuiraud1966}
\bibinfo{author}{J.~S. Darroz\`es}, \bibinfo{author}{J.~P. Guiraud},
\newblock \bibinfo{title}{G\'en\'eralisation formelle du th\'eor\`eme {H} en
  pr\'esence de parois. {Applications}},
\newblock \bibinfo{journal}{C. R. Acad. Sci. Paris A} \bibinfo{volume}{262}
  (\bibinfo{year}{1966}) \bibinfo{pages}{1368--1371}.
\bibitem[{Luchini and Bottaro(2014)}]{LuchiniBottaro2014}
\bibinfo{author}{P.~Luchini}, \bibinfo{author}{A.~Bottaro},
\newblock \bibinfo{title}{Adjoint equations in stability analysis},
\newblock \bibinfo{journal}{Annu. Rev. Fluid Mech.} \bibinfo{volume}{46}
  (\bibinfo{year}{2014}) \bibinfo{pages}{493--517}.
\bibitem[{van~den Eshof and Hochbruck(2006)}]{vandenEshofHochbruck2006}
\bibinfo{author}{J.~van~den Eshof}, \bibinfo{author}{M.~Hochbruck},
\newblock \bibinfo{title}{Preconditioning {L}anczos approximations to the
  matrix exponential},
\newblock \bibinfo{journal}{SIAM J. Sci. Comput.} \bibinfo{volume}{27}
  (\bibinfo{year}{2006}) \bibinfo{pages}{1438--1457}.
\bibitem[{Wright and Trefethen(2001)}]{Wright2001}
\bibinfo{author}{T.~G. Wright}, \bibinfo{author}{L.~N. Trefethen},
\newblock \bibinfo{title}{Large-scale computation of pseudospectra using
  {ARPACK} and eigs},
\newblock \bibinfo{journal}{SIAM J. Sci. Comput.} \bibinfo{volume}{23}
  (\bibinfo{year}{2001}) \bibinfo{pages}{591--605}.

\end{thebibliography}

\end{document}